\documentclass[onecolumn,english]{IEEEtran}
\usepackage[T1]{fontenc}
\usepackage[latin9]{inputenc}
\usepackage{array}
\usepackage{multirow}
\usepackage{amsmath}
\usepackage{amssymb}
\usepackage{graphicx}
\usepackage{esint}
\usepackage{cite}

\makeatletter

\providecommand{\tabularnewline}{\\}

\newtheorem{thm}{\protect\theoremname}
\newtheorem{defn}[thm]{\protect\definitionname}
\newtheorem{prop}[thm]{\protect\propositionname}
\newtheorem{rem}[thm]{\protect\remarkname}
\newtheorem{cor}[thm]{\protect\corollaryname}

\usepackage{mathrsfs}
\usepackage{algorithm,algpseudocode}

\usepackage{colortbl}
\definecolor{lightgray}{rgb}{0.9,0.9,0.9}
\definecolor{lightred}{rgb}{1,0.8,0.8}
\definecolor{lightgreen}{rgb}{0.8,1,0.8}
\definecolor{lightyellow}{rgb}{1,0.95,0.65}
\definecolor{lightorange}{rgb}{1,0.85,0.75}
\definecolor{lightgrey}{rgb}{0.8,0.8,0.8}

\allowdisplaybreaks[1]

\makeatother

\usepackage{babel}
\providecommand{\corollaryname}{Corollary}
\providecommand{\definitionname}{Definition}
\providecommand{\propositionname}{Proposition}
\providecommand{\remarkname}{Remark}
\providecommand{\theoremname}{Theorem}

\begin{document}
\title{Discrete Layered Entropy, Conditional Compression and a Tighter Strong Functional Representation Lemma}
\author{
\IEEEauthorblockN{Cheuk Ting Li}\\
\IEEEauthorblockA{Department of Information Engineering, The Chinese University of Hong Kong, Hong Kong, China \\
Email: ctli@ie.cuhk.edu.hk}
\thanks{This work was partially supported by two grants from the Research Grants Council of the Hong Kong Special Administrative Region, China [Project No.s: CUHK 24205621 (ECS), CUHK 14209823 (GRF)].

This paper was presented in part at the 2025 IEEE International Symposium on Information Theory (ISIT).

}}
\maketitle
\begin{abstract}
We study a quantity called discrete layered entropy, which approximates the Shannon entropy within a logarithmic gap. Compared to the Shannon entropy, the discrete layered entropy is piecewise linear, approximates the expected length of the optimal one-to-one non-prefix code, and satisfies an elegant conditioning property. These properties make it useful for approximating the Shannon entropy in linear programming and maximum entropy problems, studying the optimal length of conditional encoding, and bounding the entropy of monotonic mixture distributions. In particular, it can give a bound $I(X;Y)+\log(I(X;Y)+3.4)+1$ for the strong functional representation lemma which is optimal within $2.8$ bits, and significantly improves upon the best known bound.
\end{abstract}


\medskip{}

\allowdisplaybreaks

\section{Introduction}

In this paper, which extends \cite{li2025discrete},\footnote{The conference version \cite{li2025discrete} does not include most of the results and bounds for channel simulation (Section \ref{sec:chansim}). It does not discuss the properties of the discrete $m$-layered entropy (Section \ref{sec:mlayered}), the discrete R\'{e}nyi layered entropy (Section \ref{sec:renyi}), and the parallels between vertical and horizontal quantities (Section \ref{sec:horizontal}).} we study a quantity which we call the \emph{discrete layered entropy}
\begin{align}
\Lambda(p) &:=\sum_{i=1}^{\infty}p^{\downarrow}(i)\left(i\log i-(i-1)\log(i-1)\right), \label{eq:disclayered_intro}
\end{align}
where $p$ is a probability mass function, and $p^{\downarrow}(i)$ is the $i$-th entry of $(p(x))_{x\in\mathcal{X}}$ when sorted in descending order. 
This quantity first appeared in a lower bound of the $N$-gram entropy of the English language in a work by Shannon~\cite{shannon1951prediction}, which also showed that \eqref{eq:disclayered_intro} is a lower bound of the entropy $H(p)$ and is Schur concave.\footnote{Shannon~\cite{shannon1951prediction} used \eqref{eq:disclayered_intro} to lower bound the entropy of English in terms of the ideal predictor of the next letter given the previous $N-1$ letters. He showed two properties of \eqref{eq:disclayered_intro}: lower-bounding the entropy, and Schur concavity. We use a descriptive name ``discrete layered entropy'' instead of calling it ``Shannon lower bound'' since the latter applies specifically to the conditional distribution of the next English letter given the previous letters, and has not been considered as an entropy measure of a general non-conditional distribution.}
Also refer to \cite{savchuk1964evaluation,maixner1971some,cover1978convergent} for comments on the accuracy of this lower bound.
Moreover, 
this is a discrete analogue of the (continuous) layered entropy studied in \cite{hegazy2022randomized,ling2024rejection}. Refer to Figure \ref{fig:layered} for a plot. Compared to the Shannon entropy $H(X)$, $\Lambda(X)$ has the following properties:

\begin{figure*}
\begin{centering}
\includegraphics[scale=0.4]{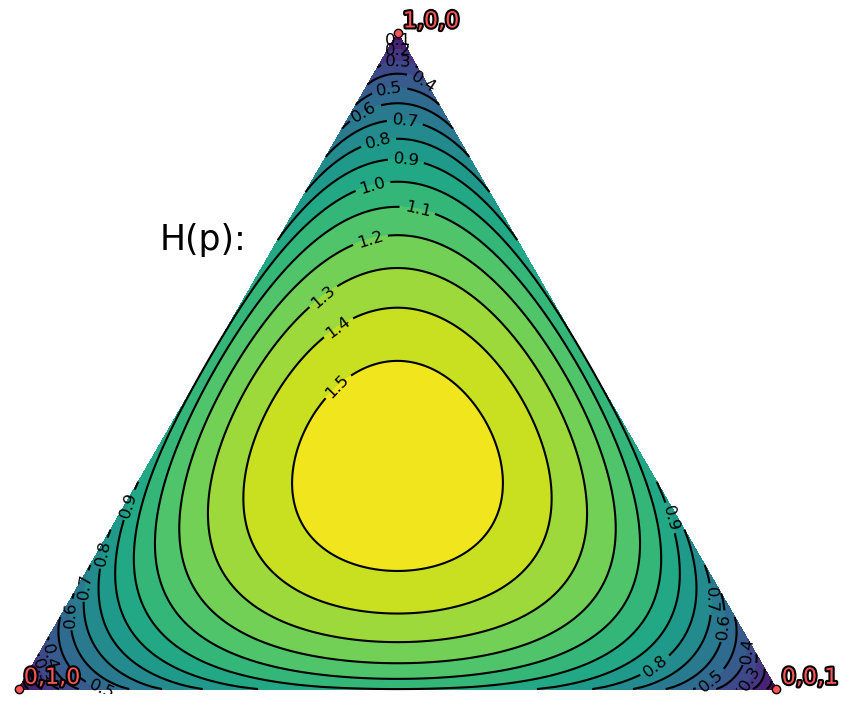}$\;\;\;$\includegraphics[scale=0.4]{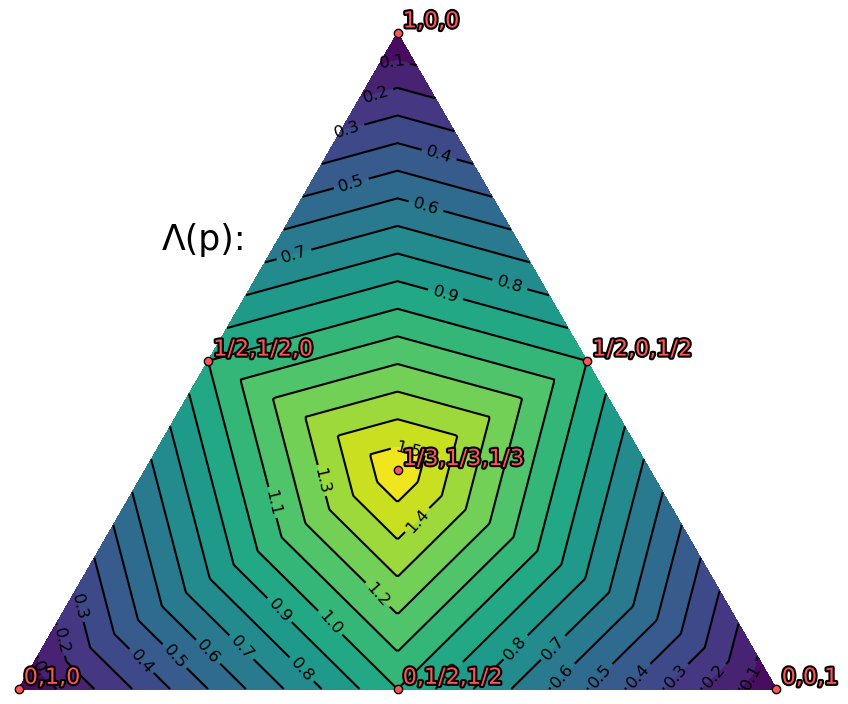}
\par\end{centering}
\caption{\label{fig:layered}Left: Contour plot of the Shannon entropy $H(p)$ for $p:\{1,2,3\}\to[0,1]$ being a ternary probability mass function. Right: Contour plot of the discrete layered entropy $\Lambda(p)$. We can see that $\Lambda(p)$ is piecewise linear and lower-bounds $H(p)$. The red points are the points where $\Lambda(p)=H(p)$. They are also the vertices of the polytope $\{(p,z)\in\mathbb{R}^{3}\times\mathbb{R}:\,0\le z\le\Lambda(p)\}$.}
\end{figure*}

\begin{itemize}
\item $\Lambda(X)$ also satisfies some basic properties of entropy, such as concavity, and $\Lambda(X)\le\log|\mathcal{X}|$ with equality if and only if $X$ is uniform. For some other properties of $H(X)$ that are not exactly satisfied by $\Lambda(X)$, relaxed versions of those properties are sometimes satisfied by $\Lambda(X)$. For example, for independent $X,Y$, the additivity property of $H$ (i.e., $H(X,Y)=H(X)+H(Y)$) is relaxed to the superadditivity property of $\Lambda$ (i.e., $\Lambda(X,Y)\ge\Lambda(X)+\Lambda(Y)$). For general $X,Y$, the subadditivity property of $H$ (i.e., $H(X,Y)\le H(X)+H(Y)$) is relaxed to the ``mixed-subadditivity'' of $\Lambda$ (i.e., $\Lambda(X,Y)\le\Lambda(X)+H(Y)$).
\item $\Lambda$ satisfies the conditioning property $\Lambda(X|Y)=\min_{U:H(X|Y,U)=0}\Lambda(U)$, making it useful for conditional encoding tasks. This is not satisfied by $H$. This will be discussed later in this section and Section \ref{sec:threecond}.
\item $\Lambda(X)$ lower-bounds $H(X)$ \cite{shannon1951prediction}, and is close to $H(X)$ within a logarithmic gap. This, together with the conditioning property, allows the derivation of a bound for the strong functional representation lemma \cite{sfrl_trans} that significantly improves upon the best known bound, to be discussed later in this section and Section \ref{sec:chansim}.
\item $H(X)$ approximates the expected length of the optimal prefix code for $X$, and hence serves as a convenient tool for analyzing prefix codes; whereas $\Lambda(X)$ approximates the expected length of the optimal one-to-one \emph{non-prefix} code \cite{alon1994lower,blundo1996new,szpankowski2011minimum} for $X$, and hence serves as a convenient tool for analyzing non-prefix codes. See Section \ref{sec:onetoone}.
\item $\Lambda$ is piecewise linear (unlike $H$), and can be incorporated into a linear program. For a maximum entropy problem where we maximize $H(p)$ for a distribution $p$ subject to linear constraints \cite{jaynes1957information,snickars1977minimum}, approximating $H(p)$ by $\Lambda(p)$ converts the problem into a linear program.   See Sections \ref{sec:piecewise_approx} and \ref{sec:mlayered}. 
\end{itemize}
\begin{table}
\caption{\label{tab:compare}Comparison between Shannon entropy and discrete layered entropy.}

\centering{}{\renewcommand*{\arraystretch}{1.5}%
\begin{tabular}{|c|c|}
\hline 
\textbf{Shannon entropy $H$} & \textbf{Discrete layered entropy $\Lambda$}\tabularnewline
\hline 
\multicolumn{2}{|c|}{Concave, Schur concave}\tabularnewline
\hline 
\multicolumn{2}{|c|}{$\Lambda(X)\le\log|\mathcal{X}|$, equality iff uniform (same for $H$)}\tabularnewline
\hline 
\multicolumn{2}{|c|}{$\Lambda(X)\le H(X)\le\Lambda(X)+\log\left(1+\frac{\Lambda(X)}{e\log e}\right)+\log e$}\tabularnewline
\hline 
Length of best prefix code & $\!\!$Length of best non-prefix code$\!\!$\tabularnewline
$\in[H(X),H(X)+1)$ & $\in(\Lambda(X)-2,\Lambda(X)]$\tabularnewline
\hline 
Not piecewise linear & Piecewise linear\tabularnewline
\hline 
$\!\!$$H(X,Y)=H(X|Y)+H(Y)$$\!\!$ & $\begin{array}{c}
\Lambda(X,Y)\le\Lambda(X|Y)+H(Y)\\
\Lambda(X,Y)\le H(X\backslash Y)+\Lambda(Y)
\end{array}$\tabularnewline
\hline 
$H(X,Y)\le H(X)+H(Y)$ & $\Lambda(X,Y)\le\Lambda(X)+H(Y)$\tabularnewline
\hline 
$H(X,Y)=H(X)+H(Y)$  & $\Lambda(X,Y)\ge\Lambda(X)+\Lambda(Y)$ \tabularnewline
if $X\perp\!\!\!\perp Y$ & if $X\perp\!\!\!\perp Y$\tabularnewline
\hline 
No conditioning property & Conditioning property\tabularnewline
$H(X\backslash Y)\ge H(X|Y)$ & $\Lambda(X\backslash Y)=\Lambda(X|Y)$\tabularnewline
\hline 
\end{tabular}}
\end{table}

\medskip{}

\subsection*{Conditional Compression}

\begin{figure}
\begin{centering}
\includegraphics[scale=0.95]{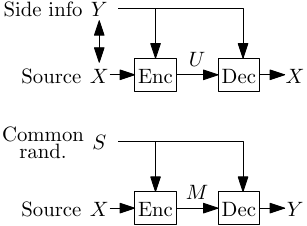}
\par\end{centering}
\caption{\label{fig:intro}Top: The conditional encoding setting. Bottom: The one-shot channel simulation setting.}
\end{figure}

We briefly discuss the conditional encoding setting where $\Lambda$ serves as a useful tool. Suppose the encoder wants to encode $X$, given the side information $Y$ that is also known to the decoder (i.e., the encoder encodes $X$ conditionally given $Y$). One method is to have the encoder find the ranking $U\in\mathbb{N}$ of $X$ among $p(x|Y)$ for all $x$ (i.e., $X$ has the $U$-th largest $p(x|Y)$), and sends $U$ to the decoder. Refer to Figure \ref{fig:intro}. This $U$ also appears as the number of guesses in the guessing problem \cite{massey1994guessing,arikan1996inequality,kosut2017asymptotics}, where we guess the value of $X$ one by one in descending order of $p(x|Y)$ given the side information $Y$ until we correctly guess $X$. This $U$ can be shown to yield the smallest possible compression, in the sense that it minimizes $H(U)$ (which is approximately the compression size using a prefix code)\footnote{If we are allowed to encode $U$ conditional on $Y$ using a conditional prefix code, then the compression size would be $\approx H(U|Y)$. Nevertheless, if we require synchronization even when the decoder does not know $Y$, then we must use an unconditional prefix code, which has a length $\approx H(U)$. Refer to Section \ref{sec:cond_varlen} for discussions.} subject to the recoverability requirement $H(X|Y,U)=0$. Hence, this $U$ can be regarded as the ``conditional compression'' of $X$ given $Y$, which contains the information in $X$ that is not in $Y$, and hence we use the notation $U=X\backslash Y$ to highlight the analogy to set difference. Refer to Definition \ref{def:cond} for a formal definition of $X\backslash Y$.

It is perhaps unsatisfying that the entropy of the conditional compression $H(X\backslash Y)$ is generally not the conditional entropy $H(X|Y)$ (we only have $H(X\backslash Y)\ge H(X|Y)$).\footnote{For example, if $Y\sim\mathrm{Unif}(\{0,1/3\})$, $X|\{Y=y\}\sim\mathrm{Bern}(y)$, then we always have $U=X+1$ since $p_{X|Y}(0|y)>p_{X|Y}(1|y)$ for every $y$, and hence $H(X\backslash Y)=H(X)>H(X|Y)$.} The conditional entropy $H(X|Y)$ does not actually correspond to the entropy of any random variable---it cannot be interpreted as the entropy $H(\cdot)$ applied to a ``conditional random variable $X|Y$''. This is in stark contrast to the joint entropy $H(X,Y)$, which is indeed $H(\cdot)$ applied to the joint random variable $(X,Y)$.

We show that $\Lambda$ satisfies the following \emph{conditioning property}:
\begin{equation}
\Lambda(X\backslash Y)=\Lambda(X|Y),\label{eq:cond_intro}
\end{equation}
where $\Lambda(X|Y):=\mathbb{E}_{Y}[\Lambda(p_{X|Y}(\cdot|Y))]$ is defined in a similar manner as $H(X|Y)$. This fact is not only aesthetically pleasing (we can indeed treat ``$X|Y$'' as a random variable given by $X\backslash Y$), but also has several implications. First, this allows us to approximate the optimal expected encoding length of the aforementioned conditional encoding task by $\Lambda(X|Y)$. Second, it makes $\Lambda(X)$ a useful tool for bounding the entropy of a mixture of distributions with monotonic probability mass functions, since \eqref{eq:cond_intro} is equivalent to saying that $\Lambda(X)$ is a linear function of the probability mass function $p_{X}$ when we restrict $p_{X}:\mathbb{N}\to\mathbb{R}$ to be a nondecreasing function. 

Moreover, we show that \eqref{eq:cond_intro} is the defining property of $\Lambda$, in the sense that the discrete layered entropy is the only function satisfying \eqref{eq:cond_intro} and $\Lambda(X)=\log k$ for $X\sim\mathrm{Unif}(\{1,\ldots,k\})$. Also, $\Lambda$ is the largest function satisfying \eqref{eq:cond_intro} and $\Lambda(X)\le H(X)$, and hence $\Lambda(X)$ is the ``best underestimate'' of $H(X)$ that satisfies \eqref{eq:cond_intro}. We show that $\Lambda(X)$ is close to $H(X)$, in the sense that
\begin{equation}
\Lambda(X)\le H(X)\le\Lambda(X)+\log\left(1+\frac{\Lambda(X)}{e\log e}\right)+\log e.\label{eq:approx_intro}
\end{equation}
We can convert between $H(X)$ and $\Lambda(X)$ depending on whether we need the properties of $H$ or $\Lambda$, incurring only a logarithmic gap for each conversion. This makes $\Lambda(X)$ a powerful tool even if we only want our final result to be in terms of the Shannon entropy. For example, for the conditional encoding task, we can approximate the optimal encoding length $H(X\backslash Y)$ using \eqref{eq:cond_intro} and \eqref{eq:approx_intro} via $H(X\backslash Y)\approx\Lambda(X\backslash Y)=\Lambda(X|Y)\approx H(X|Y)$.

\medskip{}

\subsection*{One-shot Channel Simulation}

The conditional encoding task is also the final step in one-shot channel simulation with unlimited common randomness \cite{bennett2002entanglement,harsha2010communication}, where an encoder observes a random source $X$ and sends a variable-length description $M$ (in a prefix or non-prefix code) to the decoder, in order to allow the decoder to output $Y$ which must follow the target conditional distribution $P_{Y|X}$ given $X$. We also allow the encoder and the decoder to share an arbitrary common randomness $S$ that is independent of $X$. Refer to Figure \ref{fig:intro}. The usual strategy \cite{harsha2010communication,sfrl_trans} is to have the encoder generate $Y$ dependent on $(X,S)$, and then conditionally encode $Y$ given $S$ into $M$, using approximately $H(Y|S)$ bits of communication. We can use the conditioning property of $\Lambda$ to give a tighter bound for this conditional encoding task. In particular, we show the following strengthened version of the strong functional representation lemma \cite{sfrl_trans}: for every $X,Y$, there exists $S$ independent of $X$ with $H(Y|X,S)=0$ and 
\begin{equation}
\Lambda(Y|S)<I(X;Y)+1.29,
\end{equation}
and hence, due to the fact that $\Lambda(Y|S)$ is close to $H(Y|S)$ within a logarithmic gap (Proposition \ref{prop:LX_HX_bound}),
\begin{equation}
H(Y|S)<I(X;Y)+\log(I(X;Y)+3.4)+1.\label{eq:sfrl_intro}
\end{equation}
The bound \eqref{eq:sfrl_intro} significantly improves upon previous results \cite{sfrl_trans,li2021unified,li2024pointwise}, and is optimal within $2.8$ bits (see Theorem \ref{thm:Hc_lb_continuous}). Also, note that the bound on $\Lambda(Y|S)$ (corresponding to non-prefix encoding of $M$) is simpler than that on $H(Y|S)$ (corresponding to prefix encoding), suggesting that $\Lambda(Y|S)$ and non-prefix codes might be more natural in this setting. Refer to Section \ref{sec:chansim}.
The stronger bound \eqref{eq:sfrl_intro} has been applied in \cite{zamani2026multi} to prove an extended strong functional representation lemma which is applicable to semantic communication with privacy constraints.

\medskip{}

\subsection*{Related Works on Differences between Information}

Apart from the ``$X\backslash Y$'' (which minimizes $H(U)$ subject to $H(X|Y,U)=0$) studied in this paper, there are several notions of differences between the information in $X$ and the information in $Y$ studied in the literature.

Slepian-Wolf coding \cite{slepian1973noiseless} concerns the setting where the encoder compresses a discrete memoryless source $X^{n}$ into a message $M$, such that the decoder who observes $M$ and a side information $Y^{n}$ (where $(X_{i},Y_{i})\sim p_{X,Y}$ are i.i.d., i.e., $(X^{n},Y^{n})$ is a $2$-discrete memoryless source) can recover $X^{n}$ with vanishing error probability as $n\to\infty$.\footnote{More generally, \cite{slepian1973noiseless} studies the setting where $Y^{n}$ is communicated to the decoder by another encoder, and the decoder must recover both $X^{n}$ and $Y^{n}$.} Intuitively, $M$ should be the information in $X^{n}$ that is not contained in $Y^{n}$. The optimal compression rate is $H(X|Y)$. The conditional compression setting in the introduction is different from Slepian-Wolf, in the sense that the conditional compression setting is one-shot ($n=1$), requires zero error probability, but allows the encoder to observe $Y$. Due to the one-shot nature of the setting, the compression size is generally larger than $H(X|Y)$. Also refer to \cite{el1984interactive} for a one-shot interactive setting where two terminals which hold $X$ and $Y$, respectively, carry out interactive communication in order to allow both of them to decode $X$ and $Y$.

The guessing problem \cite{massey1994guessing,mceliece1995inequality,arikan1996inequality} concerns the setting where there are two dependent random variables $X,Y$, and a player observing $Y$ who attempts to guess the value of $X$ by asking whether $X=x$ until the player correctly guesses $X$. Bounds on $\mathbb{E}[U^{\rho}]$, where $U$ is the number of guesses needed, can be given in terms of the conditional R\'{e}nyi entropy \cite{renyi1961measures,arimoto1977information}, as shown in \cite{arikan1996inequality}. The optimal $U$ is the ranking of $X$ among $p(x|Y)$ for all $x$ \cite{massey1994guessing,arikan1996inequality}, coinciding with the random variable $U$ in the conditional compression setting. Hence, the conditional compression setting with one-to-one codes \cite{alon1994lower,blundo1996new,szpankowski2011minimum} can be regarded as a guessing experiment where $\mathbb{E}[\lfloor\log U\rfloor]$ is minimized. This connection has been observed in the context of (unconditional) universal fixed-to-variable source coding \cite{kosut2017asymptotics}. See Sections \ref{sec:onetoone} and \ref{sec:renyi}. 

The task encoding problem with side information \cite{bunte2014encoding} concerns the setting where, given $(X,Y)\sim p_{X,Y}$, we would like to find an encoding function $f:\mathcal{X}\times\mathcal{Y}\to[k]$ such that $\mathbb{E}[|\{x:f(x,Y)=f(X,Y)\}|^{\rho}]$ is minimized. The function $f$ can be understood as a conditional partition of a set of tasks $\mathcal{X}$ into groups given $Y$, such that the expected $\rho$-th power of the size of a random group is minimized. Bounds can also be given in terms of the conditional R\'{e}nyi entropy \cite{bunte2014encoding}. A distributed task encoding problem has also been studied \cite{bracher2017distributed}.

Strong functional representation lemma \cite{sfrl_trans} seeks to minimize $H(X|S)$ subject to $S\perp\!\!\!\perp Y$ and $H(X|Y,S)=0$. These constraints are analogous to set differences $A\backslash B$ between sets $A,B$, which satisfies $(A\backslash B)\cap B=\emptyset$ and $A\subseteq(B\cup(A\backslash B))$. Note that $H(S)$ can be much larger than $H(X|Y)$. To review various bounds on $H(X|S)$, the result in \cite{harsha2010communication} implies that $H(X|S)\le I+(1+o(1))\log(I+1)$ can be achieved, where $I:=I(X;Y)$. \cite{braverman2014public} improved the bound to $I+\log(I+1)+O(1)$; \cite{sfrl_trans} improved it to $I+\log(I+1)+3.870$; \cite{li2021unified} improved it to $I+\log(I+1)+3.732$; and \cite{li2024pointwise} improved it to $I+\log(I+2)+2$.\footnote{After the publication of \cite{li2024pointwise,li2024channel}, the author learned through a private communication with Serhat Emre Coban that this bound has also been discussed by Emre Telatar during a research meeting.} A recent preprint \cite{hill2026rejection} shows a bound $I + \log(I+1)+2.45$. See \cite{li2024channel} for a review. In Theorem \ref{thm:l_sfrl}, we show a new bound $I+\log(I+3.4)+1$, which improves upon previous results.\footnote{Our new bound $I+\log(I+3.4)+1$ improves upon all known bounds for all $I$, except the bound $I + \log(I+1)+2.45$ in the preprint \cite{hill2026rejection}, where our bound is better for $I \ge 0.4$.}  See Figure \ref{fig:sfrl_plot}. This improved bound shows the usefulness of $\Lambda(X)$ as a technical tool.

Notions that can be interpreted as difference have been studied systematically in \cite{li2017extended}. It was argued that apart from the strong functional representation lemma, the K\"{o}rner graph entropy \cite{korner1971coding} and the maximum rate for perfect privacy \cite{makhdoumi2014information} can be treated as the difference between $X$ and $Y$. The minimum entropy coupling \cite{vidyasagar2012metric,kocaoglu2017entropic,cicalese2019minimum,li2021efficient} has also been related to the difference between information.

The set corresponding to the cell ``$X\backslash Y$'' in the I-measure \cite{yeung1991new} has been studied in \cite{down2023logarithmic,li2024decompositionisit}, though this ``$X\backslash Y$'' is not actually a random variable.

\subsection*{Notations}

Throughout this paper, all random variables are assumed to be discrete (with finite or countable support) unless otherwise stated. Entropy is in bits, and $\log$ is to the base $2$. Write $\mathbb{N}:=\{1,2,\ldots\}$, $[a:b]:=\{a,\ldots,b\}$, $[n]:=\{1,\ldots,n\}$ (let $[\infty]:=\mathbb{N}$). For a random variable $X$, write $\mathcal{X}$ for the set it lies in, $p_{X}$ for its probability mass function (pmf), and $P_{X}$ for its distribution (when $X$ can be a discrete, continuous or general random variable). The condition that $X,Y$ are independent is denoted as $X\perp\!\!\!\perp Y$. For random variables $X,Y$, we say that they are (informationally) equivalent, denoted as $X\stackrel{\iota}{=}Y$, if $H(X|Y)=H(Y|X)=0$. Denote the constant random variable as $\emptyset$ (so $X\stackrel{\iota}{=}\emptyset$ means that $X$ is a constant). Write $X^{n}=(X_{1},\ldots,X_{n})$. The min-entropy is defined as $H_{\infty}(X):=-\log\max_{x}p_{X}(x)$.

For a pmf $p$ over $\mathcal{X}$, write $p^{\downarrow}$ for a pmf over $\mathbb{N}$, where $p^{\downarrow}(i)$ is the $i$-th entry of $(p(x))_{x\in\mathcal{X}}$ when sorted in descending order ($p^{\downarrow}(i)=0$ if $i>|\mathcal{X}|$). Given pmfs $p,q$, we say that $p$ majorizes $q$, written as $p\succeq q$, if $\sum_{i=1}^{k}p^{\downarrow}(i)\ge\sum_{i=1}^{k}q^{\downarrow}(i)$ for every $k\in\mathbb{N}$ \cite{marshall1979inequalities}. A function $\Lambda$ mapping pmfs to real numbers is Schur concave if $p\succeq q$ implies $\Lambda(p)\le\Lambda(q)$. For example, Shannon entropy is Schur concave.

\section{Discrete Layered Entropy}

We now define the central quantity of this paper.
\begin{defn}
[Discrete layered entropy] The \emph{discrete layered entropy} of a probability mass function $p$ is defined as
\begin{equation}
\Lambda(p):=\sum_{i=1}^{\infty}p^{\downarrow}(i)\left(i\log i-(i-1)\log(i-1)\right).\label{eq:LX_alt}
\end{equation}
Recall that $p^{\downarrow}(i)$ is the $i$-th entry of $(p(x))_{x\in\mathcal{X}}$ when sorted in descending order, and we assume $0\log0=0$. We write $\Lambda(X):=\Lambda(p_{X})$. The \emph{conditional discrete layered entropy} of a random variable $X$ given another random variable $Y$ is
\[
\Lambda(X|Y):=\mathbb{E}_{Y}\left[\Lambda(p_{X|Y}(\cdot|Y))\right].
\]
\end{defn}
We use the name ``discrete layered entropy'' since $\Lambda(p)$ is the discrete analogue of the (continuous) layered entropy studied in \cite{hegazy2022randomized,ling2024rejection}, which is shown in \eqref{eq:layered_def} and Remark \ref{rem:continuous}.\footnote{The notation ``$\Lambda(p)$'' is chosen for two reasons---both ``layered'' and ``lambda'' start with ``la'', and the shape ``$\Lambda$'' mimics the shape of the function $\Lambda(p)$ (a piecewise-linear concave function) when $p$ is binary.} The layered entropy is also related to the lower bound of the excess functional information \cite{sfrl_trans}, the channel simulation divergence \cite{goc2024causal,flamich2025redundancy} (see Section \ref{sec:chansim_lb}), and the cumulative residual entropy \cite{rao2004cumulative}.\footnote{The cumulative residual entropy \cite{rao2004cumulative} $\mathcal{E}(Y) := -\int_0^{\infty} \mathbb{P}(|Y|>t) \log \mathbb{P}(|Y|>t) \mathrm{d}t$ (for scalar $Y$) is related to $\Lambda(X)$, in the sense that $\Lambda(X) = \log|\mathcal{X}|-|\mathcal{X}| \mathcal{E}(p_X(\bar{X}))$ where $\bar{X} \sim \mathrm{Unif}(\mathcal{X})$, due to \eqref{eq:layered_def}. Nevertheless, since $\mathcal{E}(Y)$ concerns the magnitude of $Y$, its general properties are considerably different from $\Lambda(X)$ which is invariant under relabeling.} We give several alternative definitions for $\Lambda(p)$. The proof is in Appendix \ref{subsec:pf_alt_def}.

\medskip{}

\begin{prop}
[Alternative definitions]\label{prop:alt_def} We have
\begin{enumerate}
\item (Integral form)
\begin{equation}
\Lambda(p)=\int_{0}^{\infty}p^{\downarrow}(\lceil t\rceil)\log(et)\mathrm{d}t.\label{eq:integral_def}
\end{equation}
\item (Layered form)
\begin{equation}
\Lambda(p)=\int_{0}^{1}|\{x:\,p(x)>t\}|\cdot\log|\{x:\,p(x)>t\}|\mathrm{d}t.\label{eq:layered_def}
\end{equation}
\item (Concave envelope of min-entropy)
\begin{equation}
\Lambda(X)=\max_{p_{Y|X}}H_{\infty}(X|Y),\label{eq:minent_def}
\end{equation}
where $H_{\infty}(X|Y):=\mathbb{E}_{Y}[-\log\max_{x}p_{X|Y}(x|Y)]$ is the conditional min-entropy. In other words, $\Lambda$ is the upper concave envelope of $H_{\infty}$.\footnote{This means $\Lambda(p)=\mathrm{min}_{\tilde{\Lambda}}\tilde{\Lambda}(p)$ where the minimum is over concave functions $\tilde{\Lambda}$ satisfying $\tilde{\Lambda}(p)\ge H_{\infty}(p)$ for all $p$.}
\item (Concave envelope of log cardinality)
\begin{equation}
\Lambda(X)=\max H(X|Y),\label{eq:alt_cond}
\end{equation}
where the maximum is over $p_{Y|X}$ such that $p_{X|Y}(\cdot|y)$ is a uniform distribution for every $y$, i.e., $p_{X|Y}(x_{1}|y)=p_{X|Y}(x_{2}|y)$ for every $x_{1},x_{2},y$ such that $p_{X|Y}(x_{1}|y),p_{X|Y}(x_{2}|y)>0$. Equivalently,
\[
\Lambda(X)=\max\mathbb{E}[\log|\mathcal{A}|],
\]
where the maximum is over random sets $\mathcal{A}\subseteq\mathcal{X}$ jointly distributed with $X$ such that $X|\mathcal{A}\sim\mathrm{Unif}(\mathcal{A})$ (i.e., conditional on $\mathcal{A}=a$, $X$ is uniform over $a$).\footnote{This is the discrete analogue of the alternative definition of continuous layered entropy in \cite{ling2024rejection}. Equivalently, $\Lambda(p)=\mathrm{min}_{\tilde{\Lambda}}\tilde{\Lambda}(p)$, where the minimum is over all concave functions $\tilde{\Lambda}$  satisfying $\tilde{\Lambda}(X)\ge\log|\mathcal{X}|$ when $X$ is uniformly distributed over $|\mathcal{X}|$. }
\item (Linear programming form)
\begin{equation}
\Lambda(X)=\max\mathbb{E}[\log K],\label{eq:lp_def}
\end{equation}
where the maximum is over joint probability mass functions $p_{X,K}$ over $\mathcal{X}\times[|\mathcal{X}|]$ with an $X$-marginal that coincides with $p_{X}$, and satisfying that $p_{X,K}(x,k)\le p_{K}(k)/k$ for all $x,k$, where $p_{K}(k)$ is the $K$-marginal of $p_{X,K}$. This means $\Lambda(X)$ can be formulated as a maximization in a linear program when $\mathcal{X}$ is finite.
\end{enumerate}
\end{prop}
\medskip{}

Another alternative definition of $\Lambda(X)$ will be given in Theorem \ref{thm:useful_def}. We then show some basic properties of $\Lambda(X)$, some of which has appeared in \cite{shannon1951prediction}. The proof is given in Appendix \ref{subsec:pf_basic} for the sake of completeness.

\medskip{}

\begin{prop}
[Basic properties] \label{prop:basic}We have, for every random variables $X\in\mathcal{X}$, $Y\in\mathcal{Y}$,
\begin{enumerate}
\item $H_{\infty}(X)\le\Lambda(X)\le H(X)$. For each inequality, equality holds if and only if $X$ is uniformly distributed. ($\Lambda(X)\le H(X)$ was proved in \cite{shannon1951prediction}.)
\item (Concavity) $\Lambda(p)$ is a concave function over probability mass functions $p$. Equivalently, $\Lambda(X|Y)\le\Lambda(X)$.
\item (Schur concavity) $\Lambda(p)$ is Schur concave, and hence $\Lambda(Y)\le\Lambda(X)$ if $H(Y|X)=0$. (This was proved in \cite{shannon1951prediction}.)
\item (Monotone linearity) $\Lambda(p)$ is a linear function over the convex space of nondecreasing probability mass functions $p:\mathbb{N}\to\mathbb{R}$. Equivalently, if $X\in\mathbb{N}$, and for every fixed $y$, $x\mapsto p_{X|Y}(x|y)$ is a nondecreasing function, then $\Lambda(X|Y)=\Lambda(X)$. (This was used implicitly in \cite{shannon1951prediction}. Overall, $\Lambda(p)$ is a piecewise linear function over the space of not-necessarily-nondecreasing probability mass functions.)
\item (Superadditivity) If $X,Y$ are independent, then
\[
\Lambda(X,Y)\ge\Lambda(X)+\Lambda(Y).
\]
Equality holds if and only if at least one of $X,Y$ is uniformly distributed.
\item (Mixed chain rule) For any $X,Y$, 
\begin{equation}
\Lambda(X,Y)\le\Lambda(X|Y)+H(Y).\label{eq:bounded_increase}
\end{equation}
Hence, $\Lambda(X,Y)\le\Lambda(X)+H(Y)$.
\end{enumerate}
\end{prop}
\medskip{}

\begin{rem}
\label{rem:continuous}The (continuous) layered entropy \cite{hegazy2022randomized,ling2024rejection} of a continuous random variable $Y$ with probability density function $f_{Y}$ is defined as
\[
\lambda(Y):=\int_{0}^{\infty}\mu(\{y:f_{Y}(y)>t\})\log\mu(\{y:f_{Y}(y)>t\})\mathrm{d}t,
\]
where $\mu$ denotes the Lebesgue measure. Note the similarity between this definition and \eqref{eq:layered_def}. We can express  $\lambda(Y)$ in terms of the discrete layered entropy $\Lambda$ via
\[
\lambda(Y)=\lim_{\Delta\to0}\left(\Lambda(\lfloor Y/\Delta\rfloor)+\log\Delta\right).
\]
We can also express  $\Lambda$ in terms of  $\lambda$ via
\[
\Lambda(X)=\lambda(X+Z)
\]
for $X\in\mathbb{Z}$, where $Z\sim\mathrm{Unif}([0,1])$ is independent of $X$. Compare this with the discrete Shannon entropy $H$ and the differential entropy $h$, which satisfy $h(Y)=\lim_{\Delta\to0}(H(\lfloor Y/\Delta\rfloor)+\log\Delta)$, and $H(X)=h(X+Z)$ for $X\in\mathbb{Z}$ where $Z\sim\mathrm{Unif}([0,1])$ is independent of $X$, we can see that $\Lambda$ is to $\lambda$ as $H$ is to $h$.
\end{rem}
\medskip{}

\section{$\Lambda(X)$ as a Piecewise Linear Approximation of $H(X)$\label{sec:piecewise_approx}}

It was shown in \cite{shannon1951prediction} that $\Lambda(X)\le H(X)$. We prove a bound in the other direction, showing that
$\Lambda(X)$ is close to $H(X)$ within a logarithmic gap. The proof is in Appendix \ref{subsec:pf_LX_HX_bound}.

\medskip{}

\begin{prop}
[$\Lambda(X)\approx H(X)$]\label{prop:LX_HX_bound}For every discrete  $X$, 
\begin{equation}
\Lambda(X)\le H(X)\le\Lambda(X)+\log\left(1+\frac{\Lambda(X)}{e\eta}\right)+\eta\label{eq:LX_HX_bound1}
\end{equation}
for every $\eta>0$. In particular, taking $\eta=\log e$ gives a bound  good for large $\Lambda(X)$:
\[
\Lambda(X)\le H(X)\le\Lambda(X)+\log\left(1+\frac{\Lambda(X)}{e\log e}\right)+\log e.
\]
Taking $\eta=\sqrt{\frac{\Lambda(X)\log e}{e}}$ gives a bound  good for small $\Lambda(X)$:
\begin{align*}
\Lambda(X)\le H(X) & \le\Lambda(X)+2\sqrt{\Lambda(X)e^{-1}\log e}.
\end{align*}
\end{prop}
\medskip{}

Note that $\log(1+\Lambda/(e\eta))+\eta$ (where $\Lambda=\Lambda(X)$) is minimized at $\eta=\frac{1}{2e}(\sqrt{\Lambda^{2}+4e\Lambda\log e}-\Lambda)$. This gives the best bound for \eqref{eq:LX_HX_bound1}, but is a little unwieldy.

Proposition \ref{prop:LX_HX_bound} shows that $\Lambda$ is a good approximate of $H$, and can be used in place of $H$ when the properties of $\Lambda$ are more desirable. For example, consider the scenario where we want to find a distribution $p$ (treated as a probability vector $p\in\mathbb{R}_{\ge0}^{n}$ with $\sum_{i}p_{i}=1$), subject to the linear inequality constraint $Ap\ge b$ where $A\in\mathbb{R}^{m\times n}$, $b\in\mathbb{R}^{m}$ ($Ap\ge b$ denotes entrywise comparison). The principle of maximum entropy \cite{jaynes1957information,snickars1977minimum} suggests that we should select the entropy-maximizing distribution, i.e., we should solve the problem
\begin{equation}
\text{maximize}\;H(p)\;\text{subject to}\;Ap\ge b.\label{eq:max_ent_linear}
\end{equation}
This problem has widespread applications in transport models \cite{wilson1970entropy,brice1989derivation,jornsten1989entropy}. However, it cannot be solved via linear programming since $H(p)$ is not piecewise linear.  Moreover, $H(p)$ has undefined gradient when there is a zero entry in $p$, making it difficult to solve the problem via standard gradient-based methods. Therefore, specialized iterative algorithms have been developed to solve this problem \cite{eriksson1980note,fang1993linear}. 

On the other hand, the discrete layered entropy maximization problem
\begin{equation}
\text{maximize}\;\Lambda(p)\;\text{subject to}\;Ap\ge b\label{eq:max_layer_linear}
\end{equation}
can be solved via linear programming using \eqref{eq:lp_def}. By Proposition \ref{prop:LX_HX_bound}, the optimal values of the two problems \eqref{eq:max_ent_linear} and \eqref{eq:max_layer_linear} are close within a logarithmic gap, making $\Lambda$ a good substitute for $H$. This idea can also be applied to more complex optimization problems involving linear objective and constraints with an additional entropy term (e.g., the entropy-regularized optimal transport problem \cite{cuturi2013sinkhorn}).

Interestingly, discrete layered entropy \emph{minimization} can also be solved via linear programming if the problem is invariant under permutation of entries of $p$. More precisely, if $Ap\ge b$ if and only if $AKp\ge b$ for any permutation matrix $K\in\{0,1\}^{n}$, then the problem
\begin{equation}
\text{minimize}\;\Lambda(p)\;\text{subject to}\;Ap\ge b\label{eq:min_layer}
\end{equation}
is equivalent to the following linear program
\begin{equation}
\text{minimize}\;\sum_{i=1}^{\infty}p_{i}\left(i\log i-(i-1)\log(i-1)\right)\;\text{s.t.}\;Ap\ge b,\label{eq:min_layer_linear}
\end{equation}
by definition of $\Lambda(p)$. This is in stark contrast to Shannon entropy minimization, which is a nonconvex problem.  See \eqref{eq:mlayered_lp} for an application.

If we want tighter approximations of $H$, we can use the quantity $\Lambda^{[m]}(X):=\max_{p_{Y|X}:\,Y\in[m]}(\Lambda(X,Y)-\Lambda(Y))$, which we call the \emph{discrete $m$-layered entropy}. We can show that $\Lambda^{[m]}(X)$ is piecewise linear (and can be formulated as a linear program), and approaches $H(X)$ as $m\to\infty$. Replacing $H$ with $\Lambda^{[m]}$ in the entropy maximization problem \eqref{eq:max_ent_linear} gives a linear programming algorithm for approximately solving \eqref{eq:max_ent_linear} which terminates in polynomial time with a provable closeness guarantee, unlike previous iterative algorithms such as \cite{eriksson1980note,fang1993linear,cuturi2013sinkhorn} which are only guaranteed to asymptotically approach the optimum.\footnote{Note that minimization of $\Lambda^{[m]}(p)$ cannot be performed via linear programming even if the problem is invariant under permutation.} Properties of $\Lambda^{[m]}(X)$ are discussed in Section \ref{sec:mlayered}.

In the following sections, we will see more theoretical properties of $\Lambda$ that are not satisfied by $H$.

\medskip{}

\section{Relation to One-to-one Non-prefix Codes\label{sec:onetoone}}

A one-to-one (non-prefix) code is an injective function $f:\mathcal{X}\to\{0,1\}^{*}$ which is not subject to the prefix requirement \cite{alon1994lower,blundo1996new,szpankowski2011minimum}. The optimal expected length of a one-to-one non-prefix encoding of $X$ is \cite{alon1994lower,blundo1996new}
\[
L(X):=\sum_{i=1}^{\infty}p_{X}^{\downarrow}(i)\lfloor\log i\rfloor.
\]
This was shown by ordering the sequences in $\{0,1\}^{*}$ in ascending order of length: $\emptyset,0,1,00,01,\ldots$, and then assigning the shortest sequence $\emptyset$ to the most probable $x$, the second shortest sequence $0$ to the second most probable $x$, and so on. 

We show that $L(X)$ is approximated by $\Lambda(X)$ within $2$ bits. This is in contrast to $H(X)$ which approximates the expected length of prefix codes. The proof is in Appendix \ref{subsec:pf_nonprefix}.
\begin{prop}
\label{prop:nonprefix}We have
\[
\Lambda(X)-2<L(X)\le\Lambda(X).
\]
\end{prop}
\medskip{}

Using Propositions \ref{prop:LX_HX_bound} and \ref{prop:nonprefix}, we can show that the optimal expected length of one-to-one codes is at least $H(X)-\log(H(X)+1)-O(1)$, giving a similar (and slightly weaker) bound compared to \cite{alon1994lower,blundo1996new}. By \cite{szpankowski2011minimum}, when $X^{n}=(X_{1},\ldots,X_{n})$ is an i.i.d. sequence following $p_{X}$, as $n\to\infty$, 
\[
\Lambda(X^{n})=\begin{cases}
nH(X) & \text{if}\;X\;\text{is uniform},\\
nH(X)-\frac{\log n}{2}+O(1) & \text{if}\;\text{nonuniform}.
\end{cases}
\]

One may raise the question---why should we study $\Lambda$ instead of $L$ which is exactly the optimal length? Indeed, $L$ also satisfies some properties of $\Lambda$, such as concavity, monotone linearity and the conditional property to be discussed in Proposition \ref{prop:cond}. One reason for preferring $\Lambda$ is that $\Lambda(X)$ approximates $H(X)$ better than $L(X)$. We have $L(X)\le\Lambda(X)\le H(X)$, and $\Lambda(X)=H(X)$ whenever $X$ is uniform, whereas the expression of $L(X)$ for uniform $X$ is complicated. This makes $\Lambda(X)$ a more useful tool for approximation tasks (e.g., Theorem \ref{thm:l_sfrl}). Moreover, $\Lambda(X)$ satisfies more elegant theoretical properties compared to $L(X)$, such as Theorems \ref{thm:useful_def}, \ref{thm:axiom_cond_logx}, \ref{thm:axiom_cond_largest}, and some properties in Proposition \ref{prop:basic}.\footnote{Also, $L(X)$ is dependent on the particular base of encoding in a complicated manner. We can define $L_{b}(X):=\sum_{i=1}^{\infty}p_{X}^{\downarrow}(i)\lfloor\log_{b}i\rfloor$ to be the expected length of the optimal non-prefix encoding of $X$ using $b$-ary codes. There is no clear relation between $L_{b}(X)$ for different $b$. In comparison, a change of base of logarithm in the definition of $\Lambda(X)$ only results in a change of multiplicative constant.} These reasons suggest that $\Lambda(X)$ is a more fundamental quantity compared to $L(X)$, and should be used as a convenient theoretical tool in the analysis of non-prefix codes. This is analogous to the fact that $H(X)$ is a more fundamental quantity compared to the optimal expected length of prefix codes, and should be used as a tool in the analysis of prefix codes.

We also remark that in subsequent sections, we call a code without any prefix requirement a ``non-prefix code'' instead of a ``one-to-one'' code, since the meaning of ``one-to-one'' is quite narrow and is only accurate for lossless compression where $X$ is recovered noiselessly (so $f:\mathcal{X}\to\{0,1\}^{*}$ must be one-to-one). ``One-to-one code'' would not be accurate in more general scenarios, such as lossy compression and channel simulation without prefix requirement.

\medskip{}

\section{Conditional  Compression  and ``Three Conditional Entropies''\label{sec:threecond}}

\subsection{Conditional Compression}

Given random variables $X,Y$, we are interested in finding ``the information in $(X,Y)$ that is not in $Y$''. Intuitively, it is the smallest $U$ such that $X$ can be recovered using $(Y,U)$. Its formal definition is given below, and its operational meaning will be explained in Section \ref{sec:cond_varlen}. 

\medskip{}

\begin{defn}
[Conditional compression] \label{def:cond}We say that a random variable $U$ is a \emph{conditional compression} of $X$ given $Y$ if $U$ is a minimizer of $H(U)$ subject to the constraint $H(X|Y,U)=0$.\footnote{Equivalently, by Proposition \ref{prop:cond_pmf}, we can consider $U$ that gives the maximum distribution $p_{U}$ with respect to majorization, subject to $H(X|Y,U)=0$.} The \emph{canonical conditional compression} of $X$ given $Y$, written as $X\backslash Y$, is the conditional compression $U$ that minimizes $H(X|U)$ among all conditional compressions.
\end{defn}
\medskip{}

The distribution of a conditional compression can be found. 
\begin{prop}
\label{prop:cond_pmf}The probability mass function of any conditional compression $U$ of $X$ given $Y$ must satisfy
\[
p_{U}^{\downarrow}(i)=\mathbb{E}_{Y}\big[p_{X|Y}^{\downarrow}(i|Y)\big],
\]
for $i\in\mathbb{N}$, where $p_{X|Y}^{\downarrow}(i|y)$ is the $i$-th entry of $(p_{X|Y}(x|y))_{x\in\mathcal{X}}$ when sorted in descending order.
\end{prop}
\begin{IEEEproof}
We can find a conditional compression in the following way. For each $y$, we sort the values of $(p_{X|Y}(x|y))_{x\in\mathcal{X}}$ in descending order to obtain $p_{X|Y}(x_{y}^{(1)}|y)\ge p_{X|Y}(x_{y}^{(2)}|y)\ge\cdots$, where $x_{y}^{(1)},x_{y}^{(2)},\ldots\in\mathcal{X}$ are distinct values. Take $U\in\mathbb{N}$ to be the value that satisfies $X=x_{Y}^{(U)}$, i.e., $U$ is the ranking of $p_{X|Y}(X|Y)$ among $(p_{X|Y}(x|Y))_{x\in\mathcal{X}}$ ($p_{X|Y}(X|Y)$ is the $U$-th largest among $p_{X|Y}(x|Y)$ for $x\in\mathcal{X}$). We then have $p_{U}(i)=\mathbb{E}_{Y}[p_{X|Y}(x_{Y}^{(i)}|Y)]=\mathbb{E}_{Y}[p_{X|Y}^{\downarrow}(i|Y)]$.

To show that this is indeed a conditional compression (it minimizes $H(U)$), consider any other $V$ satisfying $H(X|Y,V)=0$. Hence, $p_{X|Y}(\cdot|y)\succeq p_{V|Y}(\cdot|y)$ for every $y$, i.e., $\sum_{i=1}^{k}p_{X|Y}^{\downarrow}(i|y)\ge\sum_{i=1}^{k}p_{V|Y}^{\downarrow}(i|y)$. This implies 
\begin{align*}
\sum_{i=1}^{k}p_{V}^{\downarrow}(i) & \le\mathbb{E}\Big[\sum_{i=1}^{k}p_{V|Y}^{\downarrow}(i|Y)\Big]\\
 & \quad\le\mathbb{E}\Big[\sum_{i=1}^{k}p_{X|Y}^{\downarrow}(i|Y)\Big]=\sum_{i=1}^{k}p_{U}(i),
\end{align*}
and $p_{U}\succeq p_{V}$, and hence $H(U)\le H(V)$. Equality holds if and only if $p_{V}^{\downarrow}(i)=p_{U}(i)$, i.e., $V$ has the same distribution as $U$ up to relabeling. Therefore, any conditional compression has the same distribution $p_{U}$ up to relabeling.
\end{IEEEproof}
\medskip{}

The reason for the notation $X\backslash Y$ is that the conditional compression shares a number of properties with set difference. Firstly, $X\backslash Y\stackrel{\iota}{=}\emptyset$ (i.e., $X\backslash Y$ is a constant) if $H(X|Y)=0$, corresponding to the fact that $A\backslash B=\emptyset$ for sets $A,B$ if $A\subseteq B$. Secondly, $X\backslash Y\stackrel{\iota}{=}X$  if $X\perp\!\!\!\perp Y$, analogous to the fact that $A\backslash B=A$ if $A\cap B=\emptyset$. This property is the reason of the tie-breaking rule (minimizing $H(X|U)$ in case of a tie of minimizing $H(U)$) in the definition of $X\backslash Y$.\footnote{Although there may be multiple  conditional compressions in case there are ties among $(p_{X|Y}(x|Y))_{x\in\mathcal{X}}$, we are usually only interested in the marginal distribution $p_{U}$ so it does not matter which one  we consider (they all have the same distribution). Nevertheless, in order to justify the notation $X\backslash Y$, we select a ``canonical'' conditional compression $X\backslash Y$ in Definition \ref{def:cond} to be the one that minimizes $H(X|U)$, to ensure that $X\backslash Y\stackrel{\iota}{=}X$ if $X\perp\!\!\!\perp Y$.}

The converse of the first property holds ($X\backslash Y\stackrel{\iota}{=}\emptyset$ if and only if $H(X|Y)=0$), but the converse of the second property is false, i.e., $X\backslash Y\stackrel{\iota}{=}X$ does not imply $I(X;Y)=0$. In fact, $H(X)-\Lambda(X)$ is the maximum violation, i.e., $H(X)-\Lambda(X)$ is the largest possible $I(X;Y)$ subject to $X\backslash Y\stackrel{\iota}{=}X$. This gives a simple alternative definition of $\Lambda(X)$ given in Theorem \ref{thm:useful_def}. By Proposition \ref{prop:LX_HX_bound}, if $H(X\backslash Y)=H(X)$, then $I(X;Y)\le H(X)-\Lambda(X)\le O(\log H(X))$, so the converse does approximately hold. The proof of Theorem \ref{thm:useful_def} is in Appendix \ref{subsec:pf_useful_def}. In Section \ref{sec:cond_varlen}, we will discuss the operational meaning of this definition.

\smallskip{}

\begin{thm}
[Alternative definition of $\Lambda(X)$]\label{thm:useful_def}We have
\begin{align*}
\Lambda(X) & =\!\min_{p_{Y|X}:\,H(X\backslash Y)=H(X)}\!\!H(X|Y)=\!\min_{p_{Y|X}:\,X\backslash Y\stackrel{\iota}{=}X}\!\!\!H(X|Y).
\end{align*}
Also, $H(X\backslash Y)=H(X)$ if and only if $\Lambda(X\backslash Y)=\Lambda(X)$.
\end{thm}
\smallskip{}

An elegant property of $\Lambda(X)$ is that $\Lambda(X|Y)=\mathbb{E}_{Y}[\Lambda(p_{X|Y}(\cdot|Y)]$ coincides with $\Lambda(X\backslash Y)$. Therefore, $\Lambda(X|Y)$ is indeed the discrete layered entropy of ``the random variable $X|Y$'' which is formally given as $X\backslash Y$.
\begin{prop}
[Conditioning property] \label{prop:cond}
\[
\Lambda(X|Y)=\Lambda(X\backslash Y).
\]
Equivalently,
\begin{equation}
\Lambda(X|Y)=\min_{p_{U|X,Y}:\,H(X|Y,U)=0}\Lambda(U).\label{eq:cond}
\end{equation}
\end{prop}
We will see in Section \ref{subsec:characterization} that the conditioning property is actually the defining property of the discrete layered entropy, in the sense that if the function $\Lambda$ satisfies the conditioning property and $\Lambda(X)=\log|\mathcal{X}|$ if $X$ is uniformly distributed, then $\Lambda$ must be the discrete layered entropy.

We now have three different notions of ``conditional entropy'', namely $H(X|Y)$, $H(X\backslash Y)$ and $\Lambda(X|Y)=\Lambda(X\backslash Y)$. Although Shannon entropy does not satisfy the conditioning property, i.e., $H(X\backslash Y)$ is generally not equal to $H(X|Y)$, we can show that $H(X\backslash Y)\approx H(X|Y)$ via the discrete layered entropy. By Propositions \ref{prop:basic} and \ref{prop:LX_HX_bound}, and Jensen's inequality, for every $\eta>0$,

\begin{align}
\Lambda(X|Y) & \le H(X|Y)\le H(X\backslash Y)\nonumber \\
 & \;\;\le\Lambda(X|Y)+\log\left(1+\frac{\Lambda(X|Y)}{e\eta}\right)+\eta.\label{eq:three_approx}
\end{align}
Therefore, the three ``conditional entropies'' are close within a logarithmic gap. We highlight the following theorem that follows directly from \eqref{eq:three_approx}.\smallskip{}

\begin{thm}
\label{thm:cond_part_ent}For every $\eta>0$,
\[
H(X|Y)\le H(X\backslash Y)\le H(X|Y)+\log\left(1+\frac{H(X|Y)}{e\eta}\right)+\eta.
\]
\end{thm}
\smallskip{}

Theorem \ref{thm:cond_part_ent} can also be stated as: for every $X,Y$, there exists $U$ such that $H(X|Y,U)=H(U|X,Y)=0$, and for every $\eta>0$, $H(U)\le H(X|Y)+\log(1+H(X|Y)/(e\eta))+\eta$. This is an example of a nonlinear existential information inequality \cite{li2023automated,li2022undecidabilityaffine,li2021first}. It is perhaps interesting that such a simple (and novel, to the best of the author's knowledge) fact about Shannon entropy can be proved via the properties of discrete layered entropy $\Lambda(X)$. It shows the usefulness of $\Lambda(X)$ as a tool for proving results about Shannon entropy.

Also, unlike set difference where $A\backslash B$ and $B$ are disjoint, we do not have $I(X\backslash Y;Y)=0$. Nevertheless, Theorem \ref{thm:cond_part_ent} implies that this property approximately holds, in the sense that $I(X\backslash Y;Y)\le\log(1+H(X|Y)/(e\eta))+\eta$ is small.\footnote{Compare $X\backslash Y$ to the strong functional representation lemma (SFRL) \cite{sfrl_trans,li2024pointwise}: for every $X,Y$, there exists $S$ such that $I(S;Y)=H(X|Y,S)=0$ and $H(X|S)\le I(X;Y)+\log(I(X;Y)+2)+2$. SFRL guarantees $I(S;Y)=0$, whereas we only have $I(X\backslash Y;Y)\approx0$. Nevertheless, SFRL does not guarantee $H(S)\approx H(X|Y)$ (it only has $I(S;X,Y)=H(X|Y)$; the construction in \cite{sfrl_trans} can have extremely large $H(S)$), whereas we have $H(X\backslash Y)\approx H(X|Y)$.}

In the next subsection, we will give operational meanings to the three ``conditional entropies'', and see how \eqref{eq:three_approx} and Theorem \ref{thm:cond_part_ent} are relevant.

\medskip{}

\subsection{Conditional Variable-Length Encoding\label{sec:cond_varlen}}

Consider a one-shot variable-length encoding setting, where there is a source symbol $X$ that is correlated with a side information $Y$. The encoder observes $X,Y$, and sends a variable-length description $M\in\{0,1\}^{*}$ (possibly within a long stream of bits) to the decoder. The decoder observes $Y,M$, and has to recover $X$ losslessly, i.e., $H(X|Y,M)=0$. The goal is to make the expected length $\mathbb{E}[|M|]$ as small as possible. 

If there is no prefix requirement on $M$, the task is straightforward---given that $Y=y$, the encoder encodes $X$ into $M$ using the best one-to-one non-prefix code designed for the distribution $p_{X|Y}(\cdot|y)$. Let $\ell_{\mathrm{n}}^{*}$ be the smallest possible $\mathbb{E}[|M|]$ in this case. By Proposition \ref{prop:nonprefix}, $\Lambda(X|Y)-2<\ell_{\mathrm{n}}^{*}\le\Lambda(X|Y)$. As usual for non-prefix codes, the decoder cannot synchronize with the encoder without additional information. That is, if $M$ is followed by other data in a stream of bits, the decoder does not know where $M$ ends, unless it already knows $|M|$ from another source (e.g., when $|M|$ is already given by the metadata of the communication protocol). Also, note that this $M$ is only ``conditionally one-to-one'', that is, the mapping from $X$ to $M$ for any fixed $Y=y$ is one-to-one. It is not required to be unconditionally one-to-one since the decoder does not have to recover $X$ solely based on $M$.

For the sake of synchronization, given that $Y=y$, this task is conventionally performed by encoding $X$ conditional on $Y=y$ into $M$, using a prefix code designed for the distribution $p_{X|Y}(\cdot|y)$. This gives a \emph{conditional prefix code}, i.e., $M\in\mathcal{C}_{Y}$, where $\mathcal{C}_{y}\subseteq\{0,1\}^{*}$ is a prefix codebook for every $y\in Y$. Let $\ell_{\mathrm{c}}^{*}$ be the smallest possible $\mathbb{E}[|M|]$ in this case. Using Huffman coding \cite{huffman1952method}, $H(X|Y)\le\ell_{\mathrm{c}}^{*}<H(X|Y)+1$. 

A shortcoming of this approach is that this ``conditional prefix code'' is not a prefix code to a party that does not know $Y$. This party only knows $M\in\bigcup_{y\in\mathcal{Y}}\mathcal{C}_{y}$ which may not be a prefix-free codebook. For example, consider $X=\{1,2,3\}$, $Y=\{1,2\}$, $p_{X|Y}(1|1)=p_{X|Y}(2|1)=p_{X|Y}(1|2)=1/2$, $p_{X|Y}(2|2)=p_{X|Y}(3|2)=1/4$. The optimal conditional prefix code $M=f(X,Y)$ would be $f(1,1)=0$, $f(2,1)=1$, $f(1,2)=0$, $f(2,2)=10$, $f(3,2)=11$, which is not overall a prefix code.   

Hence, conditional prefix codes should only be used if we are 100\% certain that the decoder knows $Y$. Without knowing $Y$, the decoder not only cannot decode $X$, but also cannot know where the bit sequence ends (i.e., it cannot know $|M|$), and becomes desynchronized with the encoder and fails to decode any subsequent information. For example, if we are going to use the code $100$ times to encode $X_{i}$ into $M_{i}$ conditional on $Y_{i}$ for $i=1,\ldots,100$ and send the concatenation of $M_{1},\ldots,M_{100}$, even if we are 99\% certain that the decoder can know each $Y_{i}$, there is still a $1-(99/100)^{50}\approx39\%$ chance that the decoder becomes desynchronized at $i=50$ and fails to decode the remaining half of the symbols $X_{51},\ldots,X_{100}$. Practically, it is difficult to ensure 100\% reliable communication due to packet loss and sudden loss of connectivity, so the use of conditional prefix codes can be risky.

Therefore, it is often beneficial to use an \emph{unconditional prefix code}, that is, $M\in\mathcal{C}$ where $\mathcal{C}\subseteq\{0,1\}^{*}$ is a prefix-free codebook that does not depend on $Y$. Let $\ell_{\mathrm{u}}^{*}$ be the smallest possible $\mathbb{E}[|M|]$. Using Huffman coding \cite{huffman1952method}, $\ell_{\mathrm{u}}^{*}$ is within $1$ bit from $\min_{U:\,H(X|Y,U)=0}H(U)=H(X\backslash Y)$. In the aforementioned scenario of encoding $X_{1},\ldots,X_{100}$, using a conditional prefix code, the decoder can know the boundaries between $M_{1},\ldots,M_{100}$ with or without $Y_{1},\ldots,Y_{100}$, and will not become desynchronized. It will only fail to decode 1\% of the symbols among $X_{1},\ldots,X_{100}$ on average (since 1\% of $Y_{1},\ldots,Y_{100}$ are lost), giving a significantly lower symbol error rate.

\begin{figure}
\begin{centering}
\includegraphics{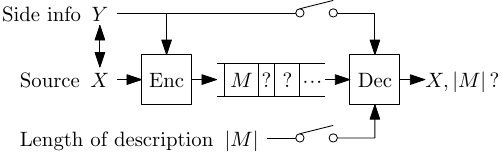}
\par\end{centering}
\bigskip{}

\begin{centering}
{\renewcommand*{\arraystretch}{1.5}%
\begin{tabular}{|c|c|c|c|c|}
\cline{2-5}
\multicolumn{1}{c|}{} & \multicolumn{2}{c|}{With $Y$} & \multicolumn{2}{c|}{Without $Y$}\tabularnewline
\cline{2-5}
\multicolumn{1}{c|}{} & $\!\!$w/ $|M|$$\!\!$ & $\!\!$w/o $|M|$$\!\!$ & $\!\!$w/ $|M|$$\!\!$ & $\!\!$w/o $|M|$$\!\!$\tabularnewline
\hline 
Non-prefix & $X,|M|$ & None & $|M|$ & None\tabularnewline
\hline 
Cond. prefix & $X,|M|$ & $X,|M|$ & $|M|$ & None\tabularnewline
\hline 
$\!\!$Uncond. prefix$\!\!$ & $X,|M|$ & $X,|M|$ & $|M|$ & $|M|$\tabularnewline
\hline 
\end{tabular}}
\par\end{centering}
\medskip{}

\caption{\label{fig:cond_encode}Top: Diagram of the conditional variable-length encoding setting, where the encoder writes a variable-length description $M$ followed by a stream of other bits transmitted to the decoder. The side information $Y$ and the description length $|M|$ are optionally given to the decoder. Bottom: Whether the decoder can recover $X$ and/or $|M|$ (i.e., keep synchronized) with or without being given $Y,|M|$, under the three settings (e.g., for conditional prefix codes, if $Y$ is available, the decoder can recover $X,|M|$).}
\end{figure}

Refer to Figure \ref{fig:cond_encode} for an illustration of the three settings. The difference between the three settings is their synchronization capabilities: non-prefix codes can synchronize if $|M|$ is provided externally; conditional prefix codes can synchronize if $|M|$ or $Y$ is provided; and unconditional prefix codes can always synchronize. We summarize the optimal expected lengths in the three settings:
\begin{itemize}
\item For non-prefix codes, $\ell_{\mathrm{n}}^{*}\approx\Lambda(X|Y)$:
\begin{equation}
\Lambda(X|Y)-2<\ell_{\mathrm{n}}^{*}\le\Lambda(X|Y).\label{eq:non_prefix}
\end{equation}
\item For conditional prefix codes, $\ell_{\mathrm{c}}^{*}\approx H(X|Y)$:
\begin{equation}
H(X|Y)\le\ell_{\mathrm{c}}^{*}<H(X|Y)+1.\label{eq:cond_prefix}
\end{equation}
\item For unconditional prefix codes, $\ell_{\mathrm{u}}^{*}\approx H(X\backslash Y)$:
\begin{equation}
H(X\backslash Y)\le\ell_{\mathrm{u}}^{*}<H(X\backslash Y)+1.\label{eq:uncond_prefix}
\end{equation}
\end{itemize}
The three ``conditional entropies'' $\Lambda(X|Y)\le H(X|Y)\le H(X\backslash Y)$ are close within a logarithmic gap, as shown in \eqref{eq:three_approx}.

\begin{figure}
\begin{centering}
\includegraphics[scale=0.95]{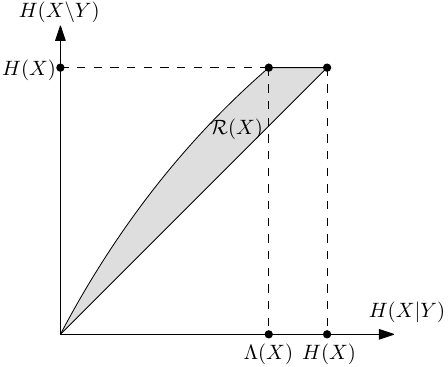}
\par\end{centering}
\caption{\label{fig:region}Illustration of $\mathcal{R}(X)=\bigcup_{p_{Y|X}}\{(H(X|Y),H(X\backslash Y))\}$ showing the extreme points $(\Lambda(X),H(X))$ and $(H(X),H(X))$.}

\end{figure}

While we know that the gap between $\Lambda(X|Y)$ and $H(X|Y)$ can be as large as logarithmic, it is not a priori clear whether the gap between $H(X|Y)$ and $H(X\backslash Y)$ can be logarithmic, or is the gap always bounded by a constant. To study the difference between $H(X|Y)$ (conditionally prefix codes \eqref{eq:cond_prefix}) and $H(X\backslash Y)$ (unconditionally prefix codes \eqref{eq:uncond_prefix}),  it is of interest to study the region of possible values of the pair $(H(X|Y),H(X\backslash Y))\in\mathbb{R}^{2}$. Given $X$, define
\[
\mathcal{R}(X):=\bigcup_{p_{Y|X}}\left\{ (H(X|Y),H(X\backslash Y))\right\} .
\]
An inner bound of $\mathcal{R}(X)$ is given by the diagonal line, that is, $(t,t)\in\mathcal{R}(X)$ for every $0\le t\le H(X)$.\footnote{To show this, consider a distribution $q$ that majorizes $p_{X}$ \cite{marshall1979inequalities} with $H(q)=t$. We can construct random variables $U,Y$ such that $U\sim q$, $U$ is independent of $Y$, and $H(X|U,Y)=H(U|X,Y)=0$. This gives $H(X|Y)=H(U|Y)=H(U)=t$. Since $p_{X|Y}(\cdot|y)$ contains the same entries (possibly reordered) as $q(\cdot)$, $U$ is a conditional compression of $X$ given $Y$, and hence $H(X\backslash Y)=H(U)=t$. } Theorem \ref{thm:cond_part_ent} gives an outer bound of $\mathcal{R}(X)$, showing that $\mathcal{R}(X)$ cannot be too far from the diagonal line. Refer to Figure \ref{fig:region}.

An interesting extreme point of $\mathcal{R}(X)$ is the minimum of $H(X|Y)$ subject to the contraint that $H(X\backslash Y)$ is maximized ($H(X\backslash Y)=H(X)$). The $Y$ that attains this minimum is the most ``conditionally useful'' (i.e., giving the shortest conditional prefix encoding length $H(X|Y)$) among ``unconditionally useless'' side information (i.e., the unconditional prefix encoding length $H(X\backslash Y)$ is the same as if $Y$ is absent). By Theorem \ref{thm:useful_def}, this minimum is given by $\Lambda(X)=\min_{p_{Y|X}:\,H(X\backslash Y)=H(X)}\!\!H(X|Y)$. Hence, we have another operational meaning of $\Lambda(X)$: \smallskip{}

\emph{If we are to design a side information $Y$ that provides no benefit to an encoder that encodes $X$ using an unconditional prefix code, how useful can it be to an encoder that uses a conditional prefix code?}\smallskip{}

It is intuitively unclear why the answer to the above question, which only concerns prefix codes, is given by $\Lambda(X)$ which is about the optimal non-prefix code.

Even more strangely, $\Lambda(X)$ is also the $\Lambda(X|Y)$ (non-prefix length \eqref{eq:non_prefix}) when $H(X|Y)$ is maximized, or when $H(X\backslash Y)$ is maximized (by Theorem \ref{thm:useful_def}). So, 
\begin{align*}
\Lambda(X) & =\min_{p_{Y|X}:\,H(X\backslash Y)=H(X)}H(X|Y)\\
 & =\min_{p_{Y|X}:\,H(X\backslash Y)=H(X)}\Lambda(X|Y)\\
 & =\min_{p_{Y|X}:\,H(X|Y)=H(X)}\Lambda(X|Y),
\end{align*}
and hence every pair of the three settings \eqref{eq:non_prefix}, \eqref{eq:cond_prefix}, \eqref{eq:uncond_prefix} have the same gap in this sense. Refer to Table \ref{tab:maxmin} for details. An intuitive explanation for this curious coincidence is left for future studies.

\begin{table}
\caption{\label{tab:maxmin}Table listing $\min\{B:\,p_{Y|X}\in\mathrm{argmax}_{p_{Y|X}}A\}$ (answers to the question ``among the  $p_{Y|X}$'s which maximize $A$, what is the smallest possible $B$?''), for $A,B$ being $\Lambda(X|Y)$, $H(X|Y)$ or $H(X\backslash Y)$.}

\begin{centering}
{\renewcommand*{\arraystretch}{1.5}%
\begin{tabular}{|c|c|c|c|c|}
\cline{3-5}
\multicolumn{1}{c}{} &  & \multicolumn{3}{c|}{what is the smallest possible...}\tabularnewline
\cline{3-5}
\multicolumn{1}{c}{} &  & $\!\!$$\Lambda(X|Y)$$\!\!$ & $\!\!$$H(X|Y)$$\!\!$ & $\!\!$$H(X\backslash Y)$$\!\!$\tabularnewline
\hline 
\multirow{3}{*}{$\!\!$$\begin{array}{c}
\text{Among}\;p_{Y|X}\text{'s}\\
\text{maximizing...}
\end{array}$$\!\!$} & $\!\!$$\Lambda(X|Y)$$\!\!$ & $\Lambda(X)$ & $\Lambda(X)$ & $H(X)$\tabularnewline
\cline{2-5}
 & $\!\!$$H(X|Y)$$\!\!$ & $\Lambda(X)$ & $H(X)$ & $H(X)$\tabularnewline
\cline{2-5}
 & $\!\!$$H(X\backslash Y)$$\!\!$ & $\Lambda(X)$ & $\Lambda(X)$ & $H(X)$\tabularnewline
\hline 
\end{tabular}}
\par\end{centering}
\end{table}

\medskip{}

\subsection{Characterizations of $\Lambda(X)$ via the Conditioning Property\label{subsec:characterization}}

Proposition \ref{thm:useful_def} gives a definition of $\Lambda(X)$ using only Shannon entropy. In this subsection, we study some more axiomatic definitions of $\Lambda(X)$ via the conditioning property $\Lambda(X|Y)=\Lambda(X\backslash Y)$ (Proposition \ref{prop:cond}). We first show that any function $\tilde{\Lambda}$ satisfying the conditioning property and $\tilde{\Lambda}(X)=\log|\mathcal{X}|$ for uniform $X$ must be the discrete layered entropy. The proof is given in Appendix \ref{subsec:pf_axiom_cond_logx}.\medskip{}

\begin{thm}
\label{thm:axiom_cond_logx}If $\tilde{\Lambda}$ is a function mapping discrete distributions to nonnegative real numbers that satisfies the conditioning property (i.e., $\tilde{\Lambda}(X\backslash Y)=\tilde{\Lambda}(X|Y):=\mathbb{E}_{Y}[\tilde{\Lambda}(p_{X|Y}(\cdot|Y))]$), $\tilde{\Lambda}(X)=\log|\mathcal{X}|$ whenever $X$ is uniformly distributed over $\mathcal{X}$, and $\tilde{\Lambda}(X)=\tilde{\Lambda}(Y)$ whenever $X\stackrel{\iota}{=}Y$ (i.e., $\tilde{\Lambda}$ is invariant under relabeling), then $\tilde{\Lambda}(X)=\Lambda(X)$.
\end{thm}
\medskip{}

Compare this with the axiomatic characterization of Shannon entropy in \cite{aczel1974shannon}: if $\tilde{H}$ satisfy the subadditivity property ($\tilde{H}(X,Y)\ge\tilde{H}(X)+\tilde{H}(Y)$), additivity property ($\tilde{H}(X,Y)=\tilde{H}(X)+\tilde{H}(Y)$ if $X\perp\!\!\!\perp Y$), continuity with respect to the probability mass function, $\tilde{H}(X)=1$ if $X\sim\mathrm{Bern}(1/2)$, and $\tilde{H}(X)=\tilde{H}(Y)$ whenever $X\stackrel{\iota}{=}Y$, then $\tilde{H}$ must be the Shannon entropy $H$. We can see that additivity, subadditivity and continuity are the defining properties of Shannon entropy, whereas conditioning is the defining property of discrete layered entropy. 

We then study another characterization of $\Lambda(X)$. Recall that $H(X)$ does not satisfy the conditioning property. Nevertheless, we can ask what is the best under-approximation of $H(X)$ that satisfy the conditioning property. The answer is given by $\Lambda(X)$. This allows us to prove tight approximation bounds using $\Lambda$ (e.g., Theorem \ref{thm:l_sfrl}). The proof is in Appendix \ref{subsec:pf_axiom_cond_largest}.

\medskip{}

\begin{thm}
\label{thm:axiom_cond_largest}If $\tilde{\Lambda}$ is a function mapping discrete distributions to nonnegative real numbers that satisfies the conditioning property, $\tilde{\Lambda}(X)\le H(X)$ for every $X$,\footnote{Actually, we only need $\tilde{\Lambda}(X)\le\log|\mathcal{X}|$ whenever $X$ is uniform.} and $\tilde{\Lambda}(X)=\tilde{\Lambda}(Y)$ whenever $X\stackrel{\iota}{=}Y$, then $\tilde{\Lambda}(X)\le\Lambda(X)$ for every $X$. Equivalently, $\Lambda(X)$ admits the following characterization:
\[
\Lambda(X)=\mathrm{max}_{\tilde{\Lambda}}\tilde{\Lambda}(X),
\]
where the maximum is over functions $\tilde{\Lambda}$ that satisfies the conditioning property, $\tilde{\Lambda}(Y)\le H(Y)$ for every $Y$, and $\tilde{\Lambda}(Y)=\tilde{\Lambda}(Z)$ whenever $Y\stackrel{\iota}{=}Z$.\medskip{}
\end{thm}

\section{Channel Simulation, Lossy Compression, and Strong Functional Representation Lemma\label{sec:chansim}}

\subsection{Three One-Shot Variable-Length Channel Simulation Settings}

In this section, we consider the one-shot variable-length channel simulation setting with unlimited common randomness \cite{harsha2010communication}. An encoder and a decoder share a common randomness $S\sim P_{S}$, where we are allowed to choose any distribution $P_{S}$ (not necessarily discrete). The encoder observes $S$ and a source $X\sim P_{X}$ (independent of $S$, not necessarily discrete), and sends a variable-length description $M\in\{0,1\}^{*}$ (possibly produced by a stochastic encoding function) to the decoder. The decoder observes $M,S$ and outputs $Y$ (not necessarily discrete, possibly produced by a stochastic decoding function). We require that $Y$ follows a certain target conditional distribution $P_{Y|X}$ given $X$. Given $P_{X}$ and $P_{Y|X}$, our goal is to find the smallest possible expected length $\mathbb{E}[|M|]$ that allows the simulation of $Y$ following $P_{Y|X}$.

\begin{figure}
\begin{centering}
\includegraphics{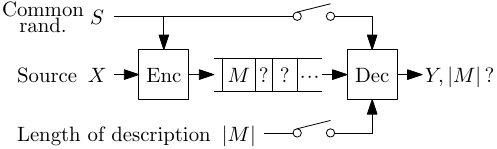}
\par\end{centering}
\bigskip{}

\begin{centering}
{\renewcommand*{\arraystretch}{1.5}%
\begin{tabular}{|c|c|c|c|c|}
\cline{2-5}
\multicolumn{1}{c|}{} & \multicolumn{2}{c|}{With $S$} & \multicolumn{2}{c|}{Without $S$}\tabularnewline
\cline{2-5}
\multicolumn{1}{c|}{} & $\!\!$w/ $|M|$$\!\!$ & $\!\!$w/o $|M|$$\!\!$ & $\!\!$w/ $|M|$$\!\!$ & $\!\!$w/o $|M|$$\!\!$\tabularnewline
\hline 
Non-prefix & $Y,|M|$ & None & $|M|$ & None\tabularnewline
\hline 
Cond. prefix & $Y,|M|$ & $Y,|M|$ & $|M|$ & None\tabularnewline
\hline 
$\!\!$Uncond. prefix$\!\!$ & $Y,|M|$ & $Y,|M|$ & $|M|$ & $|M|$\tabularnewline
\hline 
\end{tabular}}
\par\end{centering}
\medskip{}

\caption{\label{fig:chansim}Top: Diagram of one-shot variable-length channel simulation, where the encoder writes a variable-length description $M$ followed by a stream of other bits transmitted to the decoder. The common randomness $S$ and the description length $|M|$ are optionally given to the decoder. Bottom: Whether the decoder can output $Y$ and/or $|M|$ (i.e., keep synchronized) with or without being given $S,|M|$, under the three settings (e.g., for conditional prefix codes, if $S$ is available, the decoder can output $Y,|M|$).}
\end{figure}

Similar to Section \ref{sec:cond_varlen}, there are three variants of the setting in increasing order of stringency: $M$ belongs to a non-prefix code (no constraint is imposed on $M$), a conditional prefix code given $S$ ($M\in\mathcal{C}_{S}$ where $\mathcal{C}_{s}$ is a prefix-free codebook for every $s$), or an unconditional prefix code ($M\in\mathcal{C}$ where $\mathcal{C}$ is a prefix-free codebook).\footnote{Unconditional prefix codes are useful if the decoder does not always receive the common randomness. For example, if the common randomness is sent to the encoder and the decoder from a satellite, the communication may be unrealiable. It also allows the decoder to skip ahead without reading $S$ if it is uninterested in outputting $Y$.  We also remark that most existing one-shot channel simulation schemes (e.g., \cite{harsha2010communication,sfrl_trans}) are already unconditionally prefix-free, and there is little downside in considering unconditional prefix codes compared to conditional prefix codes.} Refer to Figure \ref{fig:chansim}. Let the smallest possible $\mathbb{E}[|M|]$ in these three variants be $\ell_{\mathrm{n}}^{*}$, $\ell_{\mathrm{c}}^{*}$ and $\ell_{\mathrm{u}}^{*}$, respectively.\footnote{More precisely, $\ell_{\mathrm{n}}^{*}$ is the infimum of the set of $\mathbb{E}[|M|]$ among all non-prefix schemes. Similar for $\ell_{\mathrm{c}}^{*}$ and $\ell_{\mathrm{u}}^{*}$.} The latter two settings have been discussed in \cite{li2024channel}, though the non-prefix setting does not appear to be studied previously. Note that there is no ``one-to-one'' requirement for the non-prefix setting since the decoder is not required to losslessly recover $X$. 

Similar to \eqref{eq:non_prefix}, \eqref{eq:cond_prefix} and \eqref{eq:uncond_prefix}, we have the following approximate characterizations:
\begin{itemize}
\item For non-prefix codes, defining\footnote{Even if $Y$ is continuous, $\Lambda(Y|S)$ is defined as long as $Y$ is conditionally discrete given $S$, i.e., given $S=s$ for any $s$, there are only countably many possible values of $Y$. Note that we also have $\Lambda_{\mathrm{n}}^{*}(X\to Y)=\inf_{P_{S|X,Y}:\,S\perp\!\!\!\perp X,\,H(Y|X,S)=0}\Lambda(Y|S)$, which will be explained later.}
\begin{equation}
\Lambda_{\mathrm{n}}^{*}=\Lambda_{\mathrm{n}}^{*}(X\to Y):=\inf_{P_{S|X,Y}:\,S\perp\!\!\!\perp X}\Lambda(Y|S),\label{eq:cs_lambda_n}
\end{equation}
we have
\begin{equation}
\Lambda_{\mathrm{n}}^{*}-2\le\ell_{\mathrm{n}}^{*}\le\Lambda_{\mathrm{n}}^{*}.\label{eq:cs_non_prefix}
\end{equation}
\item For conditional prefix codes, defining 
\begin{equation}
H_{\mathrm{c}}^{*}=H_{\mathrm{c}}^{*}(X\to Y):=\inf_{P_{S|X,Y}:\,S\perp\!\!\!\perp X}H(Y|S),\label{eq:cs_h_c}
\end{equation}
we have
\begin{equation}
H_{\mathrm{c}}^{*}\le\ell_{\mathrm{c}}^{*}\le H_{\mathrm{c}}^{*}+1.\label{eq:cs_cond_prefix}
\end{equation}
This bound has been observed in \cite{sfrl_trans}.
\item For unconditional prefix codes, defining
\begin{equation}
H_{\mathrm{u}}^{*}=H_{\mathrm{u}}^{*}(X\to Y):=\inf_{P_{S|X,Y}:\,S\perp\!\!\!\perp X}H(Y\backslash S),\label{eq:cs_h_u}
\end{equation}
we have
\begin{equation}
H_{\mathrm{u}}^{*}\le\ell_{\mathrm{u}}^{*}\le H_{\mathrm{u}}^{*}+1.\label{eq:cs_uncond_prefix}
\end{equation}
\end{itemize}
\medskip{}

We have $\ell_{\mathrm{n}}^{*}\le\ell_{\mathrm{c}}^{*}\le\ell_{\mathrm{u}}^{*}$, and \eqref{eq:three_approx} gives
\begin{equation}
\Lambda_{\mathrm{n}}^{*}\le H_{\mathrm{c}}^{*}\le H_{\mathrm{u}}^{*}\le\Lambda_{\mathrm{n}}^{*}+\log\left(1+\frac{\Lambda_{\mathrm{n}}^{*}}{e\eta}\right)+\eta,\label{eq:cs_three_approx}
\end{equation}
for every $\eta>0$, and hence $\ell_{\mathrm{n}}^{*}\approx\Lambda_{\mathrm{n}}^{*}\approx\ell_{\mathrm{c}}^{*}\approx H_{\mathrm{c}}^{*}\approx\ell_{\mathrm{u}}^{*}\approx H_{\mathrm{u}}^{*}$ within logarithmic gaps. We now explain the reason for \eqref{eq:cs_non_prefix} (the other two are similar). Without loss of generality, we can assume $H(Y|M,S)=0$, that is, the decoder is deterministic. This is because if the decoder is stochastic, i.e., it produces $Y$ as a function of $(M,S,W)$ where $W$ is the decoder's local randomness, then we can move $W$ to the common randomness $S$ without any downside \cite{li2024channel}. Fix a choice of $P_{S|X,Y}$. Since the encoder can always know $Y$ using $(M,S)$, the encoder has to conditionally compress $Y$ given $S$ using an expected length $\approx\Lambda(Y|S)$ for a non-prefix code. More formally, for the lower bound in \eqref{eq:cs_non_prefix}, by Proposition \ref{prop:nonprefix}, we have $\mathbb{E}[|M|]+2\ge\Lambda(M)\ge\Lambda(M|S)\ge\Lambda(Y|S)$ since $H(Y|M,S)=0$. For the upper bound, we can have the encoder generate $Y$ following $P_{Y|X,S}$ and conditionally compress $Y$ given $S$, giving $\mathbb{E}[|M|]\le\Lambda(Y|S)$.

Note that $H_{\mathrm{c}}^{*}\ge I$, where $I:=I(X;Y)$ is the mutual information, since $H(Y|S) \ge I(X;Y|S) = I(X;Y,S) \ge I(X;Y)$ when $S \perp \!\!\! \perp X$ \cite{bennet2014reverse}. The current best upper bound on $H_{\mathrm{u}}^{*}$ (and hence on $H_{\mathrm{c}}^{*}$ and $\Lambda_{\mathrm{n}}^{*}$) in terms of $I$ is proved via the Poisson functional representation scheme \cite{sfrl_trans,li2024pointwise} which uses an unconditional prefix code. It was shown in \cite{li2024pointwise} that
\begin{equation}
H_{\mathrm{u}}^{*}\le I+\log(I+2)+2.\label{eq:Hu_prev}
\end{equation}
This, combined with \eqref{eq:cs_three_approx}, gives $\Lambda_{\mathrm{n}}^{*}\approx H_{\mathrm{c}}^{*}\approx H_{\mathrm{u}}^{*}\approx I$ within logarithmic gaps. The fact that $H_{\mathrm{c}}^{*}\le I+\log(I+2)+2$ is called the \emph{strong functional representation lemma} \cite{sfrl_trans,li2024pointwise}. 

In this section, we study the non-prefix setting, which, to the best of the author's knowledge, has not been studied before. We also utilize the Poisson functional representation scheme \cite{sfrl_trans}, with an improved analysis using properties of the discrete layered entropy. In the following theorem, we show that $\Lambda_{\mathrm{n}}^{*}\le I+1.29$. This bound is not only simpler than bounds on $H_{\mathrm{u}}^{*}$ such as \eqref{eq:Hu_prev}, but is also tight enough that, combined with \eqref{eq:cs_three_approx}, can give bounds on $H_{\mathrm{c}}^{*}$ and $H_{\mathrm{u}}^{*}$ that improve upon the previous best bound \eqref{eq:Hu_prev}. The proof is in Appendix \ref{subsec:pf_l_sfrl}.

\medskip{}

\begin{thm}
[Stronger strong functional representation lemma]\label{thm:l_sfrl}We have
\[
\Lambda_{\mathrm{n}}^{*}\le I+\Lambda(\mathrm{Geom}(1/2))<I+1.29,
\]
where $\mathrm{Geom}(1/2)$ is the geometric distribution with parameter $1/2$. More explicitly, for every (not necessarily discrete) random variables $X,Y$, there exists a (not necessarily discrete) random variable $S$ such that $S$ is independent of $X$, $H(Y|X,S)=0$, and
\[
\Lambda(Y|S)\le I+\Lambda(\mathrm{Geom}(1/2)),
\]
where $I:=I(X;Y)$. Hence, for every $\eta>0$,
\begin{align}
H_{\mathrm{c}}^{*}\le H_{\mathrm{u}}^{*} & <I+\log\left(I+e\eta+ \Lambda(\mathrm{Geom}(1/2)) \right) \nonumber \\
& \;\;\;\;\; -\log(e\eta)+\eta+ \Lambda(\mathrm{Geom}(1/2)).\label{eq:sfrl_eta}
\end{align}
\end{thm}
\medskip{}

We can use \eqref{eq:sfrl_eta} to derive some convenient bounds on $H_{\mathrm{c}}^{*}$ \eqref{eq:cs_h_c} and $H_{\mathrm{u}}^{*}$ \eqref{eq:cs_h_u}. If we want the smallest outer constant, substituting $\eta=\log e$ gives
\begin{equation}
H_{\mathrm{c}}^{*}\le H_{\mathrm{u}}^{*}<I+\log(I+5.22)+0.76,\label{eq:sfrl_loge}
\end{equation}
with an outer constant $0.76$. If we want the bound to be in the form $I+\log(I+1)+c$ like \cite{sfrl_trans}, substituting $\eta=0.6235$, and using the fact that $\log(I+b)\le\log(I+1)+\log b$ for $b\ge1$, gives
\[
H_{\mathrm{c}}^{*}\le H_{\mathrm{u}}^{*}<I+\log(I+1)+2.74.
\]
This constant $2.74$ is better than the constants given by (or implied by) the results in \cite{sfrl_trans,li2021unified,li2024pointwise}. If we want a simple bound that is usually good enough, substituting $\eta=0.77$ gives
\begin{equation}
H_{\mathrm{c}}^{*}\le H_{\mathrm{u}}^{*}<I+\log(I+3.4)+1.\label{eq:sfrl_alt}
\end{equation}

\smallskip{}

\begin{figure*}
\begin{centering}
\includegraphics[scale=0.47]{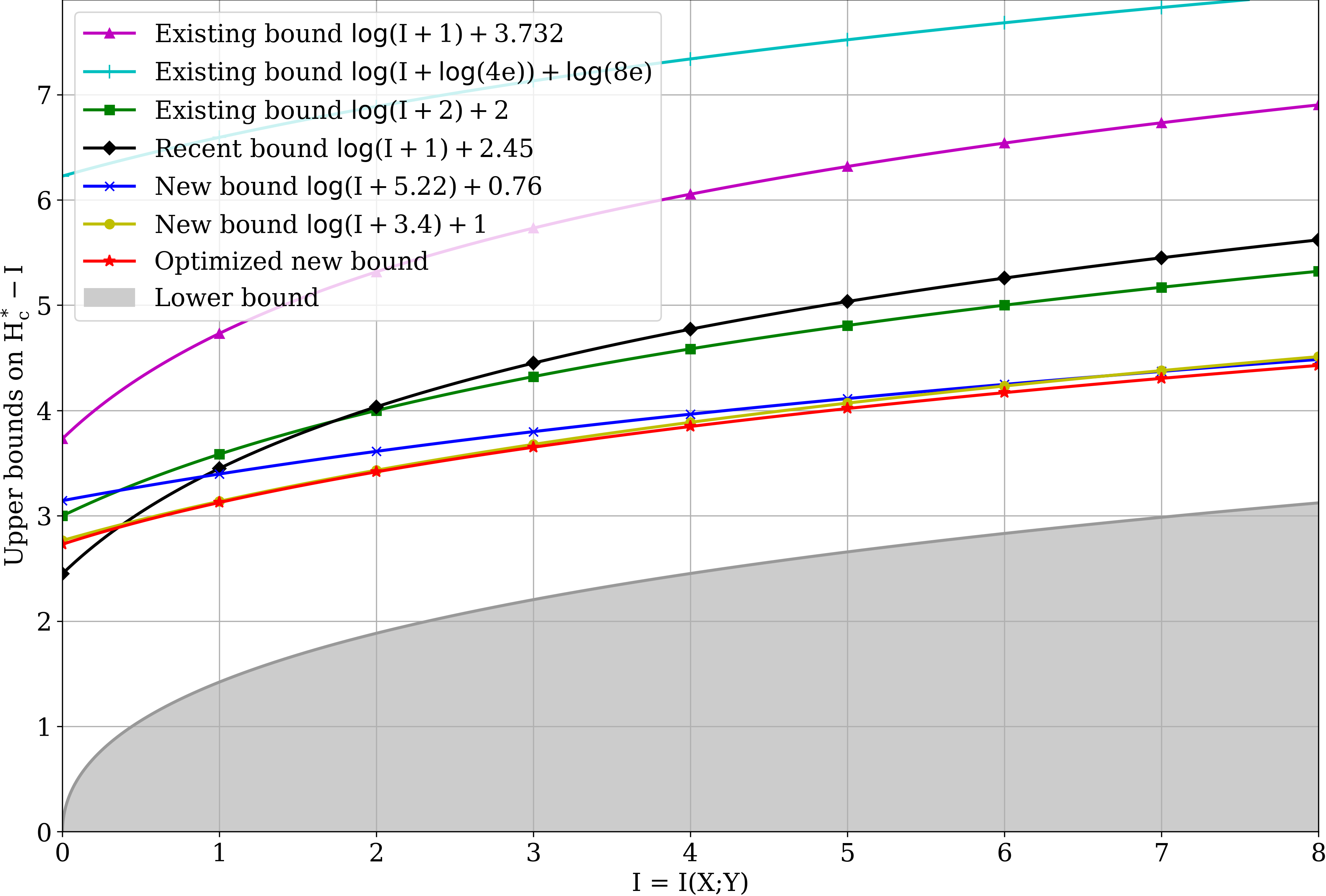}
\par\end{centering}
\caption{\label{fig:sfrl_plot}Comparison between Theorem \ref{thm:l_sfrl} and previous upper bounds \cite{li2021unified,li2024pointwise} on $H_{\mathrm{c}}^{*}-I$ for one-shot channel simulation.}
\end{figure*}

Refer to Figure \ref{fig:sfrl_plot} for various upper bounds on  $H_{\mathrm{c}}^{*}-I$. It includes the plot of the bound $\log(I+1)+3.732$  in \cite{li2021unified}; $\log(I+\log(4e))+\log(4e)$ in \cite{flamich2023adaptive} (via greedy rejection sampling \cite{harsha2010communication}); $\log(I+2)+2$ in \eqref{eq:Hu_prev}, \cite{li2024pointwise};
the bound $\log(I+1)+2.45$ in the recent preprint \cite{hill2026rejection},
our new bound \eqref{eq:sfrl_loge}; \eqref{eq:sfrl_alt}; and our new bound \eqref{eq:sfrl_eta} when optimized over $\eta$ (the optimal $\eta$ is $\frac{1}{2e}(\sqrt{\Lambda^{2}+4e\Lambda\log e}-\Lambda)$ where $\Lambda=I+1.29$). To show the tightness of our bounds, we also include the lower bound to be proved in Theorem \ref{thm:Hc_lb_continuous} (no upper bound on $H_{\mathrm{c}}^{*}-I$ can fall below that line).

From the plot, we can observe that \eqref{eq:sfrl_loge} is tighter than \eqref{eq:Hu_prev} for $I\ge0.4$, and \eqref{eq:sfrl_eta}, \eqref{eq:sfrl_alt} are tighter than \eqref{eq:Hu_prev} for every $I$,\footnote{A slight change to the analysis in \cite{li2024pointwise,li2024channel} can improve \eqref{eq:Hu_prev} to $H_{\mathrm{u}}^{*}\le I+\log(I+\eta+1)+\eta+\log(2/\eta)$ for every $\eta>0$ (call this the improved previous bound). Our  new bounds \eqref{eq:sfrl_eta}, \eqref{eq:sfrl_alt} are still tighter than the improved previous bound for every $I$.} and tighter than $\log(I+1)+2.45$ \cite{hill2026rejection} for $I \ge 0.4$. This means that in terms of theoretical bound, the current best way to perform channel simulation with conditional or unconditional prefix codes is to first use non-prefix codes, and then convert the non-prefix codeword into an unconditional prefix codeword (e.g., by the method in Appendix \ref{subsec:pf_LX_HX_bound}).

\smallskip{}

\subsection{Lossy Source Coding}

In the one-shot variable-length lossy source coding setting, the encoder observes a source $X\sim P_{X}$ (not necessarily discrete), and sends a variable-length description $M\in\{0,1\}^{*}$ (possibly produced by a stochastic encoding function) to the decoder. The decoder observes $M$ and outputs $Y$ (not necessarily discrete). We have the expected distortion constraint $\mathbb{E}[d(X,Y)]\le\mathsf{D}$, where $d:\mathcal{X}\times\mathcal{Y}\to\mathbb{R}$ is a distortion function, and $\mathsf{D}$ is the allowed expected distortion level. There are two variants of this setting: $M$ belongs to a non-prefix code (no constraint is imposed on $M$), or $M$ belongs to a prefix code. 

Lossy source coding can be performed via simulating the channel $P_{Y|X}$ which is the minimizer of the rate-distortion function $R(\mathsf{D}):=\inf_{P_{Y|X}:\,\mathbb{E}[d(X,Y)]\le\mathsf{D}}I(X;Y)$, as observed in \cite{winter2002compression}. In \cite{li2024pointwise} (which improves upon \cite{sfrl_trans}), it was proved that for the non-prefix code setting, there exists a scheme with
\begin{equation}
\mathbb{E}[|M|]\le R(\mathsf{D})+2.01,\label{eq:lossy_nonprefix_prev}
\end{equation}
and for the prefix code setting,
\begin{equation}
\mathbb{E}[|M|]\le R(\mathsf{D})+\log(R(\mathsf{D})+2)+4.01.\label{eq:lossy_prev}
\end{equation}

As a corollary of Theorem \ref{thm:l_sfrl}, we have the following bound for the prefix code setting, which improves upon \eqref{eq:lossy_prev}.\footnote{We remark that Theorem \ref{thm:l_sfrl} does not improve upon \eqref{eq:lossy_nonprefix_prev} for the non-prefix-free setting.}\smallskip{}

\begin{thm}
For one-shot lossy source coding with prefix codes, there exists a scheme with
\[
\mathbb{E}[|M|]\le R(\mathsf{D})+\log(R(\mathsf{D})+3.4)+3.
\]
\end{thm}
\smallskip{}

\begin{IEEEproof}
Fix any $P_{Y|X}$ with $\mathbb{E}[d(X,Y)]\le\mathsf{D}$, and let $I=I(X;Y)$. By Theorem \ref{thm:l_sfrl}, there exists $S\sim P_{S}$ independent of $X$ with $H(Y|X)\le I+\log(I+3.4)+0.99$. Applying the optimal conditional prefix code of $Y$ given $S$, we have $\mathbb{E}[|M|]\le H(Y|S)+1$. We also need to transmit $S$, which can be assumed to be binary by Carath\'{e}odory's theorem (see \cite[Theorem 2]{sfrl_trans}), resulting in an additional $1$-bit penalty. The result follows from minimizing over $P_{Y|X}$.
\end{IEEEproof}
\smallskip{}

\subsection{On the Optimal Bound for Channel Simulation with Non-prefix Codes}

Regarding the tightness of the constant in $\Lambda(\mathrm{Geom}(1/2))$ in $\Lambda_{\mathrm{n}}^{*}\le I+\Lambda(\mathrm{Geom}(1/2))$ in Theorem \ref{thm:l_sfrl}, we can study the optimal constant $c$ such that $\Lambda_{\mathrm{n}}^{*}\le I+c$ always holds. The optimal constant is defined as 
\begin{align}
c_{\mathrm{n}}^{*} & :=\sup_{p_{X,Y}}\left(\Lambda_{\mathrm{n}}^{*}(X\to Y)-I(X;Y)\right),
\end{align}
where the supremum is over finite discrete joint distributions $p_{X,Y}$. Theorem \ref{thm:l_sfrl} gives $c_{\mathrm{n}}^{*}\le\Lambda(\mathrm{Geom}(1/2))$.  For lower bounds, note that $c_{\mathrm{n}}^{*}\ge0$ since $\Lambda_{\mathrm{n}}^{*}(X\to Y)=\Lambda(X)$ when $X=Y$, so $\Lambda_{\mathrm{n}}^{*}(X\to Y)=I(X;Y)$ when $X=Y$ is uniformly distributed. We can actually find examples of $X,Y$ where $\Lambda_{\mathrm{n}}^{*}-I$ is bounded away from $0$. To this end, we use the following lower bound. The proof is in Appendix \ref{subsec:pf_sfrl_n_lb}.\smallskip{}

\begin{thm}
\label{thm:sfrl_n_lb}For discrete $Y$,\footnote{We assume that $Y$ is discrete for the sake of simplicity of the analysis. We believe that Theorem \ref{thm:sfrl_n_lb} is true in general.} we have $\Lambda_{\mathrm{n}}^{*}(X\to Y)\ge\underline{\Lambda}_{\mathrm{n}}^{*}(X\to Y)$, where
\begin{align*}
\underline{\Lambda}_{\mathrm{n}}^{*}(X\to Y) & :=\int_{0}^{1}\gamma(\mathbb{E}_{Y}[Q_{Y}(\tau)])\mathrm{d}\tau,
\end{align*}
\[
\gamma(t):=(1-\lceil t\rceil+t)\lceil t\rceil\log\lceil t\rceil+(\lceil t\rceil-t)\lceil t-1\rceil\log\lceil t-1\rceil,
\]
and
\[
Q_{y}(\tau):=\inf\bigg\{ t\ge0:\,\mathbb{P}\bigg(\frac{p_{Y|X}(y|X)}{p_{Y}(y)}\le t\bigg)\ge\tau\bigg\}
\]
is the quantile (inverse cdf) function of $p_{Y|X}(y|X)/p_{Y}(y)$ where $X\sim P_{X}$.\footnote{If we let $T\sim\mathrm{Unif}([0,1])$, $R_{y}:=Q_{y}(T)$, then $(R_{y})_{y}$ is called the \emph{quantile coupling} \cite{pollard2002user} of the distributions of $p_{Y|X}(y|X)/p_{Y}(y)$, and $\underline{\Lambda}_{\mathrm{n}}^{*}(X\to Y)=\mathbb{E}[\gamma(\mathbb{E}[R_{Y}|T])]$.}
\end{thm}

\smallskip{}

Consider the following example: $X\sim\mathrm{Unif}[0:a-1]$, $Z\sim\mathrm{Unif}[0:b-1]$, $Y=X+Z\;\mathrm{mod}\;a$, where $a/2\le b\le a$. Let $\theta:=b/a\in[1/2,1]$. We have $Q_{y}(\tau)=\theta^{-1}\mathbf{1}\{\tau\ge1-\theta\}$. Hence, Theorem \ref{thm:sfrl_n_lb} gives $\Lambda_{\mathrm{n}}^{*}\ge\underline{\Lambda}_{\mathrm{n}}^{*}=\theta\gamma(\theta^{-1})=2(1-\theta)$, and hence $\Lambda_{\mathrm{n}}^{*}-I\ge2(1-\theta)+\log\theta$. Letting $\theta\to(\log e)/2$, we have the lower bound $\log\log e-\log e+1>0.086$. In sum, we have
\[
0.086<c_{\mathrm{n}}^{*}\le\Lambda(\mathrm{Geom}(1/2))<1.29.
\]
It is left for future studies to find the exact value of $c_{\mathrm{n}}^{*}$.

\smallskip{}

\begin{rem}
It is perhaps unexpected that $c_{\mathrm{n}}^{*}$ is a nonzero constant. By Proposition \ref{thm:useful_def}, $c_{\mathrm{n}}^{*}$ can be defined as
\begin{equation}
c_{\mathrm{n}}^{*}=\sup_{p_{X,Y}}\Big(\inf_{p_{S,T|X,Y}}H(Y|S,T)-I(X;Y)\Big),\label{eq:cnstar_firstorder}
\end{equation}
where the infimum is over all jointly-distributed finite discrete random variables $S,T$ under the following constraint: $I(X;S)=0$, and for all $U$ (jointly distributed with $X,Y,S,T$), $H(Y|S,T,U)=0$ implies $H(U|S)\ge H(Y|S)$. 

It is surprising that such a simple formula involving only supremum, infimum and linear combinations of entropy and mutual information gives rise to a finite nonzero constant, as this makes \eqref{eq:cnstar_firstorder} a non-homogeneous equation even though it only involves linear combinations of entropy. Most examples of such formulae evaluates to $-\infty$, $\infty$ or $0$ (e.g., $\inf_{X,Y,S:\,I(X;S)=0}(H(Y|S)-I(X;Y))=0$) since most inequalities among entropy terms are homogeneous (e.g., $H(Y|S)\ge I(X;Y)$ if $I(X;S)=0$).  The author is unaware of any other formula that involves only the linear combinations of entropies among $5$ random variables that gives a finite nonzero constant.\footnote{It is possible to define various constants and functions via entropy \cite{li2021first}, though those expressions tend to involve a large number of random variables.}

There are other examples of additive constants in information theory, such as the upper bound $H(X)+1$ in the optimal prefix code \cite{shannon1948mathematical}, or the upper bound $H(X)+2$ in random number generation \cite{knuth1976complexity}. These constants ``$1$'' and ``$2$'' are ``round-off errors'' which are necessary since prefix codes and random number generation concern bit sequences which must have integer lengths. Hence, these constants are tied to the operational encoding constraints, and will change if a ternary code is used instead of a binary code. On the other hand, $\Lambda_{\mathrm{n}}^{*}\le I+c_{\mathrm{n}}^{*}$ is stated purely in terms of entropy and mutual information \eqref{eq:cnstar_firstorder}, and is not dependent on the particular (binary/ternary) encoding.\footnote{For the optimal prefix code, the expected length is bounded by $\mathbb{E}[|M_{2}|]<H(X)+1$ for binary encoding $M_{2}$, $\mathbb{E}[|M_{3}|]<(\log_{3}2)H(X)+1$ for ternary encoding $M_{3}$. We can see that these two bounds are not merely a change of a multiplicative constant. On the other hand, $\Lambda_{\mathrm{n}}^{*}\le I+c_{\mathrm{n}}^{*}$ is not dependent on the encoding. Changing the base of entropy and mutual information will merely multiply both sides by the same constant. Hence, $\mathbb{E}[|M_{2}|]<H(X)+1$ is a property of binary prefix codes, whereas $\Lambda_{\mathrm{n}}^{*}\le I+c_{\mathrm{n}}^{*}$ is a fundamental property of entropy and mutual information.} Therefore, the constant $c_{\mathrm{n}}^{*}$ is, in a certain sense, a fundamental constant about entropy.
\end{rem}
\smallskip{}

\subsection{On the Optimal Bound for Channel Simulation with Prefix Codes\label{sec:chansim_lb}}

Unlike the non-prefix case where the bound $\Lambda_{\mathrm{n}}^{*}\le I+1.29$ involves only one constant, the bound $H_{\mathrm{c}}^{*}<I+\log(I+5.22)+0.76$ for the prefix case in Theorem \ref{thm:l_sfrl} involves two constants. The outer constant $0.76$ is more important than the inner constant $5.22$ since $I+\log(I+5.22)+0.76=I+\log I+0.76+o(1)$ as $I\to\infty$. Therefore, we study the infimum of the outer constant $c$ such that there exists $b$ such that $H_{\mathrm{c}}^{*}\le I+\log(I+b)+c$ always holds. The optimal outer constant can be defined as
\begin{align}
c_{\mathrm{c}}^{*} & :=\underset{I\to\infty}{\mathrm{limsup}}\;\sup_{p_{X,Y}:\,I(X;Y)=I}\left(H_{\mathrm{c}}^{*}(X\to Y)-I-\log I\right),
\end{align}
where the supremum is over finite discrete joint distributions $p_{X,Y}$ with $I(X;Y)=I$. The results in \cite{braverman2014public} imply that $c_{\mathrm{c}}^{*}$ is finite.
Proposition \ref{prop:LX_HX_bound} gives $c_{\mathrm{c}}^{*}\le c_{\mathrm{n}}^{*}-\log\log e$, and Theorem \ref{thm:l_sfrl} gives $c_{\mathrm{c}}^{*}\le\Lambda(\mathrm{Geom}(1/2))-\log\log e\le0.76$. 

In this subsection, we study lower bounds on $c_{\mathrm{c}}^{*}$ and $H_{\mathrm{c}}^{*}$. It was shown in \cite[Proposition 1]{sfrl_trans} that $H_{\mathrm{c}}^{*}\ge\underline{H}_{\mathrm{c}}^{*}$ when $Y$ is discrete, where
\begin{align}
\underline{H}_{\mathrm{c}}^{*} & :=-\sum_{y}\int_{0}^{1}\big(\mathbb{P}(p_{Y|X}(y|X)\ge t) \nonumber \\
 & \qquad\qquad\cdot\log\mathbb{P}(p_{Y|X}(y|X)\ge t)\big)\mathrm{d}t.
\end{align}
By \eqref{eq:layered_def}, when $X$ is uniform, we can express $\underline{H}_{\mathrm{c}}^{*}$ in terms of $\Lambda(X|Y)$.

\medskip{}

\begin{prop}
\label{prop:chansim_unif_lb}If $X\sim\mathrm{Unif}(\mathcal{X})$ where $\mathcal{X}$ is finite, we have
\begin{equation}
H_{\mathrm{c}}^{*}\ge\underline{H}_{\mathrm{c}}^{*}=\log|\mathcal{X}|-\Lambda(X|Y).\label{eq:Psilb_lambda}
\end{equation}
\end{prop}
\medskip{}

\begin{IEEEproof}
We use an argument similar to \cite{goc2024causal}. Write $\ell(t):=-t\log t$. We have
\begin{align*}
\underline{H}_{\mathrm{c}}^{*} & =\sum_{y}\int_{0}^{1}\ell\big(\mathbb{P}(p_{Y|X}(y|X)\ge t)\big)\mathrm{d}t\\
 & =\sum_{y}\int_{0}^{1}\ell\left(|\{x:\,p_{Y|X}(y|x)\ge t\}|/|\mathcal{X}|\right)\mathrm{d}t\\
 & =\log|\mathcal{X}|+\frac{1}{|\mathcal{X}|}\sum_{y}\int_{0}^{1}\ell\left(|\{x:\,p_{Y|X}(y|x)\ge t\}|\right)\mathrm{d}t\\
 & =\log|\mathcal{X}|+\frac{1}{|\mathcal{X}|}\sum_{y}\int_{0}^{1}\ell\left(\left|\left\{ x:\frac{p_{X,Y}(x,y)}{1/|\mathcal{X}|}\ge t\right\} \right|\right)\mathrm{d}t\\
 & =\log|\mathcal{X}|+\sum_{y}p_{Y}(y)\int_{0}^{\infty}\ell\left(\left|\left\{ x:\,p_{X|Y}(x|y)\ge t\right\} \right|\right)\mathrm{d}t\\
 & =\log|\mathcal{X}|-\Lambda(X|Y),
\end{align*}
where the last equality is by \eqref{eq:layered_def}.
\end{IEEEproof}
\medskip{}

The following direct corollary allows us to find channels where $H_{\mathrm{c}}^{*}$ is lower-bounded. It also allows us to express $\Lambda(Z)$ in terms of $\underline{H}_{\mathrm{c}}^{*}$.

\medskip{}

\begin{cor}
\label{cor:Hc_lb}For finite discrete $Z\in[0:|\mathcal{Z}|-1]$, letting $Y\sim\mathrm{Unif}[0:|\mathcal{Z}|-1]$, $X=Y+Z\;\mathrm{mod}\;|\mathcal{Z}|$, we have
\[
H(Z)=\log|\mathcal{Z}|-I(X;Y),\;\;\Lambda(Z)=\log|\mathcal{Z}|-\underline{H}_{\mathrm{c}}^{*}.
\]
Hence,
\begin{align*}
H_{\mathrm{c}}^{*}\ge\underline{H}_{\mathrm{c}}^{*} & =\log|\mathcal{Z}|-\Lambda(Z)=I(X;Y)+H(Z)-\Lambda(Z).
\end{align*}
\end{cor}
\medskip{}

Therefore, we can give impossibility bounds for the strong functional representation lemma by finding examples of $p_{Z}$ with the right trade-off between $\log|\mathcal{Z}|$, $H(Z)$ and $\Lambda(Z)$. For example, the $p_{Z}$ in \cite[Proposition 2]{sfrl_trans} has $\log|\mathcal{Z}|=k$ (where $k\ge2$ is an integer),
\[
H(Z)=\frac{1}{2}k+\log(k+2)-\frac{3}{2}+\frac{1}{k+2},
\]
\begin{equation}
\Lambda(Z)=\frac{1}{2}k+\frac{1}{2}-\frac{1}{k+2}.\label{eq:example_Lambda}
\end{equation}
This gives $H_{\mathrm{c}}^{*}\ge I+\log(I+1)-1$ \cite{sfrl_trans}, which implies $c_{\mathrm{c}}^{*}\ge-1$. 

To give a tighter lower bound on $c_{\mathrm{c}}^{*}$, we will require continuous $X,Y$. $\underline{H}_{\mathrm{c}}^{*}$ can be generalized to the case where $Y$ is a continuous or general random variable, via the channel simulation divergence \cite{goc2024causal,flamich2025redundancy} given as 
\begin{align}
D_{\mathrm{CS}}(P\Vert Q) & :=-\int_{0}^{\infty}Q(\{x:(\mathrm{d}P/\mathrm{d}Q)(x)\ge t\})\nonumber \\
 & \qquad\quad\cdot\log Q(\{x:(\mathrm{d}P/\mathrm{d}Q)(x)\ge t\})\mathrm{d}t.\label{eq:cs_div}
\end{align}
It was shown in \cite{flamich2025redundancy} that
\begin{equation}
H_{\mathrm{c}}^{*}(X\to Y)\ge\mathbb{E}_{Y}\left[D_{\mathrm{CS}}(P_{X|Y}(\cdot|Y)\Vert P_{X})\right].\label{eq:Hc_lb_general}
\end{equation}
The right-hand side in \eqref{eq:Hc_lb_general} is referred to as the \emph{functional information} in a recent work \cite{hill2026rejection}, and coincides with $\underline{H}_{\mathrm{c}}^{*}$ when $Y$ is discrete \cite{goc2024causal}, and hence is a generalization of $\underline{H}_{\mathrm{c}}^{*}$ to general $Y$. We use \eqref{eq:Hc_lb_general} to give examples with a tighter bound on $H_{\mathrm{c}}^{*}$.

\medskip{}

\begin{thm}
\label{thm:Hc_lb_continuous}For every $t>0$, there exists jointly-continuous $P_{X,Y}$ such that $H_{\mathrm{c}}^{*}(X\to Y)\ge t$ and 
\[
I(X;Y)=t-\log(t+\log e)+\log\log e.
\]
\end{thm}
\medskip{}

\begin{IEEEproof}
Let $\alpha=t/(t+\log e)$. Consider $Z\in[0,1]$ with probability density function $f_{Z}(z)=(1-\alpha)z^{-\alpha}$, $X\sim\mathrm{Unif}(0,1)$, $Y=X+Z\;\mathrm{mod}\;1$. Direct computation gives $H_{\mathrm{c}}^{*}\ge\mathbb{E}_{Y}[D_{\mathrm{CS}}(P_{X|Y}(\cdot|Y)\Vert P_{X})]=(\alpha\log e)/(1-\alpha)=t$ and $I(X;Y)=(\alpha\log e)/(1-\alpha)+\log(1-\alpha)=t-\log(t+\log e)+\log\log e$. 
\end{IEEEproof}
\medskip{}

Theorem \ref{thm:Hc_lb_continuous} gives\footnote{Although the definition of $c_{\mathrm{c}}^{*}$ requires $X,Y$ to be discrete, we can quantize the $X,Y$ in the proof of Theorem \ref{thm:Hc_lb_continuous} to form $\hat{X}=\delta \lfloor X / \delta \rfloor$, $\hat{Y}=\delta \lfloor Y / \delta \rfloor$, and note that $\mathbb{E}_{Y}[D_{\mathrm{CS}}(P_{\hat{X}|\hat{Y}}(\cdot|\hat{Y})\Vert P_{\hat{X}})] \to t$ and $I(\hat{X};\hat{Y}) \to I(X;Y)$ as $\delta \to 0$.}
\[
-\log\log e\le c_{\mathrm{c}}^{*}\le\Lambda(\mathrm{Geom}(1/2))-\log\log e.
\]
Numerically, $-0.53<c_{\mathrm{c}}^{*}<0.76$. We conjecture that the lower bound is tight, i.e., $c_{\mathrm{c}}^{*}=-\log\log e$. This means that if we use entropy and logarithm to the base $e$ instead of base $2$, then $c_{\mathrm{c}}^{*}$ would be simply $0$. Also, note that Theorem \ref{thm:Hc_lb_continuous} means that \eqref{eq:sfrl_alt} is optimal within $\log(3.4)+1<2.8$ bits, that is, if we have to bound $H_{\mathrm{c}}^{*}$ in terms of $I(X;Y)$, then it is impossible to improve upon \eqref{eq:sfrl_alt} by more than $2.8$ bits.\medskip{}

\begin{rem}
By \eqref{eq:layered_def}, if $p$ is a distribution over a finite set $\mathcal{X}$, 
\[
\Lambda(p)=\log|\mathcal{X}|-D_{\mathrm{CS}}(p\Vert\mathrm{Unif}(\mathcal{X})).
\]
We can compare this equality with a similar equality for Shannon entropy, where the channel simulation divergence is replaced with the Kullback-Leibler divergence: 
\[
H(p)=\log|\mathcal{X}|-D_{\mathrm{KL}}(p\Vert\mathrm{Unif}(\mathcal{X})).
\]
Refer to Section \ref{sec:horizontal} for further discussion on this similarity.
\end{rem}
\medskip{}

\begin{rem}
\label{rem:oneshot_asymp}More generally, it may be of interest to find the best upper bound on $H_{\mathrm{c}}^{*}$ that can be stated solely in terms of $I$. The best bound is defined as
\[
H_{\mathrm{c}}^{**}(I):=\sup_{p_{X,Y}:\,I(X;Y)=I}H_{\mathrm{c}}^{*}(X\to Y),
\]
which is the worst-case $H_{\mathrm{c}}^{*}$ over all finite discrete $X,Y$ with mutual information $I$, or equivalently, the best upper bound in the form of $H_{\mathrm{c}}^{*}\le f(I)$ for some function $f$. We would like to characterize the asymptotic behavior of $H_{\mathrm{c}}^{**}(I)$ as $I\to\infty$. By definition of $c_{\mathrm{c}}^{*}$, we have $H_{\mathrm{c}}^{**}(I)\le I+\log I+c_{\mathrm{c}}^{*}+o(1)$.  Together with the example in \eqref{eq:example_Lambda}, we know 
\begin{equation}
H_{\mathrm{c}}^{**}(I)=I+\log I+o(\log I).\label{eq:Hcstar2_known}
\end{equation}
We conjecture that
\[
H_{\mathrm{c}}^{**}(I)=I+\log I+c_{\mathrm{c}}^{*}+o(1)
\]
as $I\to\infty$. If this is true, then $c_{\mathrm{c}}^{*}$ is the ``third-order term'' of the worst-case asymptotic behavior of one-shot channel simulation with prefix codes as the mutual information grows. Like channel coding where the second-order term \cite{hayashi2009information,polyanskiy2010channel} and the third-order term \cite{tomamichel2013tight} has been thoroughly investigated, we believe that it is worthwhile to characterize the third-order term of worst-case one-shot channel simulation with prefix codes as well.
\end{rem}
\smallskip{}

\begin{rem}
Variable-length channel simulation has also been studied in the i.i.d. asymptotic setting where $X^{n}\sim P_{X}^{\otimes n}$ is an i.i.d. source, and the channel to be simulated $P_{Y^{n}|X^{n}}=P_{Y|X}^{\otimes n}$ is a memoryless channel. This is different from the worst-case asymptotic setting in Remark \ref{rem:oneshot_asymp} since Remark \ref{rem:oneshot_asymp} only requires $I\to\infty$, i.e., it concerns the $H_{\mathrm{c}}^{*}$ of the worst case family of channels $P_{X^{(I)},Y^{(I)}}$ with $I(X^{(I)};Y^{(I)})=I$ for $I\in[0,\infty)$, without the i.i.d. restriction.  In the i.i.d. asymptotic setting, $H_{\mathrm{u}}^{*}(X^{n}\to Y^{n})=nI(X;Y)+o(n)$ \cite{bennet2014reverse} (and hence all of $\Lambda_{\mathrm{n}}^{*},H_{\mathrm{c}}^{*},H_{\mathrm{u}}^{*}$ are $nI(X;Y)+o(n)$). If $X,Y$ are discrete and $P_{Y|X}$ is non-singular,\footnote{$P_{Y|X}$ is non-singular if $\mathrm{d}P_{Y|X}/\mathrm{d}P_{Y}$ is not a deterministic function of $Y$ \cite{sriramu2024optimal}.} then it was shown in \cite{sriramu2024optimal,flamich2025redundancy} that
\begin{equation}
H_{\mathrm{c}}^{*}(X^{n}\to Y^{n})=nI(X;Y)+\frac{1}{2}\log n+o(\log n).\label{eq:asymp_iid}
\end{equation}
Compared to \eqref{eq:Hcstar2_known} with a second-order term $\log I$, the i.i.d. setting has a smaller second-order term $(1/2)\log n$, since  the central limit theorem applies to the i.i.d. setting but not the worst-case setting. The asymptotic behavior of $\Lambda_{\mathrm{n}}^{*}(X^{n}\to Y^{n})$ and $H_{\mathrm{u}}^{*}(X^{n}\to Y^{n})$ is left for future studies.
\end{rem}
\smallskip{}

\section{Better Approximations of $H$ via Discrete $m$-Layered Entropy\label{sec:mlayered}}

In Section \ref{sec:piecewise_approx}, we have seen that $\Lambda$ is a piecewise linear approximation of $H$ that is useful in linear programming tasks. We now discuss tighter piecewise linear approximations of $H$. We have proved in \eqref{eq:bounded_increase} that $\Lambda(X,Y)-\Lambda(Y)\le H(X)$. This inequality is actually tight, and we can have arbitrarily good approximations of $H(X)$ using $\Lambda(X,Y)-\Lambda(Y)$. 

\medskip{}

\begin{prop}
\label{prop:sup_gap}For any $X$, we have
\[
\sup_{p_{Y|X}}\big(\Lambda(X,Y)-\Lambda(Y)\big)=\sup_{p_{Y}}\big(\Lambda(X,Y)-\Lambda(Y)\big)=H(X),
\]
where the first supremum is over $p_{Y|X}$ ($Y$ is jointly distributed as $X$), and the second supremum is over $p_{Y}$ ($Y$ is independent of $X$).
\end{prop}
\medskip{}

\begin{IEEEproof}
The supremums are upper-bounded by $H(X)$ by \eqref{eq:bounded_increase}. To complete the proof, consider an i.i.d. sequence $(X_{i})_{i}$  following $p_{X}$. By Proposition \ref{prop:LX_HX_bound},
\begin{align}
 & \sup_{p_{Y}}(\Lambda(X_{1},Y)-\Lambda(Y))\nonumber \\
 & \stackrel{(a)}{\ge} \max_{k\in[n]}(\Lambda(X_{1},X_{2},\ldots,X_{k})-\Lambda(X_{2},\ldots,X_{k}))\nonumber \\
 & \ge n^{-1}\sum_{k=1}^{n}(\Lambda(X_{1},X_{2},\ldots,X_{k})-\Lambda(X_{2},\ldots,X_{k}))\nonumber \\
 & =n^{-1}\Lambda(X^{n})\nonumber \\
 & \stackrel{(b)}{\ge} n^{-1}H(X^{n})-n^{-1}\log\left(1+\frac{H(X^{n})}{e\log e}\right)-n^{-1}\log e\nonumber \\
 & =H(X)-\frac{1}{n}\log\left(e+\frac{nH(X)}{\log e}\right),\label{eq:delta_sum}
\end{align}
where (a) is by substituting $Y=(X_2,\ldots,X_k)$, and (b) is due to Proposition~\ref{prop:LX_HX_bound}.
Taking $n\to\infty$ completes the proof. 
\end{IEEEproof}
\medskip{}

Intuitively, Proposition \ref{prop:sup_gap} states that the Shannon entropy $H(X)$ is the largest possible increase of $\Lambda(Y)$ by including the information in $X$. Proposition \ref{prop:sup_gap} together with \eqref{eq:alt_cond} also reveals a simple alternative definition of the Shannon entropy, which may be of interest outside of the study of discrete layered entropy.

\medskip{}

\begin{prop}
We have
\begin{equation}
H(X)=\sup_{p_{Y,\mathcal{A}|X}}\mathbb{E}\left[\log\frac{|\mathcal{A}|}{|\mathcal{A}_{X}|}\right],\label{eq:HX_logA}
\end{equation}
where the supremum is over conditional probability mass functions  $p_{Y,\mathcal{A}|X}$ from $X\in\mathcal{X}$ to $Y\in\mathbb{N}$ and $\mathcal{A}\in2^{\mathcal{X}\times\mathbb{N}}$ (i.e., $\mathcal{A}\subseteq\mathcal{X}\times\mathbb{N}$ is a random set) such that $|\mathcal{A}|$ is almost surely finite, and $(X,Y)|\mathcal{A}\sim\mathrm{Unif}(\mathcal{A})$ (i.e., conditional on $\mathcal{A}=a$, $(X,Y)$ is uniformly distributed over $a$). We denote the section of $\mathcal{A}$ as $\mathcal{A}_{x}:=\{y:\,(x,y)\in\mathcal{A}\}$.
\end{prop}
\medskip{}

\begin{IEEEproof}
For the ``$\le$'' direction of \eqref{eq:HX_logA}, by \eqref{eq:alt_cond}, we can consider the $\mathcal{A}$ satisfying $\mathbb{E}[\log|\mathcal{A}|]=\Lambda(X,Y)$. Again by \eqref{eq:alt_cond}, $\mathbb{E}[\log|\mathcal{A}_{X}|]\le\Lambda(Y)$, and hence $\mathbb{E}[\log(|\mathcal{A}|/|\mathcal{A}_{X}|)]\ge\Lambda(X,Y)-\Lambda(Y)$. The result follows from Proposition \ref{prop:sup_gap}. For the ``$\ge$'' direction, We have $\mathbb{E}[\log(|\mathcal{A}|/|\mathcal{A}_{X}|)]=H(X,Y|\mathcal{A})-H(Y|\mathcal{A},X)=H(X|\mathcal{A})\le H(X)$.
\end{IEEEproof}
\medskip{}

We can now define a generalization of $\Lambda(X)$ by restricting the cardinality of $Y$ in Proposition \ref{prop:sup_gap}.

\medskip{}

\begin{defn}
[Discrete $m$-layered entropy]For a discrete random variable $X$ and integer $m\ge1$, define the\emph{ discrete }$m$\emph{-layered entropy} as
\[
\Lambda^{[m]}(X):=\max_{p_{Y|X}:\,Y\in[m]}\big(\Lambda(X,Y)-\Lambda(Y)\big).
\]
We take $\Lambda^{[\infty]}(X):=H(X)$. 
\end{defn}
\medskip{}

\begin{figure*}
\begin{centering}
\includegraphics[scale=0.36]{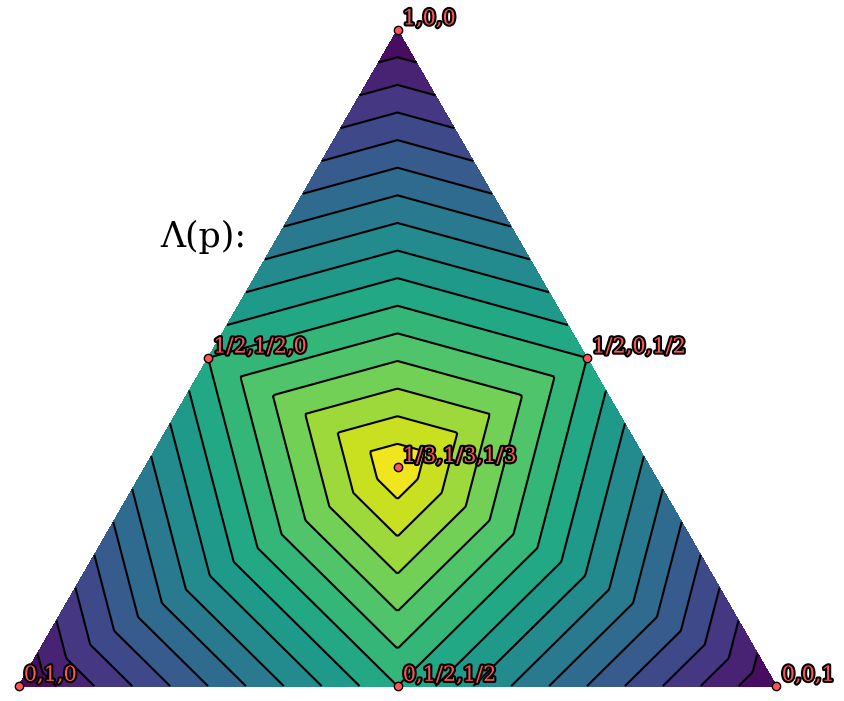}$\;\;\;$\includegraphics[scale=0.36]{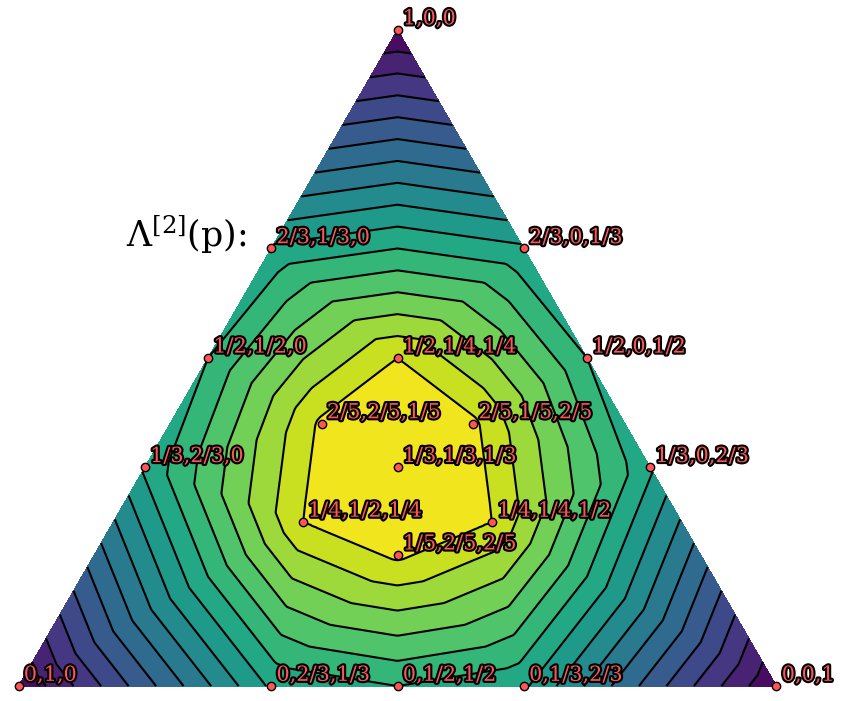}
\par\end{centering}
\vspace{-1pt}

\begin{centering}
\includegraphics[scale=0.36]{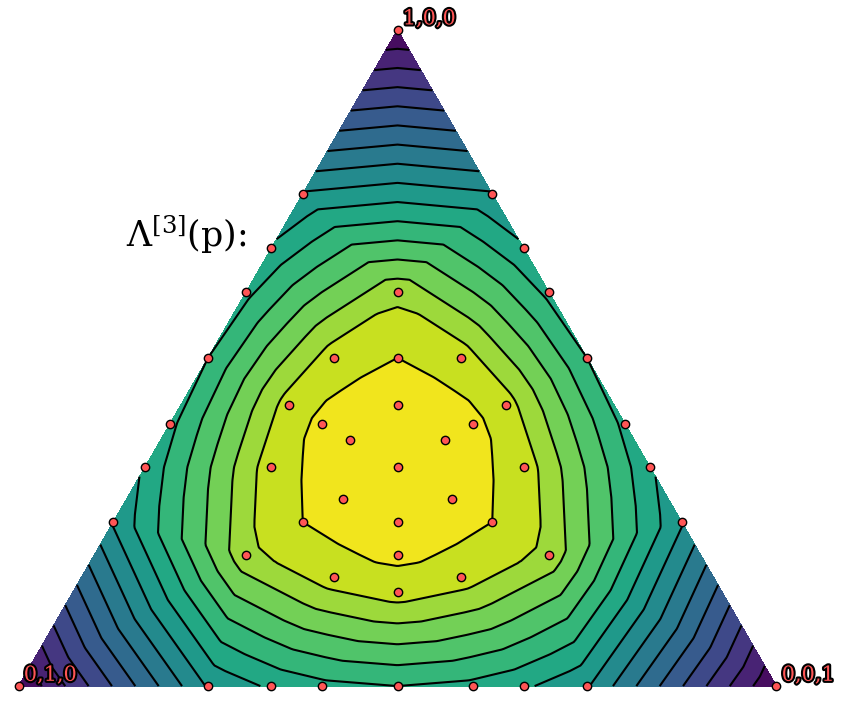}$\;\;\;$\includegraphics[scale=0.36]{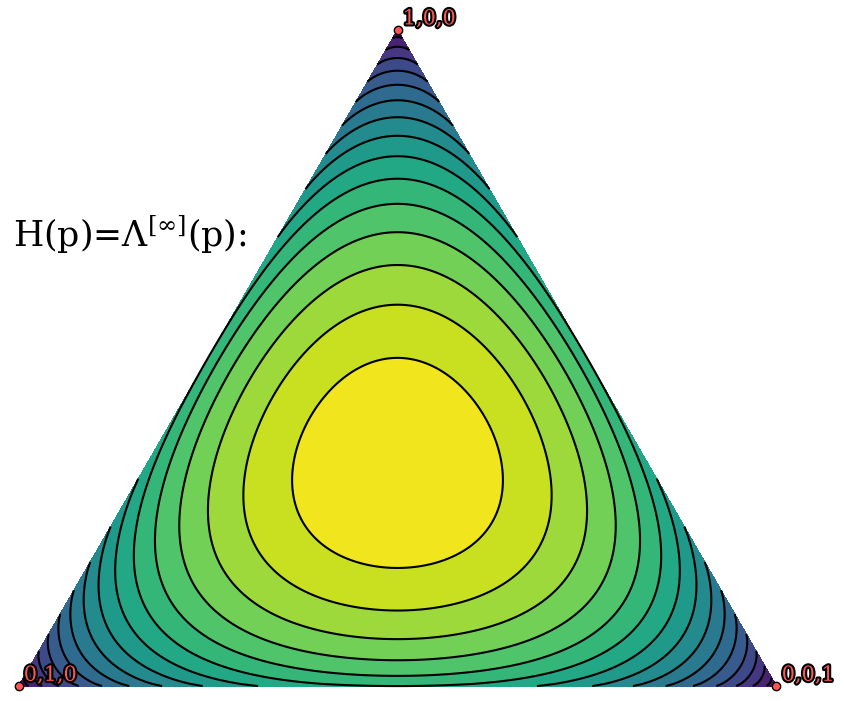}
\par\end{centering}

%
\caption{\label{fig:m_layered}Contour plot of the discrete $m$-layered entropy $\Lambda^{[m]}(p)$ for $m\in\{1,2,3,\infty\}$ and $p:\{1,2,3\}\to[0,1]$ being a ternary probability mass function. The red points are the points where $\Lambda^{[m]}(p)=H(p)$. They are also the vertices of the polytope $\{(p,z)\in\mathbb{R}^{3}\times\mathbb{R}:\,0\le z\le\Lambda^{[m]}(p)\}$.}
\end{figure*}

Refer to Figure \ref{fig:m_layered} for a plot. Note that $\Lambda^{[1]}(X)=\Lambda(X)$ and $\lim_{m\to\infty}\Lambda^{[m]}(X)=H(X)$ by Proposition \ref{prop:sup_gap}. Hence, $\Lambda^{[m]}$ gives a family of information measures that generalizes $\Lambda$ and $H$, and provides arbitrarily good approximations for $H$.

Operationally, if we regard $\Lambda(Y)$ as the (approximate, expected) encoding length of $Y$ using a one-to-one (non-prefix) code, then $\Lambda(X,Y)-\Lambda(Y)$ is the increase in encoding length if we also need to encode $X$. Since the code is non-prefix, we cannot simply concatenate the encoding of $X$ to the encoding of $Y$, and hence the increase $\Lambda(X,Y)-\Lambda(Y)$ can be greater than $\Lambda(X)$. Nevertheless, the increase cannot be greater than $H(X)$ by Proposition \ref{prop:sup_gap}. Intuitively, this is because we can always concatenate the prefix-free encoding of $X$ with the one-to-one (non-prefix-free) encoding of $Y$ to give a one-to-one encoding of $(X,Y)$, since we can decode the concatenation by decoding $X$ and then $Y$.

The value of $\Lambda^{[m]}(X)$ is the worst-case increase in encoding length to also encode $X$ if we already have to encode $Y$ using a one-to-one code, where the worst-case is taken over all $m$-ary random variables $Y$ jointly distributed with $X$. Unlike $\Lambda(X)$ which concerns the cost of a one-to-one encoding of $X$ in isolation, $\Lambda^{[m]}(X)$ concerns the cost of encoding $X$ as a part of a larger file, and $m$ specifies how large the other parts of the file can be.

We now present several properties of $\Lambda^{[m]}(X)$. In particular, the linear programming form makes $\Lambda^{[m]}(X)$ useful as an arbitrarily close approximation of $H(X)$ that can be incorporated into a linear program (see Section \ref{sec:piecewise_approx}). The proofs are in Appendix \ref{subsec:pf_m_layered}.

\medskip{}

\begin{prop}
\label{prop:m_layered}$\Lambda^{[m]}(X)$ satisfies:
\begin{enumerate}
\item $\Lambda^{[1]}(X)=\Lambda(X)$, and $\Lambda^{[m]}(X)$ is non-decreasing in $m$.
\item (Upper bound) $\Lambda^{[m]}(X)\le H(X)$. Equality holds if and only if 
\[
\frac{\max_{x\in\mathcal{X}}p_{X}(x)}{\mathrm{gcd}((p_{X}(x))_{x\in\mathcal{X}})}\le m,
\]
or equivalently, there exists a function $g:\mathcal{X}\to\{0,\ldots,m\}$ such that $p_{X}(x)=g(x)/\sum_{x'}g(x')$. 
\item (Lower bound) $\Lambda^{[m]}(X)\ge H_{\infty}(X)$. Equality holds if and only if $X$ is uniform.
\item (Approximation of Shannon entropy)  $\lim_{m\to\infty}\Lambda^{[m]}(X)=H(X)$. If $\mathcal{X}$ is finite, 
\begin{align*}
\Lambda^{[m]}(X) & \ge H(X)-\frac{\log(\ln m+e)}{\log m}\log|\mathcal{X}|\\
 & =H(X)-O\Big(\frac{\log\log m}{\log m}\Big).
\end{align*}
\item (Concavity) $\Lambda^{[m]}(X)$ is concave and Schur-concave.
\item (Linear programming form)
\begin{align}
\Lambda^{[m]}(X) & =\max\mathbb{E}\big[\log K\nonumber \\
 & \quad-\left(Y\log Y-(Y-1)\log(Y-1)\right)\big],\label{eq:mlayered_lp}
\end{align}
where the maximum is over joint probability mass functions $p_{X,Y,K}$ over $\mathcal{X}\times[m]\times[m|\mathcal{X}|]$ with $X$-marginal that coincides with $p_{X}$, and with $p_{X,Y,K}(x,y,k)\le p_{K}(k)/k$ for all $x,k$. Hence, $\Lambda^{[m]}(X)$ can be expressed as a maximization in a linear program.
\item (Mixed subadditivity) For any $X,Y$,
\begin{align*}
\Lambda^{[m]}(X,Y) & \le\Lambda^{[m]}(X)+\Lambda^{[m|\mathcal{X}|]}(Y)\\
 & \le\Lambda^{[m]}(X)+H(Y).
\end{align*}
\end{enumerate}
\end{prop}
\medskip{}

\section{Discrete R\'{e}nyi Layered Entropy\label{sec:renyi}}

R\'{e}nyi entropy \cite{renyi1961measures} is a generalization of Shannon entropy, defined as (where $\alpha\in(0,\infty)\backslash\{1\}$ is the order) 
\[
H_{\alpha}(p):=\frac{1}{1-\alpha}\log\sum_{x}(p(x))^{\alpha}.
\]
Also, define $H_{\alpha}(X):=\lim_{\alpha'\to\alpha}H_{\alpha'}(X)$ for $\alpha\in\{0,1,\infty\}$, which gives $H_{1}(p)=H(p)$, $H_{0}(p)=\log|\{x:\,p(x)>0\}|$, and $H_{\infty}(p)=-\log\max_{x}p(x)$. In this section, we generalize the discrete layered entropy to the\emph{ discrete R\'{e}nyi layered entropy} in an analogous manner.

\medskip{}

\begin{defn}
[Discrete R\'{e}nyi layered entropy] Given $\alpha\in(0,\infty)\backslash\{1\}$, the \emph{discrete R\'{e}nyi layered entropy} of order $\alpha$ of a probability mass function $p$ is defined as
\begin{equation}
\Lambda_{\alpha}(p):=\frac{1}{1/\alpha-1}\log\sum_{i=1}^{\infty}p^{\downarrow}(i)\big(i^{1/\alpha}-(i-1)^{1/\alpha}\big),\label{eq:LX_alt-1}
\end{equation}
where $p^{\downarrow}(i)$ is the $i$-th entry of $(p(x))_{x\in\mathcal{X}}$ when sorted in descending order. Also, define $\Lambda_{\alpha}(X):=\lim_{\alpha'\to\alpha}\Lambda_{\alpha'}(X)$ for $\alpha\in\{0,1,\infty\}$, which can be evaluated as $\Lambda_{0}(p)=H_{0}(p)$, $\Lambda_{\infty}(p)=H_{\infty}(p)$, and $\Lambda_{1}(p)=\Lambda(p)$.
\end{defn}
\medskip{}

\begin{figure*}
\begin{centering}
\includegraphics[scale=0.21]{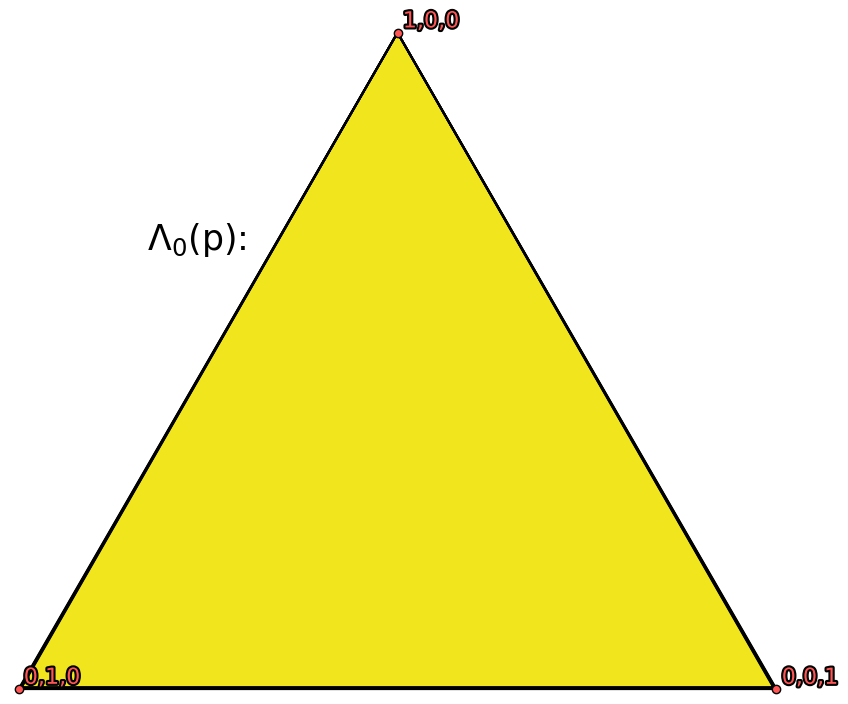}\includegraphics[scale=0.21]{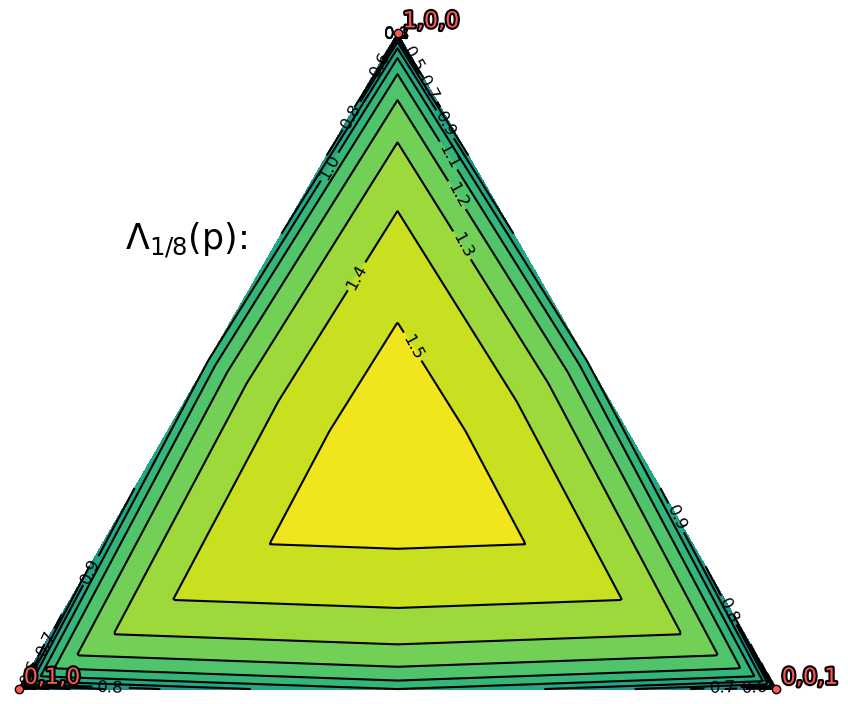}\includegraphics[scale=0.21]{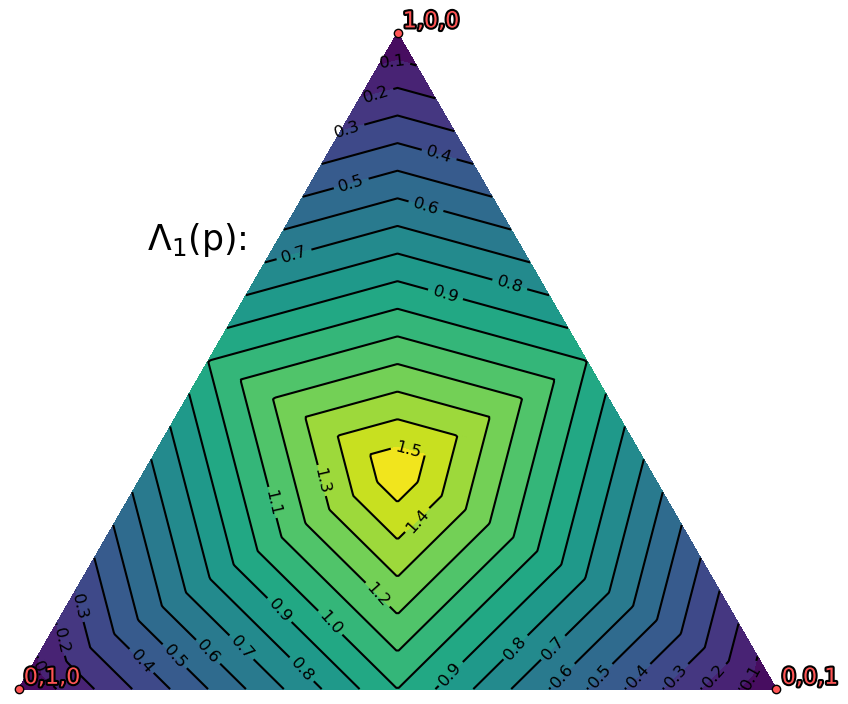}\includegraphics[scale=0.21]{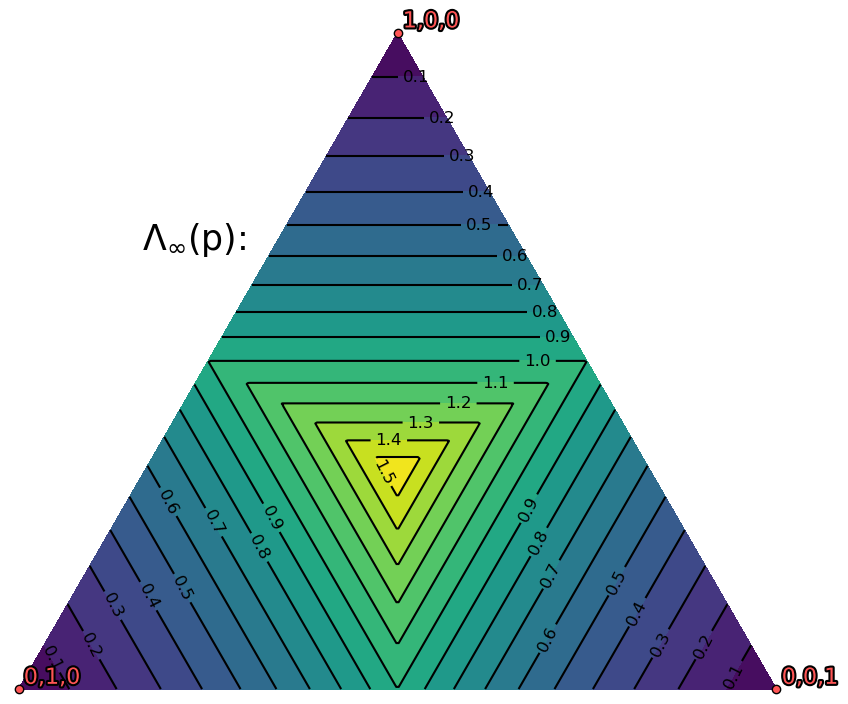}
\par\end{centering}
\begin{centering}
\par\end{centering}

\caption{\label{fig:renyi}Contour plot of the discrete R\'{e}nyi layered entropy $\Lambda_{\alpha}(p)$ for $\alpha\in\{0,1/8,1,\infty\}$ and $p:\{1,2,3\}\to[0,1]$ being a  probability mass function.}
\end{figure*}

Refer to Figure \ref{fig:renyi} for a plot. Note that $\Lambda_{\alpha}(p)$ is not directly related to $\Lambda^{[m]}(p)$ in Section \ref{sec:mlayered} (other than that they both generalizes $\Lambda(p)$). We present some properties of $\Lambda_{\alpha}(p)$. The proof is in Appendix \ref{subsec:pf_renyi}.

\begin{prop}
\label{prop:renyi}We have the following for $\alpha\in[0,\infty]$:
\begin{itemize}
\item $\Lambda_{\alpha}(X)$ is non-increasing in $\alpha$, and $\Lambda_{\alpha}(X)\le H_{\alpha}(X)$.
\item When $X$ is uniformly distributed, $\Lambda_{\alpha}(X)=H_{\alpha}(X)=\log|\mathcal{X}|$.
\item If $X\in\mathbb{N}$, then
\[
\Lambda_{\alpha}(X)\le\frac{1}{1/\alpha-1}\log\mathbb{E}\left[X^{1/\alpha}-(X-1)^{1/\alpha}\right].
\]
In particular, $\Lambda_{1/2}(X)\le\log(2\mathbb{E}[X]-1)$.
\end{itemize}
\end{prop}
\smallskip{}

\begin{rem}
The guessing problem \cite{massey1994guessing,mceliece1995inequality,arikan1996inequality} concerns the minimum expected $\rho$-th power ($\rho\ge0$) of the number of guesses needed to guess the value of $X\sim p$, given by
\[
G_{\rho}^{*}(p):=\sum_{i=1}^{\infty}p^{\downarrow}(i)\cdot i^{\rho}.
\]
Various bounds on $G_{\rho}^{*}(p)$ has been deduced. For example, for $p$ supported over a finite set $\mathcal{X}$, $G_{1}^{*}(p)\le\frac{|\mathcal{X}|-1}{2\log|\mathcal{X}|}H(p)$ \cite{mceliece1995inequality}. Also, it was shown in \cite{arikan1996inequality} that 
\begin{equation}
G_{\rho}^{*}(p)\ge(1+\ln|\mathcal{X}|)^{-\rho}2^{\rho H_{1/(\rho+1)}(p)}.\label{eq:guessing_ineq}
\end{equation}
Note that
\begin{align*}
\Lambda_{1/(\rho+1)}(p) & =\frac{1}{\rho}\log\left((\rho+1)\int_{0}^{\infty}p^{\downarrow}(\lceil t\rceil)\cdot t^{\rho}\mathrm{d}t\right)\\
 & \le\frac{1}{\rho}\log\left((\rho+1)G_{\rho}^{*}(p)\right).
\end{align*}
Also, for $\rho\ge1$, Jensen's inequality gives
\begin{align}
\Lambda_{1/(\rho+1)}(p) & =\frac{1}{\rho}\log\left((\rho+1)\int_{0}^{\infty}p^{\downarrow}(\lceil t\rceil)\cdot t^{\rho}\mathrm{d}t\right)\nonumber \\
 & \ge\frac{1}{\rho}\log\left((\rho+1)\sum_{i=1}^{\infty}p^{\downarrow}(i)\cdot(i-1/2)^{\rho}\right)\nonumber \\
 & \ge\frac{1}{\rho}\log\left((\rho+1)2^{-\rho}G_{\rho}^{*}(p)\right)\nonumber \\
 & =\frac{1}{\rho}\log\left((\rho+1)G_{\rho}^{*}(p)\right)-1.\label{eq:renyi_lb}
\end{align}
Hence, we can see that the discrete R\'{e}nyi layered entropy is related to $G_{\rho}^{*}(p)$ in the sense that $\Lambda_{1/(\rho+1)}(p)\approx\rho^{-1}\log((\rho+1)G_{\rho}^{*}(p))$ for $\rho\ge1$. Inequalities on $\Lambda_{1/(\rho+1)}(p)$ can imply inequalities on $G_{\rho}^{*}(p)$ and vice versa. For example, combining \eqref{eq:guessing_ineq} and \eqref{eq:renyi_lb} gives
\[
\Lambda_{1/(\rho+1)}(p)\ge H_{1/(\rho+1)}(p)+\frac{1}{\rho}\log(\rho+1)-\log(1+\ln|\mathcal{X}|)-1,
\]
for $\rho\ge1$. Regarding the advantage of using $\Lambda_{1/(\rho+1)}(p)$ instead of $G_{\rho}^{*}(p)$, one may argue that $\Lambda_{1/(\rho+1)}(p)$ has more elegant properties. For example, Proposition \ref{prop:renyi} gives $\Lambda_{\alpha}(X)\le H_{\alpha}(X)$ with equality if $X$ is uniform. On the other hand, the expression for $G_{\rho}^{*}(p)$ for a uniform distribution is complicated.
\end{rem}
\smallskip{}

\begin{rem}
Given \eqref{eq:minent_def}, another way to modify the R\'{e}nyi entropy to incorporate $\Lambda(X)$ is to study the concave envelope of the R\'{e}nyi entropy $\overline{H}_{\alpha}(X):=\max_{p_{Y|X}}H_{\alpha}(X|Y)$, where $H_{\alpha}(X|Y):=\mathbb{E}_{Y}[H_{\alpha}(p_{X|Y}(\cdot|Y))]$. Since $H_{\alpha}$ is already concave for $\alpha\le1$, we have $\overline{H}_{\alpha}(X)=H_{\alpha}(X)$ for $\alpha\le1$. We also have $\overline{H}_{\infty}(X)=\Lambda(X)$. Therefore, we can use $\overline{H}_{\alpha}(X)$ to interpolate between $H(X)$ and $\Lambda(X)$. The properties of $\overline{H}_{\alpha}$ are left for future studies.
\end{rem}
\smallskip{}

\section{Vertical and Horizontal Quantities\label{sec:horizontal}}

In the previous sections, we have seen numerous parallels between $H(p)$ and $\Lambda(p)$. Intuitively, $H(p)$ is a ``vertical quantity'', due to the fact that if we plot $p(x)$, then $H(p)=-\sum_{x}p(x)\log p(x)$ concerns the average negative logarithm of the height of $p(x)$; whereas $\Lambda(p)=\int_{0}^{1}|\{x:\,p(x)>t\}|\cdot\log|\{x:\,p(x)>t\}|\mathrm{d}t$ (Proposition \ref{prop:alt_def}) is a ``horizontal quantity'' since it concerns the size of the level set $|\{x:\,p(x)>t\}|$, which is the size of the intersection between the hypograph $\{(x,\tau):\,0\le\tau<p(x)\}$ and the horizontal line $\mathcal{X}\times\{t\}$ in the plot (see Table \ref{tab:horizontal_vertical} for an illustration). For similar reasons, the differential entropy $h(f)$ and the Kullback-Leibler divergence $D_{\mathrm{KL}}$ are vertical quantities, whereas the continuous layered entropy $\lambda(f)$ \cite{hegazy2022randomized,ling2024rejection} (Remark \ref{rem:continuous}) and the channel simulation divergence $D_{\mathrm{CS}}$ \cite{goc2024causal,flamich2025redundancy} (see \eqref{eq:cs_div}) are horizontal quantities. An elegant duality noted in \cite{hegazy2022randomized} states that for a continuous random variable $X$ with probability density function $f_{X}$, letting $Y|X\sim\mathrm{Unif}([0,f_{X}(X)])$ (i.e., $(X,Y)$ are uniform over $\{(x,y):\,0\le y\le f_{X}(x)\}$), we have
\[
\lambda(X)=-h(Y),\;\;h(X)=-\lambda(Y).
\]
Table \ref{tab:horizontal_vertical} list the parallels between vertical and horizontal quantities and their properties discussed in previous sections. Note that Hartley entropy $H_{0}$ and min-entropy $H_{\infty}$ are both vertical and horizontal. The intuitive reason why prefix codes are ``vertical'' and non-prefix codes are ``horizontal'' is left for future studies.

\begin{table*}
\caption{\label{tab:horizontal_vertical}Vertical and horizontal quantities, and their properties.}

\centering{}{\renewcommand*{\arraystretch}{1.5}%
\begin{tabular}{ll}
\hline 
\textbf{Vertical quantities} & \textbf{Horizontal quantities}\tabularnewline
\hline 
Shannon entropy  & Discrete layered entropy\tabularnewline
$H(p)=-\sum_{x}p(x)\log p(x)$ & $\Lambda(p)=\int_{0}^{1}|\{x:\,p(x)>t\}|\cdot\log|\{x:\,p(x)>t\}|\mathrm{d}t$\tabularnewline
$\quad$\includegraphics[scale=0.9]{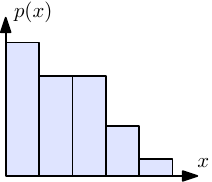} & $\quad$\includegraphics[scale=0.9]{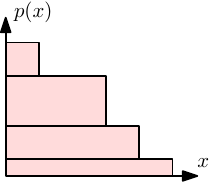}\tabularnewline
$\bullet\;\;$Avg length of best prefix code $\in[H(p),H(p)+1)$ & $\bullet\;\;$Avg length of best non-prefix code $\in(\Lambda(p)-2,\Lambda(p)]$\tabularnewline
\hline 
R\'{e}nyi entropy $H_{\alpha}(p)$ & Discrete R\'{e}nyi layered entropy $\Lambda_{\alpha}(p)$\tabularnewline
$\bullet\;\;$Shannon entropy $H(p)=\lim_{\alpha\to1}H_{\alpha}(p)$  & $\bullet\;\;$Discrete layered entropy $\Lambda(p)=\lim_{\alpha\to1}\Lambda_{\alpha}(p)$ \tabularnewline
$\bullet\;\;$Hartley entropy $H_{0}(p)=\lim_{\alpha\to0}H_{\alpha}(p)$ & $\bullet\;\;$Hartley entropy $H_{0}(p)=\lim_{\alpha\to0}\Lambda_{\alpha}(p)$\tabularnewline
$\bullet\;\;$Min-entropy $H_{\infty}(p)=\lim_{\alpha\to\infty}H_{\alpha}(p)$ & $\bullet\;\;$Min-entropy $H_{\infty}(p)=\lim_{\alpha\to\infty}\Lambda_{\alpha}(p)$\tabularnewline
\hline 
Differential entropy $h(f)$ & Continuous layered entropy $\lambda(f)$ \cite{hegazy2022randomized}\tabularnewline
$\bullet\;\;$$h(Y)=\lim_{\Delta\to0}(H(\lfloor Y/\Delta\rfloor)+\log\Delta)$ & $\bullet\;\;$$\lambda(Y)=\lim_{\Delta\to0}\left(\Lambda(\lfloor Y/\Delta\rfloor)+\log\Delta\right).$\tabularnewline
$\bullet\;\;$$H(X)=h(X+Z)$ for $X\in\mathbb{Z}$, $Z\sim\mathrm{Unif}([0,1])$ & $\bullet\;\;$$\Lambda(X)=\lambda(X+Z)$ for $X\in\mathbb{Z}$, $Z\sim\mathrm{Unif}([0,1])$\tabularnewline
\hline 
Kullback-Leibler divergence $D_{\mathrm{KL}}$ & Channel simulation divergence $D_{\mathrm{CS}}$ \cite{goc2024causal,flamich2025redundancy}\tabularnewline
$\bullet\;\;$$H(p)=\log|\mathcal{X}|-D_{\mathrm{KL}}(p\Vert\mathrm{Unif}(\mathcal{X}))$ & $\bullet\;\;$$\Lambda(p)=\log|\mathcal{X}|-D_{\mathrm{CS}}(p\Vert\mathrm{Unif}(\mathcal{X}))$\tabularnewline
\hline 
Mutual information $I$ & Functional information $I_\mathrm{F}$ \cite{hill2026rejection} / excess functional info lower bound \cite{sfrl_trans}\tabularnewline
$\bullet\;\;$ $I(X;Y)= \mathbb{E}_{Y}[D_{\mathrm{KL}}(P_{X|Y}(\cdot|Y)\Vert P_{X})]$ & $\bullet\;\;$ $I_\mathrm{F}(X;Y)= \mathbb{E}_{Y}[D_{\mathrm{CS}}(P_{X|Y}(\cdot|Y)\Vert P_{X})]$ \tabularnewline
\hline 
\end{tabular}}
\end{table*}

Interestingly, for channel simulation, even if we are minimizing $H(Y|S)$ which is a vertical quantity (when conditional prefix codes are used), the bounds (e.g., Proposition \ref{prop:chansim_unif_lb} and \eqref{eq:Hc_lb_general}) are stated in terms of horizontal quantities. In \cite{flamich2025redundancy}, the asymptotic redundancy of channel simulation with conditional prefix codes is characterized via the channel simulation divergence, showing that the channel simulation divergence bound is tight asymptotically. Very loosely speaking, we can regard the common randomness $S$, which is independent of the input $X$, as a ``rotation'' of $X$. If we draw a plot where $X$ is the x-axis, and $S$ is the y-axis, then vertical quantities with respect to $S$ becomes horizontal quantities with respect to $X$. A formal study on horizontal quantities, and how horizontal quantities can be ``rotated'' into vertical quantities and vice versa, are left for future research.

\medskip{}

\section{Acknowledgement}

This work was partially supported by two grants from the Research Grants Council of the Hong Kong Special Administrative Region, China {[}Project No.s: CUHK 24205621 (ECS), CUHK 14209823 (GRF){]}. The author would like to thank the associate editor and the anonymous reviewers of this paper and its conference version for their constructive feedback. The author would like to thank Abbas El Gamal, Shirin Saeedi Bidokhti, Aaron B. Wagner, Gergely Flamich, Sharang M. Sriramu, Serhat Emre Coban and Spencer Hill for the insightful discussions.

\smallskip{}

\appendix{}

\subsection{Proof of Proposition \ref{prop:alt_def}\label{subsec:pf_alt_def}}

$\Lambda(p)=\int_{0}^{\infty}p^{\downarrow}(\lceil t\rceil)\log(et)\mathrm{d}t$
\eqref{eq:integral_def} follows from $i\log i-(i-1)\log(i-1)=\int_{i-1}^{i}\log(et)\mathrm{d}t$.  
To prove \eqref{eq:layered_def}: $\Lambda(p)=\int_{0}^{1}|\{x:\,p(x)>t\}|\cdot\log|\{x:\,p(x)>t\}|\mathrm{d}t$,
\begin{align*}
\Lambda(p) & =\sum_{i=1}^{\infty}p^{\downarrow}(i)\left(i\log i-(i-1)\log(i-1)\right)\\
 & =\sum_{i=1}^{\infty}\int_{0}^{1}\mathbf{1}\{p^{\downarrow}(i)>t\}\left(i\log i-(i-1)\log(i-1)\right)\mathrm{d}t\\
 & =\int_{0}^{1}\sum_{i=1}^{\infty}\mathbf{1}\{p^{\downarrow}(i)>t\}\left(i\log i-(i-1)\log(i-1)\right)\mathrm{d}t\\
 & =\int_{0}^{1}|\{x:\,p(x)>t\}|\cdot\log|\{x:\,p(x)>t\}|\mathrm{d}t.
\end{align*}
We then prove
\begin{equation}
\Lambda(p)=\mathrm{min}_{\tilde{\Lambda}}\tilde{\Lambda}(p),\label{eq:envelope_def}
\end{equation}
where the minimum is over all concave functions $\tilde{\Lambda}$  satisfying $\tilde{\Lambda}(X)\ge\log|\mathcal{X}|$ when $X$ is uniformly distributed. Assume $X\in\mathbb{N}$ and $p_{X}(x)$ is in descending order without loss of generality. Note that
\[
p_{X}(x)=\sum_{i=1}^{\infty}i(p_{X}(i)-p_{X}(i+1))\cdot\frac{\mathbf{1}\{x\le i\}}{i}
\]
is a convex combination of uniform distributions $\mathrm{Unif}([i])$. Hence, for every concave $\tilde{\Lambda}$ satisfying that $\tilde{\Lambda}(\mathrm{Unif}([i]))\ge\log i$, we have
\begin{align*}
\tilde{\Lambda}(X) & \ge\sum_{i=1}^{\infty}i(p_{X}(i)-p_{X}(i+1))\cdot\log i=\Lambda(X).
\end{align*}
To show that there exists $\tilde{\Lambda}$ satisfying $\tilde{\Lambda}(Y)\ge\log|\mathcal{Y}|$ whenever $Y$ is uniform and $\tilde{\Lambda}(X)=\Lambda(X)$ for this fixed $p_{X}$, we take 
\[
\tilde{\Lambda}(p)=\sum_{i=1}^{\infty}p(i)\left(i\log i-(i-1)\log(i-1)\right)
\]
to be a linear function. We have $\tilde{\Lambda}(X)=\Lambda(X)$. For any uniform $Y$, we have 
\begin{align*}
\tilde{\Lambda}(Y) & =|\mathcal{Y}|^{-1}\sum_{y\in\mathcal{Y}}\left(y\log y-(y-1)\log(y-1)\right)\\
 & \ge|\mathcal{Y}|^{-1}\sum_{y=1}^{|\mathcal{Y}|}\left(y\log y-(y-1)\log(y-1)\right)\,=\,\log|\mathcal{Y}|.
\end{align*}
 Hence, $\Lambda(X)=\min_{\tilde{\Lambda}}\tilde{\Lambda}(X)$.

We now prove \eqref{eq:alt_cond}: $\Lambda(X)=\max H(X|Y)$, where the maximum is over $p_{Y|X}$ such that $p_{X|Y}(\cdot|y)$ is a uniform distribution for every $y$.
If $X$ is conditionally uniform given $Y$, then $H(X|Y)=\Lambda(X|Y)\le\Lambda(X)$ by concavity (the concavity of $\Lambda$ follows from \eqref{eq:envelope_def}). To show that $H(X|Y)=\Lambda(X)$ is achievable, consider a random variable $Y\in[0,\infty)$ distributed as $Y|\{X=x\}\sim\mathrm{Unif}(0,p_{X}(x))$ (if we want $Y$ to be discrete, we can discretize $Y$ by dividing $[0,\infty)$ at the points $(p_{X}(x))_{x}$). Note that $p_{X|Y}(\cdot|y)$ is the uniform distribution over $\{x:\,p_{X}(x)\ge y\}$, and hence $H(X|Y)=\Lambda(X)$ is achievable by \eqref{eq:layered_def}. 

Before we prove \eqref{eq:minent_def}, we first prove $H_{\infty}(p)\le\Lambda(p)$. Let $a:=\max_{x}p_{X}(x)$. By \eqref{eq:layered_def} and Jensen's inequality,
\begin{align*}
\Lambda(p) & =\int_{0}^{a}|\{x:\,p(x)>t\}|\cdot\log|\{x:\,p(x)>t\}|\mathrm{d}t\\
 & \ge a\cdot\frac{\int_{0}^{a}|\{x:\,p(x)>t\}|\mathrm{d}t}{a}\cdot\log\frac{\int_{0}^{a}|\{x:\,p(x)>t\}|\mathrm{d}t}{a}\\
 & =\log(1/a)\,=\,H_{\infty}(p).
\end{align*}
Equality holds if and only if $|\{x:\,p(x)>t\}|$ is constant for almost all $t\in[0,a]$, meaning that $p$ is a uniform distribution.

We now prove \eqref{eq:minent_def}: $\Lambda(X)=\max_{p_{Y|X}}H_{\infty}(X|Y)$. We have $H_{\infty}(X|Y)\le\Lambda(X|Y)\le\Lambda(X)$. Achievability of $H_{\infty}(X|Y)=\Lambda(X)$ is the same as in the proof of \eqref{eq:alt_cond}.

We finally prove \eqref{eq:lp_def}. For every $p_{X,K}$ satisfying $p_{X,K}(x,k)\le p_{K}(k)/k$, we have 
\begin{align*}
H_{\infty}(X|K) & =\mathbb{E}[-\log\max_{x}p_{X,K}(x,K)/p_{K}(K)]\ge\mathbb{E}[\log K].
\end{align*}
Hence, $\mathbb{E}[\log K]\le\Lambda(X)$ by \eqref{eq:minent_def}. For the other direction, consider random $Y\in[0,\infty)$ distributed as $Y|\{X=x\}\sim\mathrm{Unif}(0,p_{X}(x))$, and let $K=|\mathrm{supp}p_{X|Y}(\cdot|Y)|$. We have $p_{X,K}(x,k)\in\{0,\,p_{K}(k)/k\}$ and $\mathbb{E}[\log K]=H(X|Y)=\Lambda(X)$.

\subsection{Proof of Proposition \ref{prop:basic}\label{subsec:pf_basic}}

We have proved $H_{\infty}(X)\le\Lambda(X)$ in Appendix \ref{subsec:pf_alt_def}. $\Lambda(X)\le H(X)$ follows from \eqref{eq:alt_cond}. From direct evaluation, $\Lambda(X)=H(X)=\log|\mathcal{X}|$ for uniform $X$. To show $\Lambda(X)=H(X)$ implies that $X$ is uniform, by \eqref{eq:alt_cond}, if $\Lambda(X)=H(X)$, then there exists $Y$ such that $X$ is conditionally uniform given $Y$ and $H(X|Y)=H(X)$, so $Y\perp\!\!\!\perp X$, and $X$ is uniform.

The concavity of $\Lambda$ follows from \eqref{eq:envelope_def}. Schur concavity follows from concavity and the fact that $\Lambda(p)$ is invariant under labelling of $p$. Monotone linearity follows directly from the definition of $\Lambda$.

To prove the superadditivity property $\Lambda(X,Y)\ge\Lambda(X)+\Lambda(Y)$ for $X \perp \!\!\! \perp Y$, let $\Lambda(X)=H(X|Z)$ and $\Lambda(Y)=H(Y|W)$ where $Z,W$ attain the maximum in \eqref{eq:alt_cond}, and $(X,Z)\perp\!\!\!\perp(Y,W)$. Since $X$ is uniform conditional on $Z$, and $Y$ is uniform conditional on $W$, $(X,Y)$ is uniform conditional on $(Z,W)$, and hence $\Lambda(X,Y)\ge H(X,Y|Z,W)=\Lambda(X)+\Lambda(Y)$ by \eqref{eq:alt_cond}. We then show the equality case. If $\Lambda(X,Y)=\Lambda(X)+\Lambda(Y)$, then $\Lambda(X,Y)=H(X,Y|Z,W)$. Let $\mathcal{X}_{z}:=\{x:\,p_{X|Z}(x|z)>0\}$, $\mathcal{Y}_{w}:=\{y:\,p_{Y|W}(y|w)>0\}$. We have $(X,Y)|\{(Z,W)=(z,w)\}\sim\mathrm{Unif}(\mathcal{X}_{z}\times\mathcal{Y}_{w})$. For $(z_{1},w_{1})\neq(z_{2},w_{2})$, if neither of $\mathcal{X}_{z_{1}}\times\mathcal{Y}_{w_{1}}$ or $\mathcal{X}_{z_{2}}\times\mathcal{Y}_{w_{2}}$ contains the other, then 
\begin{align*}
 & \Lambda(X,Y|(Z,W)\in\{(z_{1},w_{1}),(z_{2},w_{2})\})\;>\\
 & \sum_{i=1}^{2}\frac{p_{Z,W}(z_{i},w_{i})}{p_{Z,W}(z_{1},w_{1})+p_{Z,W}(z_{2},w_{2})}\Lambda(X,Y|(Z,W)=(z_{i},w_{i}))
\end{align*}
since $p_{X,Y}(x,y|z_{1},w_{1})$ has a different ordering as $p_{X,Y}(x,y|z_{2},w_{2})$, and hence $H(X,Y|Z,W)$ does not achieve the maximum of $H(X,Y|T)$ subject to that $(X,Y)$ is conditionally uniform given $Y$. Therefore, for every $(z_{1},w_{1})\neq(z_{2},w_{2})$, one of $\mathcal{X}_{z_{1}}\times\mathcal{Y}_{w_{1}}$ or $\mathcal{X}_{z_{2}}\times\mathcal{Y}_{w_{2}}$ contains the other. This is impossible if there are two possible sets for $\mathcal{X}_{z}$ and two possible sets for $\mathcal{Y}_{w}$ (in this case, we can take $\mathcal{X}_{z_{1}}\nsubseteq\mathcal{X}_{z_{2}}$ and $\mathcal{Y}_{w_{2}}\nsubseteq\mathcal{Y}_{w_{1}}$). Hence, there is only one possible set for $\mathcal{X}_{z}$ (implying $X$ is uniform), or there is only one possible set for $\mathcal{Y}_{w}$ (implying $Y$ is uniform).

To prove the mixed chain rule property $\Lambda(X,Y)\le\Lambda(X|Y)+H(Y)$, consider any jointly distributed $X,Y$. Using \eqref{eq:alt_cond}, let $U$ be such that $p_{X,Y|U}(\cdot,\cdot|u)=\mathrm{Unif}(\mathcal{S}_{u})$ is a uniform distribution over $\mathcal{S}_{u}\subseteq\mathcal{X}\times\mathcal{Y}$ for all $u\in\mathcal{U}$, and $\Lambda(X,Y)=H(X,Y|U)$. Note that $p_{X|Y,U}(\cdot|y,u)=\mathrm{Unif}(\{x:(x,y)\in\mathcal{S}_{u}\})$ is uniform as well. Hence, by \eqref{eq:alt_cond}, $\Lambda(X|Y)\ge H(X|Y,U)$. Therefore, $\Lambda(X,Y)-\Lambda(X|Y)\le H(X,Y|U)-H(X|Y,U)=H(Y|U)\le H(Y)$.

\subsection{Proof of Proposition \ref{prop:LX_HX_bound}\label{subsec:pf_LX_HX_bound}}

We now prove Proposition \ref{prop:LX_HX_bound}: for every discrete  $X$, $\Lambda(X)\le H(X)\le\Lambda(X)+\log(1+\Lambda(X)/(e\eta))+\eta $ for every $\eta>0$. 
Assume $X\in\mathbb{N}$, and $p_{X}(x)$ is sorted in descending order. Fix $\lambda>1$. Let
\[
\psi_{\lambda}(x):=c^{-1}2^{-\lambda(x\log x-(x-1)\log(x-1))}
\]
be a probability mass function over $\mathbb{N}$, where $c:=\sum_{x=1}^{\infty}2^{-\lambda(x\log x-(x-1)\log(x-1))}$. We have
\begin{align*}
H(X) & \le\sum_{x=1}^{\infty}p_{X}(x)\log\frac{1}{\psi_{\lambda}(x)}=\lambda\Lambda(X)+\log c.
\end{align*}
To bound $c$, we have
\begin{align*}
c & =1+\sum_{x=2}^{\infty}2^{-\lambda(x\log x-(x-1)\log(x-1))}\\
 & =1+\sum_{x=2}^{\infty}2^{-\lambda\int_{x-1}^{x}\log(et)\mathrm{d}t}\\
 & \le1+\sum_{x=2}^{\infty}\int_{x-1}^{x}2^{-\lambda\log(et)}\mathrm{d}t\\
 & =1+\int_{1}^{\infty}(et)^{-\lambda}\mathrm{d}t\\
 & =1+\frac{e^{-\lambda}}{\lambda-1}\,<\,1+\frac{e^{-1}}{\lambda-1}.
\end{align*}
The bound \eqref{eq:LX_HX_bound1} follows from taking $\lambda=1+\eta/\Lambda(X)$.

Operationally, if we want to encode $X\in\mathbb{N}$ with $p_{X}(x)$ in descending order using a prefix code, we use the Shannon code \cite{shannon1948mathematical} for the distribution $\psi_{\lambda}(x)$ with $\lambda=1+\eta/\Lambda(X)$, which guarantees that the expected length is upper-bounded by $\Lambda(X)+\log(1+\Lambda(X)/(e\eta))+\eta$. If we are instead given a non-prefix codeword $M\in\{0,1\}^{*}$ and want to transform it to a prefix codeword, we can first convert it to a positive integer $X$ by taking $X$ to be ``$1M$'' treated as a binary representation (e.g., if $M=01$, then $X=101_{2}=5$), and the use the aforementioned Shannon code to encode $X$.

\smallskip{}

\subsection{Proof of Proposition \ref{prop:nonprefix}\label{subsec:pf_nonprefix}}

We now prove Proposition \ref{prop:nonprefix}:
$\Lambda(X)-2<L(X)\le\Lambda(X)$. 
Without loss of generality, assume $X\in\mathbb{N}$ and $p_{X}(x)$ is in descending order. We have $L(X)=\mathbb{E}[\lfloor\log X\rfloor]$, and $\Lambda(X)=\mathbb{E}[X\log X-(X-1)\log(X-1)]$. Since $x\log x-(x-1)\log(x-1)\ge x\log x-(x-1)\log x=\log x$, we have $\mathbb{E}[\lfloor\log X\rfloor]\le\Lambda(X)$. To show the lower bound, note that any $p_{X}$ in descending order is a convex combination of $\mathrm{Unif}([k])$ for $k\in\mathbb{N}$. Hence, it suffices to show $L(X)\ge\Lambda(X)-2$ for $X\sim\mathrm{Unif}([k])$. In this case,
\begin{align*}
kL(X) & =\sum_{x=1}^{k}\lfloor\log x\rfloor\\
 & =\sum_{i=0}^{\lfloor\log k\rfloor-1}2^{i}\cdot i+(k-2^{\lfloor\log k\rfloor}+1)\lfloor\log k\rfloor\\
 & =2^{\lfloor\log k\rfloor}\left(\lfloor\log k\rfloor-2\right)+2+(k-2^{\lfloor\log k\rfloor}+1)\lfloor\log k\rfloor\\
 & =-2\cdot2^{\log k-t}+2+(k+1)(\log k-t)\\
 & =k\log k+\log k-t(k+1)-2^{1-t}k+2,
\end{align*}
where $t:=\log k-\lfloor\log k\rfloor\in[0,1]$. Note that $t(k+1)+2^{1-t}k$ is a convex function in $t$, which is maximized at $t=0$ or $t=1$. Hence, $t(k+1)+2^{1-t}k\le\max\{2k,k+1+k\}=2k+1$, and 
\begin{align*}
kL(X) & =k\log k+\log k-t(k+1)-2^{1-t}k+2\\
 & \ge k\log k+\log k-2k+1\,>\,k(\Lambda(X)-2).
\end{align*}

\subsection{Proof of Theorem \ref{thm:useful_def}\label{subsec:pf_useful_def}}

Note that $H(X\backslash Y)=H(X)$ if and only if $X\backslash Y\stackrel{\iota}{=}X$ due to the tie-breaking rule in Definition \ref{def:cond} (if $H(X\backslash Y)=H(X)$, then $X$ is a conditional compression, so we must choose $X\backslash Y\stackrel{\iota}{=}X$ since it minimizes $H(X|U)$).

To show
\begin{equation}
\min_{p_{Y|X}:\,H(X\backslash Y)=H(X)}H(X|Y)\le\Lambda(X),\label{eq:pf_useful_def_ineq1}
\end{equation}
consider a random variable $Y\in[0,\infty)$ distributed as $Y|\{X=x\}\sim\mathrm{Unif}(0,p_{X}(x))$ (if we want $Y$ to be discrete, we can discretize $Y$ by dividing $[0,\infty)$ at the points $(p_{X}(x))_{x}$). Note that $p_{X|Y}(\cdot|y)$ is the uniform distribution over $\{x:\,p_{X}(x)\ge y\}$, and hence $H(X|Y)=\Lambda(X)$ by \eqref{eq:layered_def}. By Proposition \ref{prop:cond_pmf}, the probability mass function of $X\backslash Y$ is
\begin{align}
 & p_{X\backslash Y}^{\downarrow}(i)\nonumber \\
 & =\mathbb{E}_{Y}\big[p_{X|Y}^{\downarrow}(i|Y)\big]\nonumber \\
 & =\int_{0}^{\infty}|\{x:\,p_{X}(x)\ge y\}|\cdot p_{X|Y}^{\downarrow}(i|y)\mathrm{d}y\nonumber \\
 & =\int_{0}^{\infty}|\{x:\,p_{X}(x)\ge y\}|\frac{\mathbf{1}\{|\{x:\,p_{X}(x)\ge y\}|\ge i\}}{|\{x:\,p_{X}(x)\ge y\}|}\mathrm{d}y\nonumber \\
 & =\int_{0}^{\infty}\mathbf{1}\{|\{x:\,p_{X}(x)\ge y\}|\ge i\}\mathrm{d}y\;=\;p_{X}^{\downarrow}(i),\label{eq:layer_y}
\end{align}
and hence $H(X\backslash Y)=H(X)$. Therefore, \eqref{eq:pf_useful_def_ineq1} holds.

It is left to show 
\begin{equation}
\min_{p_{Y|X}:\,H(X\backslash Y)=H(X)}H(X|Y)\ge\Lambda(X).\label{eq:pf_useful_def_ineq2}
\end{equation}
Without loss of generality, assume $X\in\mathbb{N}$ and $p_{X}(x)$ is sorted in descending order. Consider any $Y$ with $H(X\backslash Y)=H(X)$. Let $U=X\backslash Y$. By Proposition \ref{prop:cond_pmf},
\[
\sum_{i=1}^{k}p_{U}^{\downarrow}(i)=\mathbb{E}\Big[\sum_{i=1}^{k}p_{X|Y}^{\downarrow}(i|Y)\Big]\ge\sum_{i=1}^{k}p_{X}(i),
\]
and hence $p_{U}\succeq p_{X}$ and $H(U)\le H(X)$. Since equality holds ($H(U)=H(X)$), we must have $p_{U}^{\downarrow}(i)=p_{X}(i)$, and 
\[
\mathbb{E}\Big[\sum_{i=1}^{k}p_{X|Y}^{\downarrow}(i|Y)\Big]=\sum_{i=1}^{k}p_{X}(i).
\]
This implies $p_{X|Y}(x|y)$ must be nonincreasing with $x$ for every fixed $y$, i.e., $p_{X|Y}(x|y)=p_{X|Y}^{\downarrow}(x|y)$ (otherwise if there exists $k$ such that $\sum_{i=1}^{k}p_{X|Y}^{\downarrow}(i|y)>\sum_{i=1}^{k}p_{X|Y}(i|y)$, then $\mathbb{E}[\sum_{i=1}^{k}p_{X|Y}^{\downarrow}(i|Y)]>\sum_{i=1}^{k}p_{X}(i)$). By monotone linearity (Proposition \ref{prop:basic}), $H(X|Y)\ge\Lambda(X|Y)=\Lambda(X)$. Therefore, \eqref{eq:pf_useful_def_ineq2} holds. Note that this also shows that $\Lambda(X|Y)=\Lambda(X)$ if $H(X\backslash Y)=H(X)$.

\subsection{Proof of Theorem \ref{thm:axiom_cond_logx}\label{subsec:pf_axiom_cond_logx}}

We will prove Theorem \ref{thm:axiom_cond_logx}: 
if $\tilde{\Lambda}$ satisfies the conditioning property $\tilde{\Lambda}(X\backslash Y)=\tilde{\Lambda}(X|Y):=\mathbb{E}_{Y}[\tilde{\Lambda}(p_{X|Y}(\cdot|Y))]$, $\tilde{\Lambda}(X)=\log|\mathcal{X}|$ whenever $X$ is uniformly distributed over $\mathcal{X}$, and $\tilde{\Lambda}(X)=\tilde{\Lambda}(Y)$ whenever $X\stackrel{\iota}{=}Y$, then $\tilde{\Lambda}(X)=\Lambda(X)$.

Assume $\tilde{\Lambda}$ satisfies the requirements.
Consider any $X$. Consider a random variable $Y\in[0,\infty)$ distributed as $Y|\{X=x\}\sim\mathrm{Unif}(0,p_{X}(x))$ (if we want $Y$ to be discrete, we can discretize $Y$ by dividing $[0,\infty)$ at the points $(p_{X}(x))_{x}$). Note that $p_{X|Y}(\cdot|y)$ is the uniform distribution over $\{x:\,p_{X}(x)\ge y\}$. We have shown in \eqref{eq:layer_y} that $p_{X}^{\downarrow}(i)=p_{X\backslash Y}^{\downarrow}(i)$, and hence 
\begin{align}
\tilde{\Lambda}(X) & \stackrel{(a)}{=} \tilde{\Lambda}(X\backslash Y)\nonumber \\
 & \stackrel{(b)}{=} \tilde{\Lambda}(X|Y)\nonumber \\
 & =\mathbb{E}_{Y}\big[\tilde{\Lambda}(p_{X|Y}(\cdot|Y))\big]\nonumber \\
 & =\mathbb{E}_{Y}\left[\tilde{\Lambda}(\mathrm{Unif}(\{x:\,p_{X}(x)\ge y\}))\right]\label{eq:L_tilde_exp}\\
 & \stackrel{(c)}{=}\mathbb{E}_{Y}\left[\log|\{x:\,p_{X}(x)\ge y\}|\right]\nonumber \\
 & =\int_{0}^{\infty}|\{x:\,p_{X}(x)\ge y\}|\log|\{x:\,p_{X}(x)\ge y\}|\mathrm{d}y\nonumber \\
 & \stackrel{(d)}{=}\Lambda(X), \nonumber 
\end{align}
where (a) is because $p_{X}^{\downarrow}(i)=p_{X\backslash Y}^{\downarrow}(i)$, (b) is by the conditioning property, (c) is by the assumption that $\tilde{\Lambda}(X)=\log|\mathcal{X}|$ whenever $X$ is uniformly distributed over $\mathcal{X}$, and (d) is by \eqref{eq:layered_def}.

\smallskip{}

\subsection{Proof of Theorem \ref{thm:axiom_cond_largest}\label{subsec:pf_axiom_cond_largest}}

We will prove Theorem \ref{thm:axiom_cond_largest}: if $\tilde{\Lambda}$ satisfies the conditioning property, $\tilde{\Lambda}(X)\le H(X)$ for every $X$, and $\tilde{\Lambda}(X)=\tilde{\Lambda}(Y)$ whenever $X\stackrel{\iota}{=}Y$, then $\tilde{\Lambda}(X)\le\Lambda(X)$ for every $X$.

Assume $\tilde{\Lambda}$ satisfies the requirements.
Consider any $X$, and define $Y$ as in the proof of Theorem \ref{thm:axiom_cond_logx}. By \eqref{eq:L_tilde_exp},
\begin{align*}
\tilde{\Lambda}(X) & =\mathbb{E}_{Y}\left[\tilde{\Lambda}(\mathrm{Unif}(\{x:\,p_{X}(x)\ge y\}))\right]\\
 & \le\mathbb{E}_{Y}\left[H(\mathrm{Unif}(\{x:\,p_{X}(x)\ge y\}))\right]\\
 & =\mathbb{E}_{Y}\left[\log|\{x:\,p_{X}(x)\ge y\}|\right]\,=\,\Lambda(X).
\end{align*}

\medskip{}

\subsection{Proof of Proposition \ref{prop:m_layered}\label{subsec:pf_m_layered}}

By definition, we have $\Lambda^{[1]}(X)=\Lambda(X)$, and $\Lambda^{[m]}(X)$ is non-decreasing in $m$. Proposition \ref{prop:sup_gap} gives $\Lambda^{[m]}(X)\le H(X)$. For the equality case, assume $\Lambda^{[m]}(X)=H(X)$, which means that there exists $Y\in[m]$ such that $\Lambda(X,Y)-\Lambda(Y)=H(X)$. By \eqref{eq:alt_cond}, we can let $Z$ be a random variable such that $p_{X,Y|Z}(\cdot,\cdot|z)=\mathrm{Unif}(\mathcal{S}_{z})$ is a uniform distribution over $\mathcal{S}_{z}\subseteq\mathcal{X}\times\mathcal{Y}$ for all $z\in\mathcal{Z}$, and $\Lambda(X,Y)=H(X,Y|Z)$. Note that $p_{Y|X,Z}(\cdot|x,z)=\mathrm{Unif}(\mathcal{S}_{z,x})$ (where $\mathcal{S}_{z,x}:=\{y:\,(x,y)\in\mathcal{S}_{z}\}\subseteq[m]$) is uniform as well. Hence, by \eqref{eq:alt_cond}, $\Lambda(Y)\ge H(Y|X,Z)$. Therefore, 
\begin{align*}
H(X)=\Lambda(X,Y)-\Lambda(Y) & \le H(X,Y|Z)-H(Y|X,Z)\\
 & =H(X|Z)\,\le\,H(X).
\end{align*}
Both inequalities above must be equalities. Hence, $Z$ is independent of $X$. We have, for any $z$,
\begin{align*}
p_{X}(x) & =p_{X|Z}(x|z)=\sum_{y=1}^{m}p_{X,Y|Z}(x,y|Z)=\frac{|\mathcal{S}_{z,x}|}{|\mathcal{S}_{z}|}.
\end{align*}
Taking $g(x)=|\mathcal{S}_{z,x}|\le m$, we have $p_{X}(x)=g(x)/\sum_{x'}g(x')$. For the other direction, if $p_{X}(x)=g(x)/\sum_{x'}g(x')$ for some $g:\mathcal{X}\to\{0,\ldots,m\}$, taking $Y|\{X=x\}\sim\mathrm{Unif}([g(x)])$, it is straightforward to verify that $\Lambda(X,Y)-\Lambda(Y)=H(X)$.

The lower bound $\Lambda^{[m]}(X)\ge H_{\infty}(X)$ follows directly from Proposition \ref{prop:basic} and the monotonicity of $\Lambda^{[m]}(X)$ with respect to $m$. From the proof of Proposition \ref{prop:sup_gap}, we have $\lim_{m\to\infty}\Lambda^{[m]}(X)=H(X)$.

By \eqref{eq:delta_sum}, we have
\[
\Lambda^{[m]}(X)\ge H(X)-\frac{1}{n}\log\left(e+\frac{nH(X)}{\log e}\right),
\]
as long as $m\ge|\mathcal{X}|^{n-1}$. Taking $n=\lfloor(\log m)/(\log|\mathcal{X}|)\rfloor+1$,
\begin{align*}
\Lambda^{[m]}(X) & \ge H(X)-\frac{\log|\mathcal{X}|}{\log m}\log\left(e+\frac{H(X)}{\log|\mathcal{X}|}\cdot\frac{\log m}{\log e}\right)\\
 & \ge H(X)-\frac{\log|\mathcal{X}|}{\log m}\log(\ln m+e).
\end{align*}
The linear programming form follows from \eqref{eq:lp_def} and \eqref{eq:min_layer_linear}. The concavity (and hence the Schur concavity) of $\Lambda^{[m]}(X)$ follows from the linear programming form. For the mixed subadditivity, we have
\begin{align*}
\Lambda^{[m]}(X,Y) & =\max_{p_{Z|X,Y}:\,Z\in[m]}\big(\Lambda(X,Y,Z)-\Lambda(Z)\big)\\
 & \le\max_{p_{Z|X,Y}:\,Z\in[m]}\big(\Lambda(X,Z)-\Lambda(Z)\big)\\
 & \;\;\;\;+\max_{p_{Z|X,Y}:\,Z\in[m]}\big(\Lambda(X,Y,Z)-\Lambda(X,Z)\big)\\
 & \le\Lambda^{[m]}(X)+\Lambda^{[m|\mathcal{X}|]}(Y).
\end{align*}

\medskip{}

\subsection{Proof of Proposition \ref{prop:renyi}\label{subsec:pf_renyi}}

First prove that $\Lambda_{\alpha}(X)$ is non-increasing in $\alpha$. We have
\begin{align*}
\frac{\mathrm{d}\Lambda_{1/\beta}(X)}{\mathrm{d}\beta} & =\frac{\mathrm{d}}{\mathrm{d}\beta}\frac{1}{\beta-1}\log\sum_{i=1}^{\infty}p^{\downarrow}(i)\big(i^{\beta}-(i-1)^{\beta}\big)\\
 & =\frac{\mathrm{d}}{\mathrm{d}\beta}\frac{1}{\beta-1}\log\sum_{i=1}^{\infty}i^{\beta}(p^{\downarrow}(i)-p^{\downarrow}(i+1))\\
 & =\frac{1}{(\beta-1)^{2}}\Bigg(\!\frac{\sum_{i=1}^{\infty}i^{\beta}(p^{\downarrow}(i)-p^{\downarrow}(i+1))\log i^{\beta-1}}{\sum_{i=1}^{\infty}i^{\beta}(p^{\downarrow}(i)-p^{\downarrow}(i+1))}\\
 & \quad\quad\quad\quad\quad\quad-\log\sum_{i=1}^{\infty}i^{\beta}(p^{\downarrow}(i)-p^{\downarrow}(i+1))\Bigg)\\
 & =\frac{1}{(\beta-1)^{2}}\sum_{i=1}^{\infty}p_{\beta}(i)\log\frac{p_{\beta}(i)}{q(i)}\\
 & =\frac{1}{(\beta-1)^{2}}D_{\mathrm{KL}}(p_{\beta}\Vert q)\;\ge0,
\end{align*}
where
\[
p_{\beta}(x):=\frac{x^{\beta}(p^{\downarrow}(x)-p^{\downarrow}(x+1))}{\sum_{i=1}^{\infty}i^{\beta}(p^{\downarrow}(i)-p^{\downarrow}(i+1))},
\]
\[
q(x):=x(p^{\downarrow}(x)-p^{\downarrow}(x+1)).
\]

We then prove $\Lambda_{\alpha}(X)\le H_{\alpha}(X)$. We already have $\Lambda_{1}(X)\le H_{1}(X)$ in Proposition \ref{prop:basic}. We first prove $\Lambda_{\alpha}(X)\le H_{\alpha}(X)$ for $\alpha<1$. It is equivalent to
\begin{equation}
\sum_{i=1}^{\infty}p^{\downarrow}(i)\big(i^{1/\alpha}-(i-1)^{1/\alpha}\big)\le\Big(\sum_{i=1}^{\infty}(p^{\downarrow}(i))^{\alpha}\Big)^{1/\alpha}.\label{eq:pf_renyi_ineq}
\end{equation}
Note that any non-increasing probability mass function (pmf) over $\mathbb{N}$ is a convex combination of $\mathrm{Unif}(\{1,\ldots,k\})$ for $k\ge1$. Also, the right-hand side of \eqref{eq:pf_renyi_ineq} is concave in $p^{\downarrow}$. Hence, it suffices to verify \eqref{eq:pf_renyi_ineq} when $p$ is the pmf of $\mathrm{Unif}(\{1,\ldots,k\})$, where \eqref{eq:pf_renyi_ineq} holds since both sides are $k^{1/\alpha-1}$. Then, we prove $\Lambda_{\alpha}(X)\le H_{\alpha}(X)$ for $\alpha>1$. It is equivalent to
\begin{equation}
\sum_{i=1}^{\infty}p^{\downarrow}(i)\big(i^{1/\alpha}-(i-1)^{1/\alpha}\big)\ge\Big(\sum_{i=1}^{\infty}(p^{\downarrow}(i))^{\alpha}\Big)^{1/\alpha}.\label{eq:pf_renyi_ineq-1}
\end{equation}
The same arguments for \eqref{eq:pf_renyi_ineq} hold, except now the right hand side is convex in $p^{\downarrow}$, so the inequality is flipped. The remaining properties follow from direct computation.

\smallskip{}

\subsection{Proof of Theorem \ref{thm:l_sfrl}\label{subsec:pf_l_sfrl}}

We now prove Theorem \ref{thm:l_sfrl}: $\Lambda_{\mathrm{n}}^{*}\le I(X;Y) +\Lambda(\mathrm{Geom}(1/2))$.
We invoke a result in \cite{li2021unified} (also see \cite[Lemma 12]{li2024channel}): for every $X,Y$, there exists $S,K$ such that $S$ is independent of $X$, $H(K|X,S)=H(Y|K,S)=0$, and $K$ is conditionally a geometric random variable given $(X,Y)$:
\begin{align}
 & K|\{(X,Y)=(x,y)\}\sim\mathrm{Geom}(\rho(x,y)),\label{eq:sfrl_geom}
\end{align}
where
\begin{align*}
\rho(x,y) & :=\Big(\mathbb{E}_{Y'\sim P_{Y}}\Big[\max\big\{2^{\iota_{X;Y}(x;y)},\,2^{\iota_{X;Y}(x;Y')}\big\}\Big]\Big)^{-1}\\
 & \ge(2^{\iota_{X;Y}(x;y)}+1)^{-1},
\end{align*}
where $\iota_{X;Y}(x;y):=\log\frac{\mathrm{d}P_{Y|X}(\cdot|x)}{\mathrm{d}P_{Y}}(y)$ is the information density. Let 
\[
\phi(a):=\Lambda(\mathrm{Geom}(1/a)).
\]
Since the geometric distribution $\mathrm{Geom}(1/a)$ first-order stochastically dominates $\mathrm{Geom}(1/a')$ for $a>a'$, $\phi(a)$ is nondecreasing. By the same arguments as Appendix \ref{subsec:pf_l_sfrl},
\begin{align*}
\Lambda(Y|S) & \stackrel{(a)}{\le}\Lambda(K|S)\\
 & \stackrel{(b)}{\le}\Lambda(K)\\
 & \stackrel{(c)}{=}\Lambda(K|X,Y)\\
 & =\mathbb{E}\left[\phi(\mathbb{E}[K|X,Y])\right]\\
 & \le\mathbb{E}\big[\phi(2^{\iota_{X;Y}(X;Y)}+1)\big]\\
 & =I(X;Y)+\mathbb{E}\big[\phi(2^{\iota_{X;Y}(X;Y)}+1)-\iota_{X;Y}(X;Y)\big]\\
 & =I(X;Y)+\mathbb{E}\big[g(2^{-\iota_{X;Y}(X;Y)})\big],
\end{align*}
where (a) is because $H(Y|K,S)=0$ and the Schur concavity of $\Lambda$ (Proposition \ref{prop:basic}), (b) is by the concavity of $\Lambda$ (Proposition \ref{prop:basic}), (c) is by monotone linearity (Proposition \ref{prop:basic}) since $p_{K|X,Y}(\cdot|x,y)$ is always nonincreasing, and
\begin{align*}
g(t) & :=\phi(1/t+1)+\log t=\Lambda\Big(\mathrm{Geom}\Big(\frac{t}{t+1}\Big)\Big)+\log t
\end{align*}
for $t>0$. If there are $a,b$ such that
\begin{equation}
g(x)\le a+(t-1)b\label{eq:g_linear_bd}
\end{equation}
for $t>0$, then
\begin{align}
\Lambda(Y|S) & \le I(X;Y)+\mathbb{E}\big[g(2^{-\iota_{X;Y}(X;Y)})\big]\nonumber \\
 & \le I(X;Y)+a+b\mathbb{E}\big[2^{-\iota_{X;Y}(X;Y)}-1\big]\nonumber \\
 & \le I(X;Y)+a,\label{eq:LYS_bd_a}
\end{align}
since $\mathbb{E}[2^{-\iota_{X;Y}(X;Y)}]\le1$ (note that $b>0$ since $\lim_{t\to\infty}g(t)=\infty$). By Proposition \ref{prop:LX_HX_bound}, for every $\eta>0$, 
\[
H(Y\backslash S)\le I+\log\left(I+e\eta+a\right)-\log(e\eta)+\eta+a.
\]
We now discuss different choices of $a,b$ in \eqref{eq:g_linear_bd}.

\smallskip{}

\subsubsection{$a=\log3<1.59$ }

We first discuss a simple bound that can be proved using the discrete R\'{e}nyi layered entropy. By Proposition \ref{prop:renyi},
\begin{align}
g(t) & =\Lambda\Big(\mathrm{Geom}\Big(\frac{t}{t+1}\Big)\Big)+\log t\nonumber \\
 & \le\Lambda_{1/2}\Big(\mathrm{Geom}\Big(\frac{t}{t+1}\Big)\Big)+\log t\nonumber \\
 & \le\log\Big(2\cdot\frac{t+1}{t}-1\Big)+\log t\nonumber \\
 & =\log\left(t+2\right)\label{eq:gt_ub_simple}\\
 & \le\log3+(t-1)(\log e)/3.\nonumber 
\end{align}
By \eqref{eq:LYS_bd_a}, $\Lambda(Y|S)\le I(X;Y)+\log3$. This simple bound is already enough to improve upon the strong functional representation lemma in \cite{sfrl_trans,li2021unified,li2024pointwise} for $I\ge2$. 

\smallskip{}

\subsubsection{$a=\Lambda(\mathrm{Geom}(1/2))<1.29$}

We now prove a stronger bound:
\begin{equation}
g(t)\le\Lambda(\mathrm{Geom}(1/2))+(t-1)g'(1).\label{eq:conj_assume-1}
\end{equation}
This holds if $g$ is concave. Together with \eqref{eq:LYS_bd_a} this would imply $\Lambda(Y|S)\le I(X;Y)+\Lambda(\mathrm{Geom}(1/2))$. A simple plot suggests that \eqref{eq:conj_assume-1} indeed holds, and $g$ is indeed concave. For the sake of mathematical rigor, we now prove \eqref{eq:conj_assume-1} in a rather mechanical manner. We consider four cases of $t$: 

Case 1: $t\in[0.3,4]\backslash[0.975,1.025]$. Let $p:\mathbb{N}\to[0,1]$ be the probability mass function of $\mathrm{Geom}(z)$. Let $m\in\mathbb{N}$. Let
\begin{align*}
\gamma_{i} & :=(p(i)-p(i+1))i=z^{2}(1-z)^{i-1}i.
\end{align*}
 We express $p$ as the mixture
\begin{align*}
p(x) & =\sum_{i=1}^{m}\gamma_{i}\bar{p}_{i}(x)+\left(1-\sum_{i=1}^{m}\gamma_{i}\right)\tilde{p}_{m}(x),
\end{align*}
where $\bar{p}_{i}$ is the probability mass function of $\mathrm{Unif}(\{1,\ldots,i\})$, and
\[
\tilde{p}_{m}(x)=\frac{p(\max\{x,m+1\})}{1-\sum_{i=1}^{m}\gamma_{i}}.
\]
Since $\tilde{p}$ is non-increasing, by the monotone linearity of $\Lambda$ and Proposition \ref{prop:renyi},
\begin{align*}
 & \Lambda(\mathrm{Geom}(z))\\
 & =\sum_{i=1}^{m}\gamma_{i}\log i+\left(1-\sum_{i=1}^{m}\gamma_{i}\right)\Lambda(\tilde{p}_{m})\\
 & \le\sum_{i=1}^{m}\gamma_{i}\log i+\left(1-\sum_{i=1}^{m}\gamma_{i}\right)\log\left(2\mathbb{E}_{X\sim\tilde{p}_{m}}[X]-1\right)\\
 & \le\sum_{i=1}^{m}\gamma_{i}\log i+\Big(1-\sum_{i=1}^{m}\gamma_{i}\Big)\log\left(2\mathbb{E}_{X\sim p}[X|X\ge m+1]-1\right)\\
 & =\sum_{i=1}^{m}\gamma_{i}\log i+\left(1-\sum_{i=1}^{m}\gamma_{i}\right)\log\left(2z^{-1}+2m-1\right).
\end{align*}
Hence, taking $z=t/(t+1)$,
\begin{align}
g(t) & =\Lambda\left(\mathrm{Geom}\left(\frac{t}{t+1}\right)\right)+\log t\nonumber \\
 & \le\sum_{i=1}^{m}\gamma_{i}\log(it)+\left(1-\sum_{i=1}^{m}\gamma_{i}\right)\left(\log\left((m+\frac{1}{2})t+1\right)+1\right).\label{eq:gt_bd_rational}
\end{align}
We then lower-bound the right-hand side of \eqref{eq:conj_assume-1}. We have $\Lambda(\mathrm{Geom}(1/2))\ge\sum_{i=1}^{30}\gamma_{i}\log i$. Let 
\begin{align*}
\beta_{i} & :=i\log i-(i-1)\log(i-1)=\int_{i-1}^{i}\log(e\tau)\mathrm{d}\tau.
\end{align*}
To upper-bound $g'(1)$, by direct evaluation, for $m\ge2$,
\begin{align}
\frac{\mathrm{d}}{\mathrm{d}t}\Lambda\left(\mathrm{Geom}\left(\frac{t}{t+1}\right)\right)\Big|_{t=1} & =\sum_{i=2}^{\infty}2^{-i-1}\left(2-i\right)\beta_{i}\nonumber \\
 & \le\sum_{i=2}^{m}2^{-i-1}\left(2-i\right)\beta_{i}.\label{eq:dL_ub}
\end{align}
To lower-bound $g'(1)$,
\begin{align}
 & \frac{\mathrm{d}}{\mathrm{d}t}\Lambda\left(\mathrm{Geom}\left(\frac{t}{t+1}\right)\right)\Big|_{t=1}\nonumber \\
 & =\sum_{i=2}^{\infty}2^{-i-1}\left(2-i\right)\beta_{i}\nonumber \\
 & \ge\sum_{i=2}^{m}2^{-i-1}\left(2-i\right)\beta_{i}+\sum_{i=m+1}^{\infty}2^{-i-1}\left(2-i\right)i\nonumber \\
 & =\sum_{i=2}^{m}2^{-i-1}\left(2-i\right)\beta_{i}-2^{-m-1}(m^{2}+2m+2).\label{eq:dL_lb}
\end{align}
Hence, the right-hand side of \eqref{eq:conj_assume-1} can be lower-bound via \eqref{eq:dL_ub} for $t\le1$, or via \eqref{eq:dL_lb} for $t>1$. At this point, it might be sufficient to simply plot \eqref{eq:gt_bd_rational} for a fixed $m$ (e.g., $m=18$) together with the above lower bound of the right-hand side of \eqref{eq:conj_assume-1} to verify that \eqref{eq:conj_assume-1} indeed holds over this range of $t$. For $m=18$, there are only $18$ terms in the summation, so floating-point inaccuracy is negligible. Nevertheless, it is more appropriate to prove this rigorously using exact rational arithmetic, eliminating the possibility of floating-point error. Our goal is to upper-bound \eqref{eq:gt_bd_rational} by a rational function of $t$. To this end, we use the continued fraction bound in \cite{khovanskii1963application}:
\begin{align}
\ln(1+x)\le & \psi_{k}(x):=x/(\alpha_{1}+1x/(2+1x/(\alpha_{2}+2x/(2\nonumber \\
 & \;+2x/(\alpha_{3}+3x/(2+3x/(\cdots/\alpha_{k}\cdots)))))))\label{eq:ln_cf}
\end{align}
where $k\in\mathbb{N}$, $\alpha_{i}:=2i-1$. We then have
\begin{align}
g(t) & \le\frac{-1}{\psi_{k}(-1/2)}\bigg(\sum_{i=1}^{m}\gamma_{i}\psi_{k}(it-1)\nonumber \\
 & \;\;\;+\left(1-\sum_{i=1}^{m}\gamma_{i}\right)\left(\psi_{k}\left((m+\frac{1}{2})t\right)+1\right)\bigg),\label{eq:gt_bd_rational2}
\end{align}
which is a rational function of $t$. We can now prove that \eqref{eq:conj_assume-1} holds over $t\in[0.3,4]\backslash[0.975,1.025]$ by verifying that the difference between \eqref{eq:gt_bd_rational2} (with $k=5$, $m=18$) and the right-hand side of \eqref{eq:conj_assume-1} (lower bound via \eqref{eq:dL_ub} or \eqref{eq:dL_lb}, bounded via \eqref{eq:ln_cf} with $k=8$, $m=20$) has no zeros in the range $t\in[0.3,4]\backslash[0.975,1.025]$. This can be performed algorithmically and rigorously via Sturm's theorem (which checks whether a polynomial has a root over an interval). This was proved using SymPy \cite{meurer2017sympy}.\footnote{The Python code for proving \eqref{eq:conj_assume-1} using exact rational arithmetic is available at: https://github.com/cheuktingli/DiscreteLayeredEntropy}

Case 2: $t\ge4$. In this range, \eqref{eq:conj_assume-1} is implied by \eqref{eq:gt_ub_simple} via \eqref{eq:dL_lb}.

Case 3: $t\in(0,0.3)$. By Jensen's inequality, for $i\ge2$,
\begin{align*}
 & \int_{i-1}^{i}\left(1-z\right)^{\tau-i+1}\log(e\tau)\mathrm{d}\tau\\
 & \ge\beta_{i}\left(1-z\right)^{\beta_{i}^{-1}\int_{i-1}^{i}(\tau-i+1)\log(e\tau)\mathrm{d}\tau}\\
 & \ge\beta_{i}\left(1-z\right)^{\beta_{2}^{-1}\int_{1}^{2}(\tau-1)\log(e\tau)\mathrm{d}\tau}\;\ge\;\beta_{i}\left(1-z\right)^{0.542}.
\end{align*}
Hence,
\begin{align*}
 & \frac{(1-z)^{0.542}}{z}\Lambda(\mathrm{Geom}(z))\\
 & =\sum_{i=2}^{\infty}\beta_{i}\left(1-z\right)^{0.542+i-1}\\
 & \le\int_{1}^{\infty}\left(1-z\right)^{\tau}\log(e\tau)\mathrm{d}\tau\\
 & \le\int_{0}^{\infty}\left(1-z\right)^{\tau}\log(e\tau)\mathrm{d}\tau\\
 & \;\;\;+\int_{0}^{1/e}\left(1-z\right)^{\tau}\log\frac{1}{e\tau}\mathrm{d}\tau-\int_{1/e}^{1}\left(1-z\right)^{\tau}\log(e\tau)\mathrm{d}\tau\\
 & \le\int_{0}^{\infty}\left(1-z\right)^{\tau}\log(e\tau)\mathrm{d}\tau\\
 & \;\;\;+\int_{0}^{1/e}\log\frac{1}{e\tau}\mathrm{d}\tau-\left(1-z\right)\int_{1/e}^{1}\log(e\tau)\mathrm{d}\tau\\
 & =\int_{0}^{\infty}\left(1-z\right)^{\tau}\log(e\tau)\mathrm{d}\tau+\frac{z\log e}{e}\\
 & =\frac{1}{\ln(1/(1-z))}\left(\int_{0}^{\infty}e^{-\tau}\log(e\tau)\mathrm{d}\tau-\log\ln\frac{1}{1-z}\right)+\frac{z\log e}{e}\\
 & \le\frac{0.61-\log\ln\frac{1}{1-z}}{\ln(1/(1-z))}+\frac{z\log e}{e}.
\end{align*}
For $t\le0.3$, letting $z=t/(t+1)\le0.231$, we have $(1-z)^{0.542}\ge 1 - 0.58 z$. Hence,
\begin{align}
g(t) & \le\frac{1}{1 - 0.58 z}\left(\frac{0.61-\log\ln\frac{1}{1-z}}{z^{-1}\ln(1/(1-z))}+\frac{z^{2}\log e}{e}\right)+\log\frac{z}{1-z}\label{eq:g_ub_ln}\\
 & \le \frac{1}{1 - 0.58 z}\left(0.61+\frac{z^{2}\log e}{e}\right)-0.67z\log z-\log(1-z).\label{eq:g_ub_ln2}
\end{align}
For $t\le0.021$, \eqref{eq:conj_assume-1} is implied by (\ref{eq:g_ub_ln2}) and $-0.67z\log z\le0.078$ for $t\le0.021$. For $0.021<t<0.3$, \eqref{eq:conj_assume-1} is implied by (\ref{eq:g_ub_ln}). This can be proved algorithmically in a similar manner as the previous case. 


Case 4: $t\in[0.975,1.025]$. This case is difficult since equality is attained in \eqref{eq:conj_assume-1} when $t=1$, so there is no room for approximation error. The strategy is to show $g(t)$ is concave in this range, which implies \eqref{eq:conj_assume-1}. We consider the second derivative of $g(t)$. Since $\beta_{i}\le i$, we have
\begin{align*}
 & \frac{\mathrm{d}^{2}}{\mathrm{d}t^{2}}\Lambda\left(\mathrm{Geom}\left(\frac{t}{t+1}\right)\right)\\
 & =\sum_{i=2}^{\infty}\left(\frac{1}{t+1}\right)^{i+2}i\left((i-1)t-2\right)\beta_{i}\\
 & \le\sum_{i=2}^{m}\left(\frac{1}{t+1}\right)^{i+2}i\left((i-1)t-2\right)\beta_{i}\\
 & \;\;\;+\sum_{i=m+1}^{\infty}\left(\frac{1}{1.8}\right)^{i+2}i^{2}\left((i-1)1.2-2\right)\\
 & =\sum_{i=2}^{m}\left(\frac{1}{t+1}\right)^{i+2}i\left((i-1)t-2\right)\beta_{i}\\
 & \;\;\;+2^{-6}\left(\frac{9}{5}\right)^{-m-2}\left(96m^{3}+392m^{2}+1116m+1845\right).
\end{align*}
Again using the bound \eqref{eq:ln_cf} with $k=14$, $m=70$, we can show that $\mathrm{d}^{2}g(t)/\mathrm{d}t^{2}\le-0.013$ for $t\in[0.975,1.025]$ via exact rational arithmetic.\footnote{The code is available at: https://github.com/cheuktingli/DiscreteLayeredEntropy} This completes the proof.

\subsection{Proof of Theorem \ref{thm:sfrl_n_lb}\label{subsec:pf_sfrl_n_lb}}

Let $\beta_{i}:=i\log i-(i-1)\log(i-1)$. We have $\gamma(t)=\int_{0}^{t}\beta_{\lceil\tau\rceil}\mathrm{d}\tau$. Fix any $S$ independent of $X$, and let $Y=f(S,X)$. By Proposition \ref{prop:cond_pmf},
\[
\Lambda(Y|S)=\sum_{i=1}^{\infty}\mathbb{E}\left[p_{Y|S}^{\downarrow}(i|S)\right]\beta_{i}.
\]
Fix any $m\in\mathbb{N}$. Let $\mathcal{A}_{s}\subseteq\mathcal{Y}$, $|\mathcal{A}_{s}|=m$ be the set of $y$'s with the $m$ largest $p_{Y|S}(y|s)$'s. We have
\begin{align*}
 & \sum_{i=1}^{m}\mathbb{E}\left[p_{Y|S}^{\downarrow}(i|S)\right]\\
 & =\mathbb{E}\Big[\sum_{y\in\mathcal{A}_{S}}p_{Y|S}(y|S)\Big]\\
 & =\sum_{y\in\mathcal{Y}}\mathbb{E}\left[\mathbf{1}\{y\in\mathcal{A}_{S}\}p_{Y|S}(y|S)\right]\\
 & \stackrel{(a)}{=}\sum_{y\in\mathcal{Y}}\mathbb{P}\left(y\in\mathcal{A}_{S},\,f(S,X)=y\right)\\
 & =\sum_{y\in\mathcal{Y}}\mathbb{E}\left[\mathbb{P}\left(y\in\mathcal{A}_{S},\,f(S,X)=y\,\big|\,X\right)\right]\\
 & =\sum_{y\in\mathcal{Y}}\mathbb{E}\left[\min\left\{ \mathbb{P}(y\in\mathcal{A}_{S}),\,p_{Y|X}(y|X)\right\} \right]\\
 & \stackrel{(b)}{=} \sum_{y\in\mathcal{Y}}\int_{0}^{1}\min\{\mathbb{P}(y\in\mathcal{A}_{S}),\,p_{Y}(y)Q_{y}(\tau)\}\mathrm{d}\tau\\
 & \le\int_{0}^{1}\min\Big\{\sum_{y\in\mathcal{Y}}\mathbb{P}(y\in\mathcal{A}_{S}),\,\sum_{y\in\mathcal{Y}}p_{Y}(y)Q_{y}(\tau)\Big\}\mathrm{d}\tau\\
 & =\int_{0}^{1}\min\left\{ m,\mathbb{E}[Q_{Y}(\tau)]\right\} \mathrm{d}\tau,
\end{align*}
where (a) is because $Y=f(S,X)$ and $S \perp\!\!\! \perp X$, (b) is because $Q_y(\tau)$ is the quantile (inverse cdf) function of $p_{Y|X}(y|X)/p_Y(y)$.
Hence,
\begin{align*}
 & \Lambda(Y|S)\\
 & =\sum_{i=1}^{\infty}\mathbb{E}\left[p_{Y|S}^{\downarrow}(i|S)\right]\beta_{i}\\
 & =\sum_{m=1}^{\infty}\left(1-\sum_{i=1}^{m}\mathbb{E}\left[p_{Y|S}^{\downarrow}(i|S)\right]\right)\left(\beta_{m+1}-\beta_{m}\right)\\
 & \ge\sum_{m=1}^{\infty}\left(1-\int_{0}^{1}\min\left\{ m,\mathbb{E}[Q_{Y}(\tau)]\right\} \mathrm{d}\tau\!\right)\left(\beta_{m+1}-\beta_{m}\right)\\
 & \stackrel{(c)}{=}\sum_{m=1}^{\infty}\left(\int_{0}^{1}\max\left\{ \mathbb{E}[Q_{Y}(\tau)]-m,0\right\} \mathrm{d}\tau\!\right)\left(\beta_{m+1}-\beta_{m}\right)\\
 & \stackrel{(d)}{=}\int_{0}^{1}\int_{0}^{\mathbb{E}[Q_{Y}(\tau)]}\beta_{\lceil t\rceil}\mathrm{d}t\mathrm{d}\tau\;=\;\int_{0}^{1}\gamma(\mathbb{E}[Q_{Y}(\tau)])\mathrm{d}\tau,
\end{align*}
where (c) is because $\int_{0}^{1}\mathbb{E}[Q_{Y}(\tau)]\mathrm{d}\tau = \sum_y p_Y(y) \int_{0}^{1} Q_{y}(\tau)\mathrm{d}\tau  = \sum_y p_Y(y) \mathbb{E}[p_{Y|X}(y|X)/p_Y(y)] = 1$, and (d) is by summation by parts. 

\smallskip{}

\bibliographystyle{IEEEtran}
\bibliography{ref}

\end{document}